\documentclass[useAMS,usenatbib]{mn2e}
\usepackage[dvips]{graphicx}
\usepackage{setspace}
\usepackage{multirow}
\usepackage{color}
\usepackage{amssymb}
\usepackage{amsmath}

\newcommand{\apj}{ApJ}
\newcommand{\apjl}{ApJL}
\newcommand{\mnras}{MNRAS}
\newcommand{\aj}{AJ}
\newcommand{\apjs}{ApJS}
\newcommand{\nat}{Nature}
\newcommand{\araa}{ARA\&A}
\newcommand{\aap}{A\&A}

\newcommand{\pasp}{PASP}

\title[Hot- and cold-mode
    accretion in radio galaxies at $z\sim 1$]{Black-hole masses, accretion rates and hot- and cold-mode
    accretion in radio galaxies at $z\sim 1$} \author[C. A. C. Fernandes et
  al. ]{C. A. C. Fernandes$^{1,2,3,4}$\thanks{E-mail: cfernandes@on.br},
  M. J. Jarvis$^{4,5}$, A. Mart\'{i}nez-Sansigre$^{6}$,
  S. Rawlings$^{4}$,\newauthor J. Afonso$^{2,3}$,
  M.J.~Hardcastle$^{7}$, M. Lacy$^{8}$, J. A. Stevens$^{7}$, and
  E. Vardoulaki$^{4,9}$ \\ 
$^{1}$Observat\'orio Nacional, Rua General Jos\'e Cristino, 77, 20921-400, Rio de Janeiro -RJ, Brazil\\
$^{2}$Observat\'orio Astron\'omico de Lisboa,
  Faculdade de Ci\^encias, Universidade de Lisboa, Tapada da Ajuda,
  1349-018 Lisbon, Portugal \\ 
$^{3}$Centro de Astronomia e
  Astrof\'isica da Universidade de Lisboa, Lisbon, Portugal
  \\ 
$^{4}$University of Oxford, Subdepartment of Astrophysics, Denys
  Wilkinson Building, Keble Road, Oxford OX1 2DL, UK\\ 
$^{5}$Department of Physics, University of the Western Cape, Private
Bag X17, Bellville 7535, South Africa\\
$^{6}$Institute of Cosmology and Gravitation, University of Portsmouth, Dennis
  Sciama Building, Burnaby Road, Portsmouth, PO1 3FX,
  UK\\ 
$^{7}$Centre for Astrophysics Research, STRI, University of
  Hertfordshire,Hatfield, AL10 9AB, UK\\ 
$^{8}$National Radio
  Astronomy Observatory, 520 Edgemont Road, Charlottesville, VA 22903,
  USA\\
$^{9}$Department of Physics, University of Crete, GR-71003, Heraklion, Greece}

\begin{document}

\date{}

\pagerange{\pageref{firstpage}--\pageref{lastpage}} \pubyear{}

\maketitle

\label{firstpage}

\begin{abstract}
Understanding the evolution of accretion activity is fundamental to
our understanding of how galaxies form and evolve over the history of
the Universe.  We analyse a complete sample of 27 radio galaxies which
includes both high-excitation (HEGs) and low excitation galaxies
(LEGs), spanning a narrow redshift range of $0.9< z < 1.1$ and
covering a factor of $\sim 1000$ in radio luminosity. Using data from
the {\em Spitzer Space Telescope} combined with ground-based optical
and near-infrared imaging, we show that the host galaxies have masses
in the range of $10.7 < \log_{10}(M / M_{\odot}) < 12.0$ with HEGs and
LEGs exhibiting no difference in their mass distributions.  We also
find that HEGs accrete at significantly higher rates than LEGs, with
the HEG/LEG division lying at an Eddington ratio of $\lambda \sim
0.04$, which is in excellent agreement with theoretical predictions of
where the accretion rate becomes radiatively inefficient, thus
supporting the idea of HEGs and LEGs being powered by different modes
of accretion. Our study also shows that at least up to $L_{\rm
  151MHz}\sim 3\times 10^{27}\rm \,W\,Hz^{-1}\,sr^{-1}$, HEGs and LEGs
are indistinguishable in terms of their radio properties. From this
result we infer that, at least for the lower radio luminosity range,
another factor besides accretion rate must play an important role in
the process of triggering jet activity.

\end{abstract}


\section{Introduction}

It is now widely accepted that black holes reside in the centre of
nearly all galaxies
\citep[e.g.][]{1996gfa..book.....B,1998Natur.395A..14R,
  1998AJ....115.2285M}. Accretion onto these black holes is believed
to be the mechanism responsible for powering Active Galactic Nuclei
\citep[AGN; e.g.][]{1993ARA&A..31..473A,1995PASP..107..803U}.

The black holes in these galaxies can accrete matter, either from the
interstellar medium close to the event horizon or from thick disks of
gas. In order to be accreted, gas needs to fall in with an angular
momentum that does not exceed that needed for disk capture. In the
vicinity of the sphere of influence of the black hole, the Keplerian
motion of the accreting material is exceedingly high, thus processes
that redistribute the angular momentum of the accreting gas determine
the accretion rate. Galactic mergers, gravitational torques from bars
and other structures can efficiently channel gas into the nuclear
regions
\citep[e.g.][]{1989Natur.338...45S,1990Natur.345..679S,1995ApJ...448...41H}. Closer
to the sphere of influence of the black hole, however, other
mechanisms, such as turbulence driven by magnetic instabilities, are
efficient sources of angular momentum transport
\citep[e.g.][]{1995ApJ...445..767M,1996ApJ...463..656S}.

There are different modes of accretion. The `standard' accretion mode
occurs when matter is accreted onto the black hole through a
radiatively efficient process. This is generally associated with
optically thick and geometrically thin accretion disks
\citep[e.g.][]{1973A&A....24..337S}, and thought to be associated with
quasar activity. This process is likely to be prevalent where there is
a plentiful supply of cold gas, which can accrete efficiently towards
the black hole, and as such the fuel, in the form of the cold gas,
should also be able to condense and provide the fuel for star
formation \citep[e.g.][]{2007MNRAS.376.1849H}. Indirect evidence for
such a scenario comes from the similar evolution of the AGN activity
in the Universe and the star-formation rate density, both increasing
rapidly from $z\sim 0 \rightarrow 1$ and steadily turning over at
$z>3$ \citep[e.g.][]{2009ApJ...696..396S}. Furthermore, such a close
connection between AGN and star-formation activity may provide a
relatively straightforward explanation of the local correlation
between galaxy mass and black-hole mass \citep{1998AJ....115.2285M,2000ApJ...539L...9F,2000ApJ...539L..13G,2004ApJ...604L..89H}.

Accretion in an AGN may also be radiatively inefficient. When mass
accretion is too low ($\dot{M}\ll \dot{M}_{\rm Edd}$), the inflowing
gas may not be able to radiate the gravitational potential energy
liberated due to accretion. Magnetohydrodynamic simulations show that
the energy dissipated during accretion could heat up only the ions in
the disk
\citep[e.g.][]{1999ApJ...520..248Q,2000ApJ...541..811M,2002ApJ...577..524Q}. On
the other hand, radiative losses, synchrotron radiation, and
Comptonization of low-energy photons take energy from the electrons,
cooling them. The characteristic time-scale for the hot ions and the
cold electrons to achieve thermal equilibrium through coupling depends
on the gas temperature in the vicinity of the black hole, and is
inversely proportional to the number density of ions. When the
accretion rate is low, the ionic density is also small, and this
time-scale is of the same order as, or larger than, the inflow
time. Thus a two-temperature plasma is expected to form when
$\dot{M}\ll \dot{M}_{\rm Edd}$. In this regime, advection dominated
accretion flows (ADAFs) can be formed. This radiatively inefficient
process is currently taken as the most likely explanation for
accretion flows around black holes at low accretion rates. The
critical accretion rate below which it becomes radiatively inefficient
is estimated to be $\dot{M}_{\rm crit}\simeq \alpha^{2}\dot{M}_{\rm
  Edd}$ \citep{1995ApJ...452..710N}, where $\alpha$ is a dimensionless
parameter that measures the efficiency of angular momentum transport
in disks. With the standard value of $\alpha=0.25$
\citep[e.g.][]{1997ApJ...489..865E}, the critical accretion rate is
situated around $\dot{M}_{\rm crit}\sim0.06$ normalised to the
Eddington accretion rate.  Due to the inability of the gas to cool
down radiatively in this regime, the gas can be driven to a higher
temperature, and thus cause a vertical thickening of the disk. ADAFs
are thus thought to be characteristic of hot thick disks.

What causes the different modes of radiatively efficient and
inefficient accretion remains uncertain. It has been
suggested that the nature of the gas being accreted might determine
the type of accretion, with cold gas producing a stable accretion
disk and thus a radiatively efficient accretion, and hot gas producing
flows that would result in an ADAF
\citep[e.g.][]{2007MNRAS.376.1849H}. The black hole spin has also been
suggested to play a role in determining the accretion mode
\citep[e.g.][]{2011MNRAS.414.1937M}.


Radio galaxies are a class of AGN that can be divided into two
sub-categories according to the ratio between the intensity of high
and low excitation emission lines in their optical spectra: low
excitation galaxies (LEGs) and high excitation galaxies (HEGs). The
two distinct classes were first noted by \citet{1979MNRAS.188..111H}
and further categorised by \citet{1994ASPC...54..201L} and
\citet{1997MNRAS.286..241J}. \citet{1997MNRAS.286..241J} defined LEGs
as objects that obey the following requirements: have [O{\sc
    iii}]-line equivalent widths $< 10$\AA, have a ratio $\rm [O{\sc
    II}]/[O{\sc III}]>1$, or both.


It is widely accepted that LEGs could be the result of radiatively
inefficient accretion processes
\citep[e.g.][]{2007MNRAS.376.1849H,2010A&A...509A...6B,2014MNRAS.440..269M}. \citet{2012MNRAS.421.1569B}
compared a large number of local radio-loud AGN and concluded that the
HEG and LEG populations show different accretion rate distributions,
consistent with the idea that radiative efficiency is determined
purely on accretion rate.

This is a subject that impacts on the most
relevant issues for models of AGN and their roles in galaxy evolution. For instance, if the HEG/LEG
distinction is indeed related to the accretion process, then the fact
that their radio properties are similar requires that the
power being channeled into the jets is independent of the accretion rate. 

Moreover, given that radiatively efficient accretion is related to the
`quasar mode' form of feedback processes in AGN, and radiatively
inefficient accretion associated with `radio mode' AGN feedback in
semi-analytic models
\citep[e.g.][]{Croton2006}, the HEG and LEG characterisation could be used to
diagnose the different modes at play in each system. LEGs would for
instance be the perfect laboratories to investigate the `radio mode'
form of feedback, currently taken as being the principal mechanism
responsible for shutting down star formation in the most massive
systems at $z<1$.

In this paper we investigate the HEG/LEG division in radio-loud
sources and their different accretion processes, extending the study
of \citet{2012MNRAS.421.1569B} to high-redshift sources
($z\sim1$). Moving to high redshift is not only complementary in terms
of looking back time but also in terms of luminosity as it provides us
with the opportunity to study the rarer more luminous objects in each
class.

Section~\ref{sec:data} describes the data
selection. Section~\ref{sec:sed_method} describes the SED fitting
method used. In Section~\ref{sec:sed_fits} and \ref{sec:ind} the SED
fits are shown and the physical values extracted from them are
presented. The results are discussed in Section~\ref{sec:discuss}, and
conclusions are presented in Section~\ref{sec:conc}. Throughout this
paper we adopt the following values for the cosmological parameters:
$\rm H_0=70\,km\,s^{-1}\,Mpc^{-1}$, $\rm \Omega_M=0.3$ and $\rm
\Omega_\Lambda=0.7$.



\section{Data}\label{sec:data}

We use {\em Spitzer Space Telescope} observations and multiple data
collected from the literature, ranging from mid-infrared to optical
wavelengths, of the sample of radio galaxies presented by
\citet{2011MNRAS.411.1909F}. This sample consists of 27 radio sources
selected from the spectroscopically complete 3CRR
\citep{1983MNRAS.204..151L}, 6CE
\citep{1997MNRAS.291..593E,2001MNRAS.322..523R}, 6C*
\citep{2001MNRAS.326.1563J,2001MNRAS.326.1585J}, 7CRS
\citep{1999MNRAS.308.1096L,2003MNRAS.339..173W} and TOOT
\citep{HillRawlings2003,2010MNRAS.401.1709V} surveys to have narrow
lines and a redshift of $0.9\lesssim z\lesssim 1.1$. The narrow-line
selection was intended to ensure the exclusion of quasars from the
sample (though 3C343 has since been classified as a
quasar and 3C22 is classified as a weak quasar). \\

\subsection{Spitzer data}

The {\em Spitzer} data consist of observations with MIPS $24\,\mu \rm
m$ and IRAC $3.6$, $4.5$, $5.8$ and $8.0\, \mu \rm m$ bands. The MIPS
and IRAC observations took place in August 2006 and August 2007, as
described by \citet{2011MNRAS.411.1909F}. The photometric measurements for the IRAC
data were performed using a 7~arcsec diameter aperture with aperture
corrections of 1.112, 1.113, 1.125 and 1.218 for IRAC channels 1, 2, 3 and
4 respectively, unless there was a nearby source in which case we used
a 3.5~arcsec diameter aperture with appropriate aperture
corrections. For MIPS we extracted the photometry in a 13~arcsec
diameter aperture with aperture correction of 1.167. A summary of
these data is shown in Table\,\ref{table:spitzer}.

\begin{table*}
  \caption{{\em Spitzer} photometry for our sample of radio galaxies.\textbf{Column 1} gives the name of the object;
    \textbf{Columns 2, 4, 6, 8, 10} give the flux density at 
    $3.6$, $4.5$, $5.8$, $8.0$ and $24\,\mu \rm m$, respectively;
  \textbf{Columns 3, 5, 7, 9, 11} give the respective flux density errors.} \centering
\begin{tabular}{l c c c c c c c c c c}
\noalign{\smallskip}
\hline\hline
\noalign{\smallskip}
Object & $S_{3.6\,\mu \rm m}$ &  Err$_{S_{3.6\,\mu \rm m}}$ & $S_{4.5\,\mu \rm m}$ &  Err$_{S_{4.5\,\mu \rm m}}$ & $S_{5.8\,\mu \rm m}$ &  Err$_{S_{5.8\,\mu \rm m}}$ & $S_{8.0\,\mu \rm m}$ &  Err$_{S_{8.0\,\mu \rm m}}$ & $S_{24\,\mu \rm m}$ &  Err$_{S_{24\,\mu \rm m}}$ \\[0.5ex]
  &  $\mu \rm Jy$ &  $\mu \rm Jy$ &  $\mu \rm Jy$ &  $\mu \rm Jy$ &  $\mu \rm Jy$ &  $\mu \rm Jy$ &  $\mu \rm Jy$ &  $\mu \rm Jy$ &  $\mu \rm Jy$ &  $\mu \rm Jy$   \\
\noalign{\smallskip}
\hline
\noalign{\smallskip}
  3C280 & 277.26 & 38.69 & 524.69 & 87.58 & 1004.40 & 126.74 & 2200.45 &
  204.88 & 9229.46 & 258.67\\
  3C268.1 & 35.45 & 13.94 & 78.20 & 33.91 & 108.84 & 43.09 & 209.22 & 64.37 & 940.96 & 123.57\\
  3C356 & 108.00 & 11.00 & 110.00 & 11.00 & 122.00 & 14.00 & 434.00 & 47.00 & 4060.00 & 192.00\\
  3C184 & 129.88 & 44.40 & 145.48 & 46.09 & 242.00 & 55.00 & 288.00 & 76.00 & 742.00 & 184.00\\
  3C175.1 & 119.60 & 25.54 & 109.55 & 40.21 & 92.03 & 42.56 & 124.70 & 53.55 & 836.90 & 166.56\\
  3C22 & 1399.41 & 86.86 & 2431.72 & 188.43 & 3647.24 & 239.86 & 5852.97 & 333.61 & 13744.13 & 310.40\\
  3C289 & 38.91 & 14.62 & 44.00 & 25.53 & 26.81 & 25.20 & 37.15 & 26.55 & 3650.16 & 168.21\\
  3C343 & 151.35 & 28.62 & 142.75 & 45.74 & 211.85 & 60.29 & 683.25 & 114.58 & 7294.40 & 230.19\\
  6CE1256+3648 & 62.50 & 18.44 & 61.01 & 29.99 & 55.72 & 32.24 & 133.86 & 52.70 & 1351.94 & 124.20\\
  6CE1217+3645 & 118.37 & 25.34 & 120.80 & 42.11 & 125.26 & 51.99 & 150.98 & 53.52 & 313.64 & 104.94\\
  6CE1017+3712 & 26.52 & 12.15 & 31.88 & 21.83 & 43.08 & 30.52 & 129.35 & 53.11 & 1136.21 & 131.80\\
  6CE0943+3958 & 72.68 & 19.90 & 87.66 & 35.95 & 141.66 & 49.31 & 244.95 & 70.77 & 1996.79 & 153.62\\
  6CE1257+3633 & 81.58 & 21.05 & 70.04 & 32.11 & 36.46 & 27.63 & 110.35 & 48.18 & 850.67 & 108.30\\
  6CE1019+3924 & 159.69 & 29.41 & 117.47 & 41.57 & 75.93 & 38.14 & 23.26 & 28.80 & 334.56 & 181.18\\
  6CE1011+3632 & 74.20 & 20.11 & 79.07 & 34.14 & 89.07 & 40.89 & 200.03 & 64.69 & 1326.94 & 150.39\\
  6CE1129+3710 & 81.67 & 21.08 & 62.57 & 30.39 & 40.67 & 35.62 & 72.84 & 41.39 & 856.71 & 119.07\\
  6C*0128+394 & 85.50 & 21.58 & 92.74 & 36.95 & 65.62 & 36.02 & 74.97 & 42.26 & 134.85 & 86.14\\
  6CE1212+3805 & 74.06 & 20.10 & 61.81 & 30.24 & 30.77 & 31.14 & 55.14 & 37.46 & 257.42 & 92.98\\
  6C*0133+486 & 53.85 & 17.20 & 64.35 & 30.86 & 50.99 & 41.43 & 75.81 & 37.92 & 82.15 & 75.92\\
  5C6.24 & 111.17 & 24.57 & 102.82 & 38.89 & 86.34 & 41.21 & 118.65 & 50.69 & 728.22 & 119.30\\
  5C7.23 & 54.77 & 17.32 & 48.11 & 26.75 & 67.04 & 35.75 & 41.12 & 34.62 & 623.52 & 160.69\\
  5C7.82 & 114.37 & 24.92 & 83.15 & 34.99 & 45.29 & 31.63 & 55.92 & 37.71 & 422.33 & 154.47\\
  5C7.242 & 123.35 & 25.87 & 104.96 & 39.33 & 69.57 & 36.58 & 41.92 & 34.64  & 1994.72 & 161.08\\
  5C7.17 & 30.88 & 13.10 & 22.64 & 18.50 & 56.68 & 33.88 & 5.14 & 22.51 & 1371.30 & 176.80\\
  TOOT1267 & 114.48 & 24.91 & 95.55 & 37.47 & 88.35 & 39.65 & 141.41 & 53.90 & 1079.06 & 121.87\\
  TOOT1140 & 50.48 & 16.63 & 64.88 & 30.92 & 47.60 & 31.18 & 28.71 & 28.25 & 136.24 & 80.29\\
  TOOT1066 & 49.37 & 16.50 & 35.92 & 23.09 & -3.85 & 14.10 & 16.21 & 22.64 & 298.16 & 86.86\\
\noalign{\smallskip}
\hline
\end{tabular}
\label{table:spitzer}
\end{table*}

\subsection{Literature data}

To increase the number of data points in the near-infrared and optical
bands, we used as many photometric values as possible from the
literature. The values available were extracted with a range of
different apertures and, where possible, we chose apertures of $8\, \rm
arcsec$, or larger, which approximately encompass the whole of the
emission from the galaxies in our sample.

\subsubsection{Aperture corrections}

For consistency, when magnitude values were only available with
smaller apertures we applied a correction to transform them into
magnitudes with a $8\, \rm arcsec$ aperture or larger. Given that the
majority of the photometric data in the literature is presented in
apertures of 8 or 9 arcsec diameter, and that this is a good
approximation to the total flux of the source, we preferred to convert
all the smaller aperture magnitudes to their equivalent value at 8 or
9 arcsec.

All the magnitudes measured using smaller apertures that we converted
to 8 or 9~arcsec were measured either in 4 or 5~arcsec apertures, thus
the corrections are small \citep[e.g. ][]{2009MNRAS.394.2197B}. To
find relations between 4 and 5~arcsec and 8 or 9~arcsec, we gathered a
large sample of galaxies ($\sim 40$ galaxies), from the papers
referenced in Tables~1 to 4 of the Appendix, which had photometry
available for several different apertures.  With this sample, we
computed linear relations between the magnitudes at 4~arcsec and
9~arcsec and between 5~arcsec and 8~arcsec. We then used these linear
fits to determine all magnitudes in 8 or 9~arcsec apertures.


We also applied a Galactic extinction correction to all the magnitude
values that had not been corrected based on the maps of
\citet{1998ApJ...500..525S}.


\subsubsection{Line emission correction}

Following \cite{2001MNRAS.326.1585J} we also subtracted the flux of
the dominant emission lines $\rm H\alpha$, [O{\sc{II}}], [Ne{\sc{V}}],
Mg{\sc{II}}, [Ne{\sc{IV}}] and C{\sc{II}}], from J, R, F702W, F606W,
  V, B, and g bands, as these lines can contribute a significant
  percentage to the total flux of the band where they lie. Where the
  line fluxes were not available we used the [O{\sc II}] fluxes from
  \citet{2011MNRAS.411.1909F} and determined the remaining emission
  lines using the line ratios given by \citet{1993ARA&A..31..639M} and
  \citet{1997MNRAS.292..758B}.  When the flux of H$\alpha$ is not
  known, we used the average ratio estimated by
  \citet{1993ARA&A..31..639M} for high redshift radio galaxies
  ($z\lesssim2$) from the 3CR survey.

To correct for emission line contamination we then determined the
location of each emission line for the redshift of the given source
and account for the shape of the filter at that wavelength.  Given the
uncertainties affecting the aperture, emission-line contamination and
Galactic extinction corrections, we assume an additional $10$ per cent error
in the flux density for each magnitude value that had any of these
corrections applied. All these corrections are detailed in
Tables~\ref{table:JHK}, \ref{table:HST} \ref{table:UBVRI}, and
\ref{table:ugriz} of the Appendix.


\section{SED fitting}\label{sec:sed_method}

The photometric data span a wavelength range from $0.36\,\mu\rm m$ to
$24\,\mu \rm m$. At optical wavelengths, the emission from radio
galaxies - obscured AGN - is dominated by the stellar emission of the
host galaxy. At mid-infrared wavelengths, dust in the `torus' region
that re-emits the radiation from the central AGN is the main source of
emission, at least for those sources with a relatively powerful and
radiatively efficient AGN. This can be thought of as a composite of
the central emission passing through a screen of dust. We follow the
approach of \citet{2007MNRAS.379L...6M,2009ApJ...706..184M} and model
the radio galaxies with a galaxy template that dominates the emission
in the optical part of the spectrum and a quasar template,
extinguished by a dust extinction law, to fit the emission in the
infrared spectral region.

\subsection{Galaxy template}

To replicate the host galaxy emission, we consider a Bruzual and
Charlot stellar synthesis model \citep[][hereafter
  BC03]{2003MNRAS.344.1000B}, as well as the Maraston et
al. ‘fuel-consumption’ stellar population synthesis model
\citep[][hereafter M05]{2005MNRAS.362..799M}. For both of these we
assume a Salpeter initial mass function and solar metallicity.

Since radio galaxies are almost exclusively hosted by elliptical
galaxies or recent merger remnants
\citep[e.g.][]{2003MNRAS.340.1095D,2010ApJ...713...66F}, we expect
them to have a strong short episode of star formation in the beginning
of their formation and for the star formation rate to quickly drop
subsequently. Therefore, we consider that a single simple stellar
population suffices to reproduce the host galaxy's star formation
history, and a synthesis of SSPs is not required. To select the
stellar age of our template we note that our objects all have
$z\sim1$, when, according to the adopted cosmology, the Universe was
$\sim5.7\,\rm Gyr$ old. We, thus, constrain the range of possible ages
with an upper limit for the stellar population age of $6\,\rm Gyr$. As
for a lower limit, previous studies have shown that for
passively-evolving early-type galaxies in cluster environments, the
bulk of stars form at $z\gtrsim3$ and in low-density environments at
$z\gtrsim1.5-2$
\citep[e.g.][]{1992MNRAS.254..601B,1998MNRAS.299.1193B,2006ARA&A..44..141R}. This
yields a stellar age of $\sim5.6\,\rm Gyr$ for cluster environments
and $\sim1.5-2.5\,\rm Gyr$ for field galaxies. Even though our sample
consists of powerful radio galaxies, which tend to inhabit
high-density environments as suggested by various lines of
observations
\citep[e.g.][]{HillLilly1991,Wold2001,Hardcastle2004,Kauffmann2008,Falder2010},
including X-ray observations that show X-ray cavities identified with
clusters of galaxies to be spatially coincident with non-thermal radio
emission from radio-loud AGN (e.g. \citealt{2002MNRAS.331..369F},
\citealt{2004ApJ...607..800B}), we adopt a conservative lower limit
for the stellar age of $\sim0.5\,\rm Gyr$. We thus use templates with
$0.5\,\rm Gyr$, $1\,\rm Gyr$, $2\,\rm Gyr$, $3\,\rm Gyr$, $4\,\rm
Gyr$, $5\,\rm Gyr$, and $6\,\rm Gyr$ of age for the SED fitting of the
galaxies in our sample.



\subsection{Quasar and dust template}\label{sec:torus}

To model the infrared emission of the radio galaxies, we need to
account for the dusty torus absorbing the radiation from the central
source and re-emitting it at longer wavelengths. This can be done, as
an approximation, by assuming that a radio galaxy is equivalent to a
quasar with a layer of obscuring dust in front of it, where this layer
can have diverse column densities depending on how obscured the galaxy
is \citep[e.g.][]{2007MNRAS.379L...6M,2009ApJ...706..184M}. Indeed,
\citet{2008ApJ...688..122H} find that they can reproduce a diversity
of rest-frame $1.6-10\,\rm \mu m$ SEDs of radio galaxies by combining
various amounts of extinction of AGN light with host galaxy
starlight. This representation is consistent with the
orientation-dependent unified scheme and it is intended as a
simplification only, as a more careful approximation should fully
account for radiative transfer effects in a dusty medium \citep[e.g][]{2008ApJ...685..147N}. This
approach would, however, imply a large number of free parameters with
too few data points to constrain them, resulting in a high degeneracy,
making it difficult to infer physically meaningful quantities. Thus, we
build, as a simplification, a composite model of a quasar and a screen
of dust.

As a quasar template, we use the radio-quiet quasar SED by
\citet{1994ApJS...95....1E}, which was constructed based on the mean
energy distribution of a sample of 29 radio-quiet quasars. We
normalise this template by forcing the bolometric luminosity to be 1.
For this, we used the median value of the B-band luminosity, $L_{\rm
  B}$, which, according to Table 17 of \citet{1994ApJS...95....1E},
should be \textbf{$L_{\rm bol}=10.7 L_{\rm B}$.}

More recent works such as \cite{2013ApJ...777..164M} have shown a
composite model of a jet, a truncated thin accretion disk and an ADAF
to be a good fit for low-luminosity AGNs, and therefore possibly
better suited for LEGs. \cite{2013ApJ...777..164M} however report an
insufficient IR-emission produced by the truncated thin disk in order
to match the observations. Given these issues, we opt to use the model
by \citet{1994ApJS...95....1E} for all the objects in the sample for
consistency.

For a dust template, we used the extinction laws derived by
\citet{1992ApJ...395..130P}, which replicate how the extinction caused
by the different types of dust varies with wavelength. These templates are described by the following summation: 
\begin{equation}\label{eq:pei}
\frac{A_{\lambda}}{A_{\rm B}}(\lambda)=\sum\limits_{i=1}^6\frac{a_{i}}{(\lambda/\lambda_{i})^{n_{i}}+(\lambda/\lambda_{i})^{-n_{i}}+b_{i}}   ~~~~~~(Pei, 1992), 
\end{equation}
where $a_{i}$, $b_{i}$ and $n_{i}$ vary for each term and for each
dust type (see Table~4 of \citealt{1992ApJ...395..130P} for a full
description). The six terms involved represent the background,
far-ultraviolet and far-infrared extinctions and the $\rm 2175\,\AA$,
$9.7\,\mu$m and $18\mu$m features. For reddened quasars, SMC dust
has been found to be appropriate
\citep[e.g.][]{2004AJ....128.1112H,2005ApJ...627L.101W}, since dust in
the host galaxy of high-redshift galaxies tends to have lower
metallicity and thus can be better approximated by SMC dust
type. However, for more obscured quasars, and in particular for the
majority of the objects in our sample, we find that SMC dust does not
provide as good a fit as MW dust. This could be explained by the fact
that the dust intersecting the line of sight of the central emission
in obscured galaxies comes from the inner region due to the galaxy's
edge-on orientation, and the central regions of galaxies are usually
more metal rich, as there is in general a metallicity gradient in
galaxies
\citep[e.g.][]{1993MNRAS.262..650D,1994MNRAS.270..523C,2009ApJ...691L.138S}. Therefore,
we adopted a MW dust type for all the galaxies in our sample.

\subsection{Method of fitting}


We fit our model to the photometric data points by applying an
  extinction curve to the intrinsic quasar light template and then
  adding it to the galaxy light model:
\begin{equation}
\rm S_{\nu,model}=S_{\nu,QSOmodel}+S_{\nu,GALmodel},
\end{equation}
with
\begin{equation}
\rm S_{\rm \nu,QSOmodel} = S_{\rm \nu, QSOtemplate}\times 10^{-\frac{A_{\rm V}\times A_{\lambda}(\lambda)}{2.5}},
\end{equation}
where $S_{\rm \nu, QSOtemplate}$ is the quasar light template and
$S_{\rm \nu,QSOmodel}$ is the quasar light model already affected by
extinction.

To convert the extinction law $\frac{A_{\lambda}}{A_{\rm
      B}}(\lambda)$ to magnitudes ($A_{\lambda}$) we multiply equation
  \eqref{eq:pei} by the term $1/R_{V}+1$, where $R_{V}$ is the ratio
  of total-to-selective extinction defined by the equation:
\begin{equation}\label{eq:Rv}
R_{\rm V}=\frac{A_{\rm V}}{E_{\rm B-V}},
\end{equation}
where $E_{\rm B-V}=A_{\rm B}-A_{\rm V}$ is the colour excess. 

We multiply each template
by a coefficient which we allow to vary along a range of physically
motivated values. 

The extinction law is multiplied by the visual extinction coefficient
$A_{\rm V}$ spanning values of $0 < A_{\rm V} < 400$. The resolution
we use to run through these values varies from galaxy to galaxy
depending on how precise the fit is, and we make it finer where the
dispersion of the fit is smaller, as detailed below. 

After multiplying the extinction law in flux units by the transmitted
quasar light, we force it to be as close as possible to the $24\,\rm
\mu m$ flux density value. This means we do not allow the bolometric
luminosity of the model to vary freely. For each given $A_{V}$, it is
fixed by the observed flux density at $24\,\rm \mu m$. The reasoning
for this is to constrain the fit, as we found that without this step,
the higher degeneracy of the fits often provided poorer matches to the
data in terms of reduced-$\chi^{2}$.

The galaxy template, $\rm S_{\nu,GALtemplate}$, is multiplied by a mass
normalisation factor, $M_{\rm gal}$, which we allow to vary between
$10^{10}$ and $10^{13}\,\rm M_{\sun}$, an interval that comprises the
typical values of stellar mass content in early-type galaxies and
certainly the masses of the majority of radio galaxies studied to date
\citep[e.g.][]{2007ApJS..171..353S}: 
\begin{equation} 
\rm S_{\nu,GALmodel}=\frac{S_{\nu,GALtemplate}}{M_{\odot}}\times M_{gal}
\end{equation}

We then perform a grid search in $A_{\rm V}$ and $M_{\rm
    gal}$ to obtain the best fit.



Given that the data from the literature were gathered from several
different instruments, we opted to use a single set of filter
transmission curves: for $J$, $H$ and $K_{s}$ bands we used the
filters from the Visible and Infrared Survey Telescope for Astronomy
\citep[VISTA; see e.g. ][]{2013MNRAS.428.1281J} For the HST F606W,
F702W and F814W passbands, we used the respective HST filter response
profiles; for $U$, $B$, $V$, $R$ and $I$ bands we used the filters
used on the auxiliary-port camera (ACAM) mounted on the William
Herschel Telescope (WHT); finally, for the $u$, $g$, $r$, $i$ and $z$
bands, given that most data at these wavelengths were extracted from
the SDSS public release, we used the SDSS bandpass filters. For the
small number of photometric data points that were not observed with
the exact same filters we use, the errors are smaller or of the same
order of magnitude as the uncertainties on the data.

Using the filter responses, we evaluate the flux density that
the model produces for each band using the following:

\begin{equation}\label{eq:filter}
\nu S_{\nu, \rm model}=\frac{\int_{\lambda_i}^{\lambda_f}\nu S_{\nu, \rm unfiltered}\times  T(\lambda) d\lambda}{\int_{\lambda_i}^{\lambda_f}T(\lambda) d\lambda},
\end{equation}
where $S_{\nu, \rm unfiltered}$ is the flux density of the model
before it has been convolved with the filter response; $S_{\nu, \rm
  model}$ is the flux density of the model after the filter response
has been taken into account; $\lambda_i$ and $\lambda_f$ are the
wavelength where the filter response starts and ends respectively;
and $T(\lambda)$ is the filter transmission curve. 


To determine the flux $\nu S_{\nu,\rm data}$ of the data points, we
multiply the flux density values by the mean frequency of each band. The
mean frequency is determined by computing the effective wavelength of
each band, given by:

\begin{equation}
\lambda_{\rm eff}=\frac{\int_{\lambda i}^{\lambda f} \lambda T(\lambda) d\lambda}{\int_{\lambda i}^{\lambda f} T(\lambda) d\lambda},   
\end{equation}
and converting it to $\nu_{\rm eff}$. \\

We determine the $\chi^{2}$ distribution of each model using,
\begin{equation}
\chi^2(A_{\rm V},M_{\rm gal})=\sum_{n}\left(\frac{\nu S_{\nu, \rm model}(A_{\rm V},M_{\rm gal})-\nu S_{\nu, \rm data}}{\sigma}\right)^{2},
\end{equation}
where $n$ is the number of flux/magnitude data points available, and
$\sigma$ is the error associated with each data point. 
We repeat this process for all the different stellar models we are
considering (ages 0.5, 1, 2, 3, 4, 5, and 6\,Gyr for both BC03 and M05
models) and choose the model that has the lowest $\chi^{2}$ amongst
these.\\

Sources that are detected below a 2$\sigma$ level in the imaging
  data are plotted as upper limits at a 2$\sigma$ level. We use their
  measured photometric flux for the SED fitting and place their error
  bars between zero and the detection limit for the $\chi^{2}$
  evaluation. However, photometric bands for which both the observed
flux density and the model are lower than the flux density limit make
no contribution to the $\chi^{2}$.



\section{SED fits}\label{sec:sed_fits}

\begin{figure*}
\includegraphics[width=0.99\columnwidth]{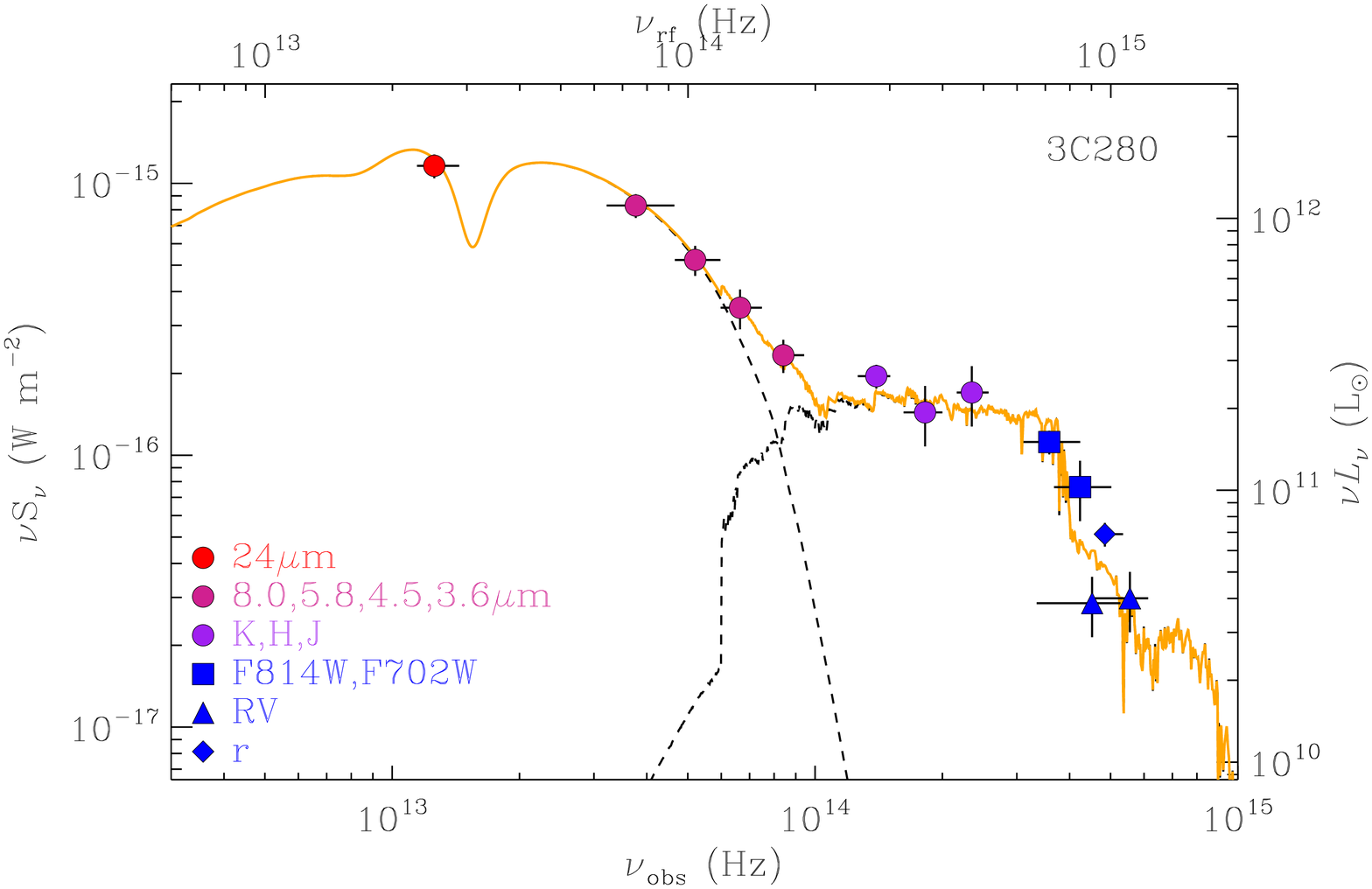} 
\includegraphics[width=0.99\columnwidth]{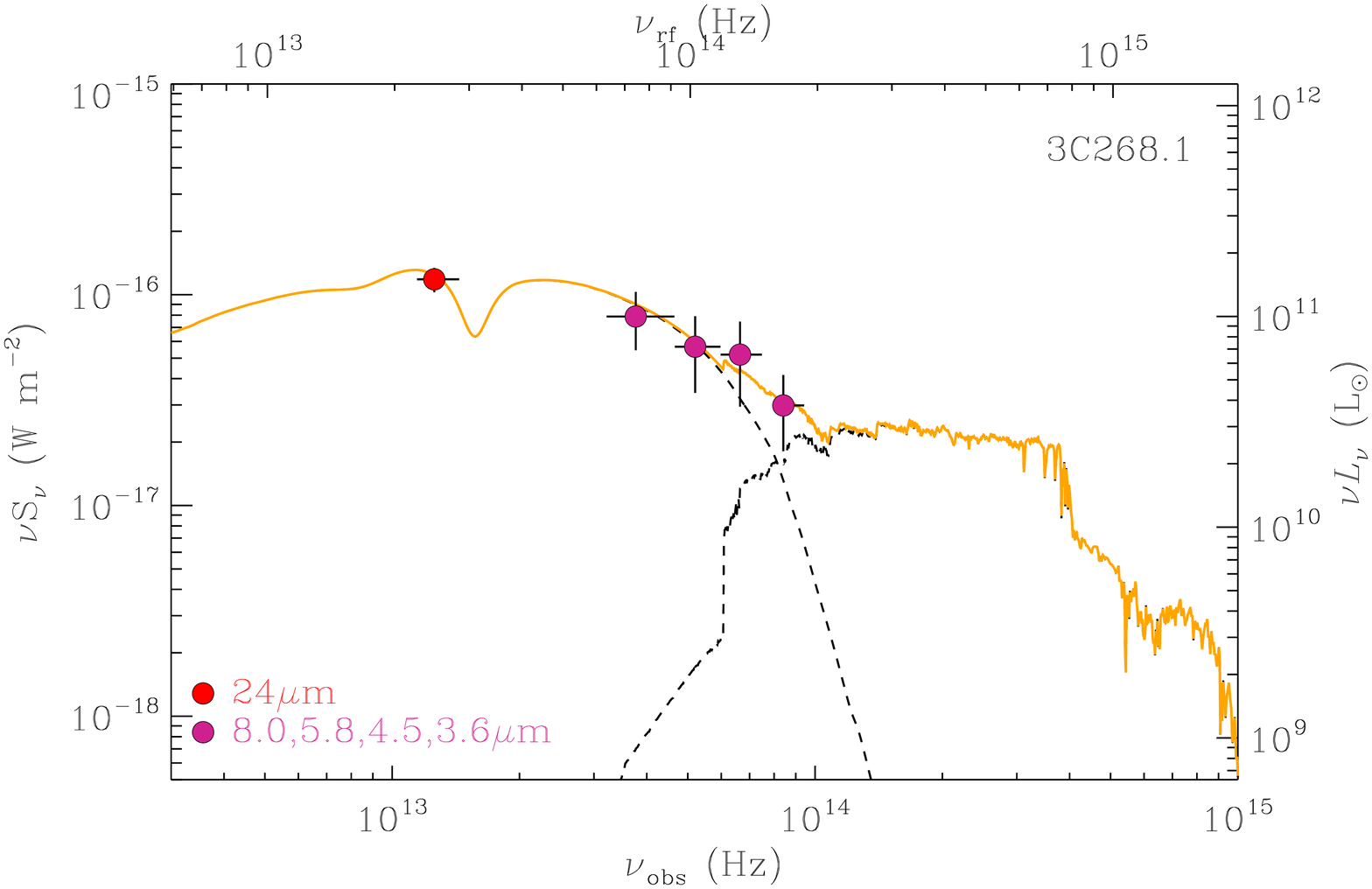} \\
\includegraphics[width=0.99\columnwidth]{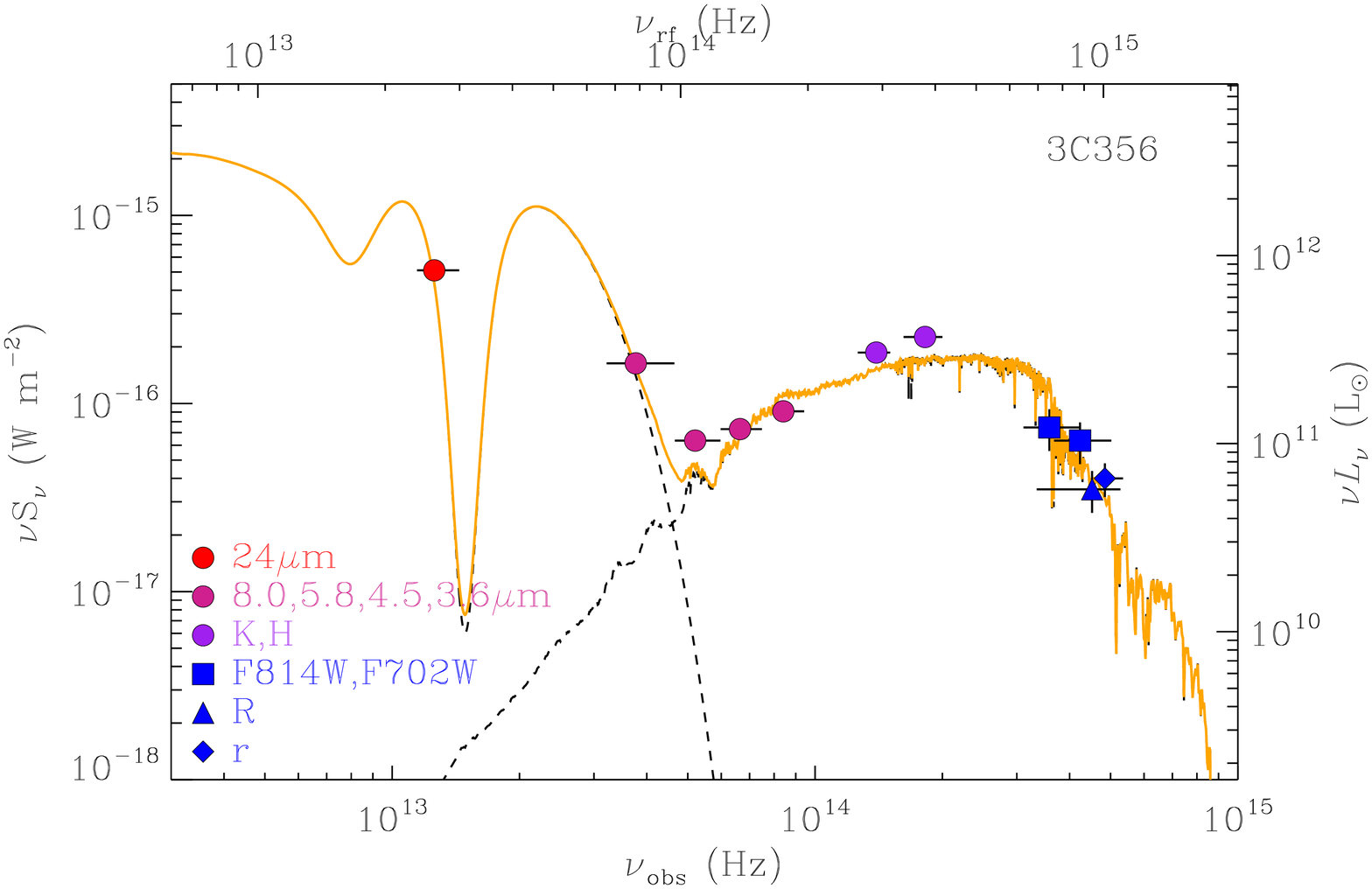}  
\includegraphics[width=0.99\columnwidth]{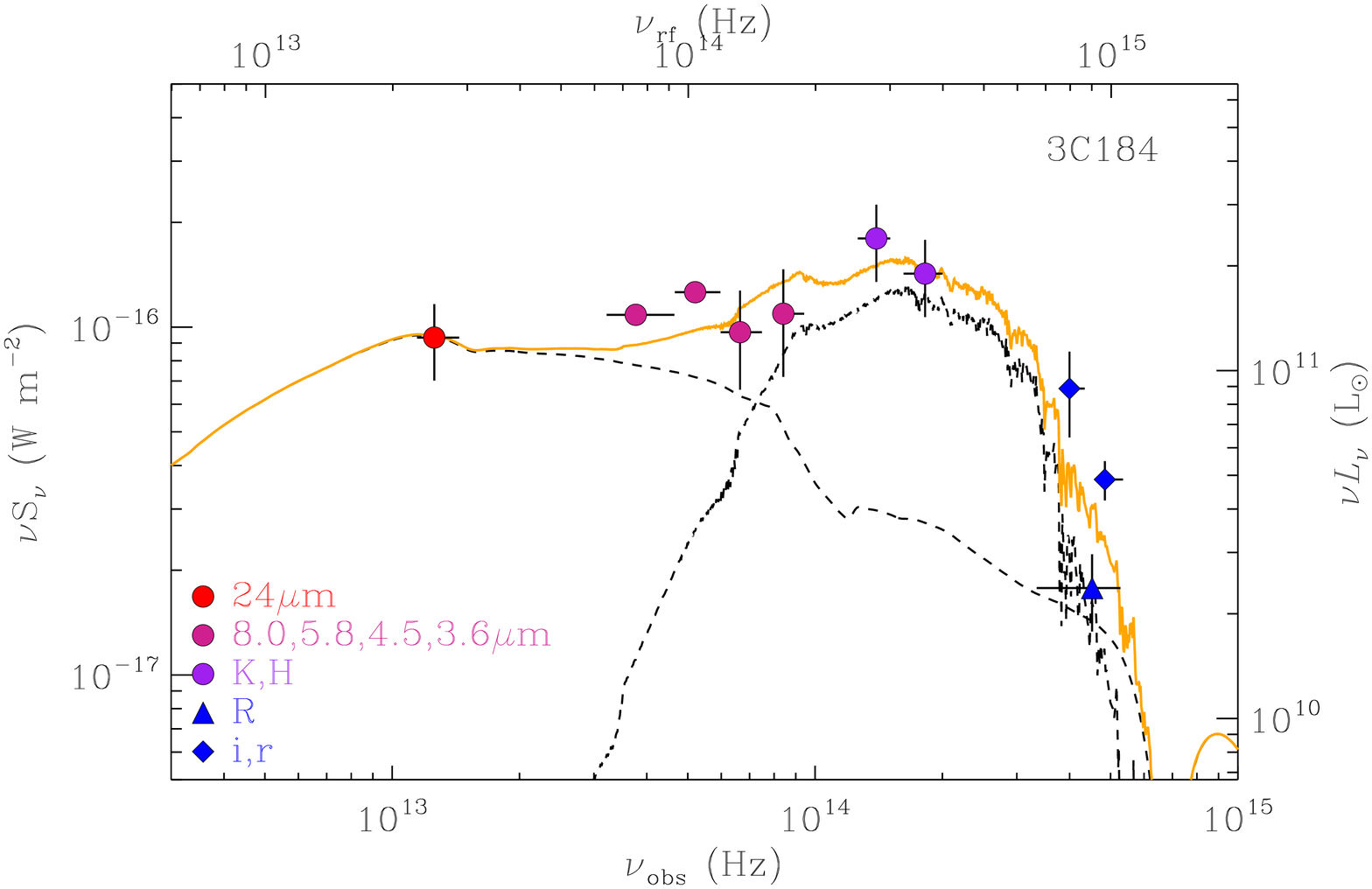} \\
\includegraphics[width=0.99\columnwidth]{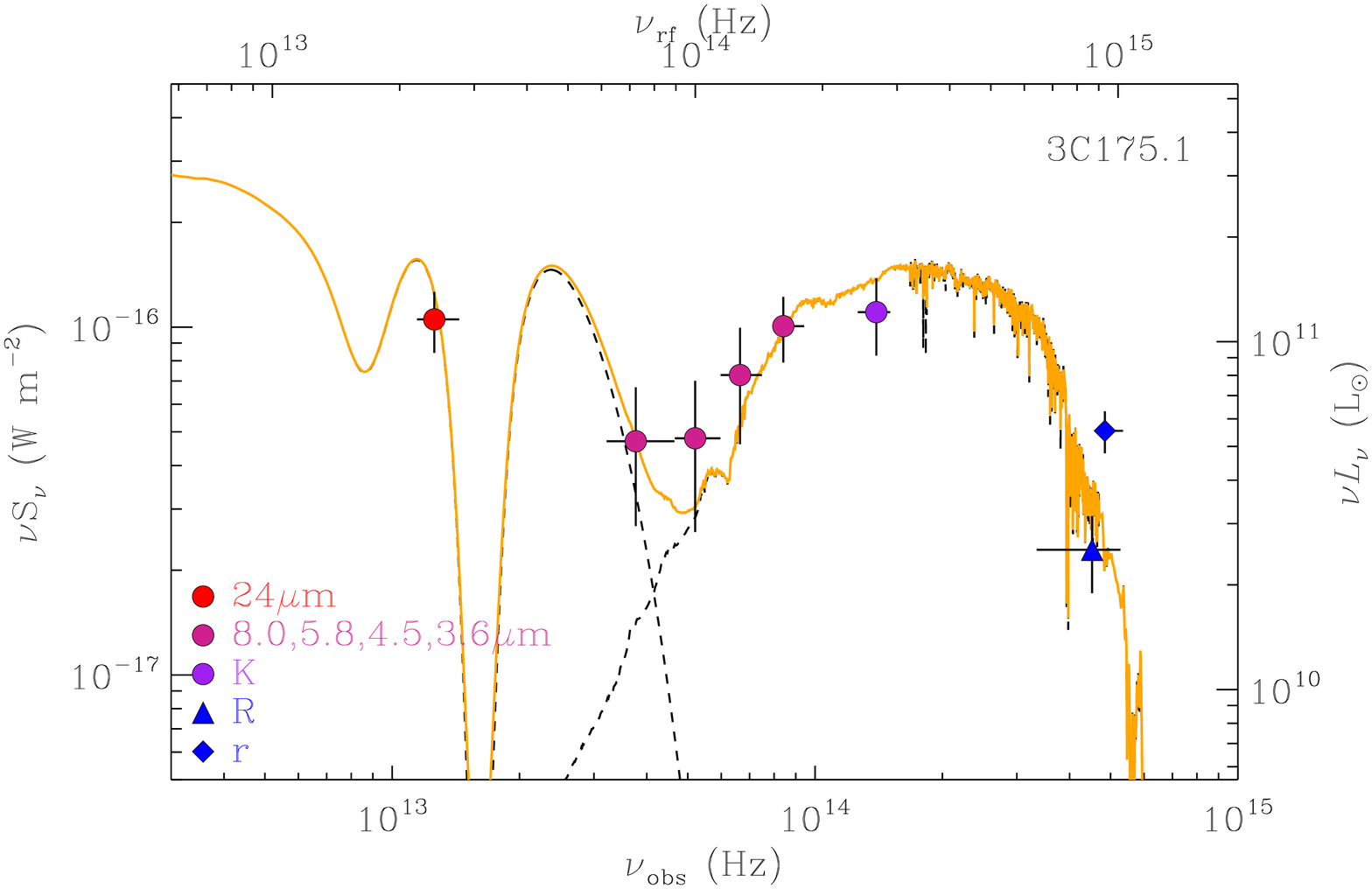}  
\includegraphics[width=0.99\columnwidth]{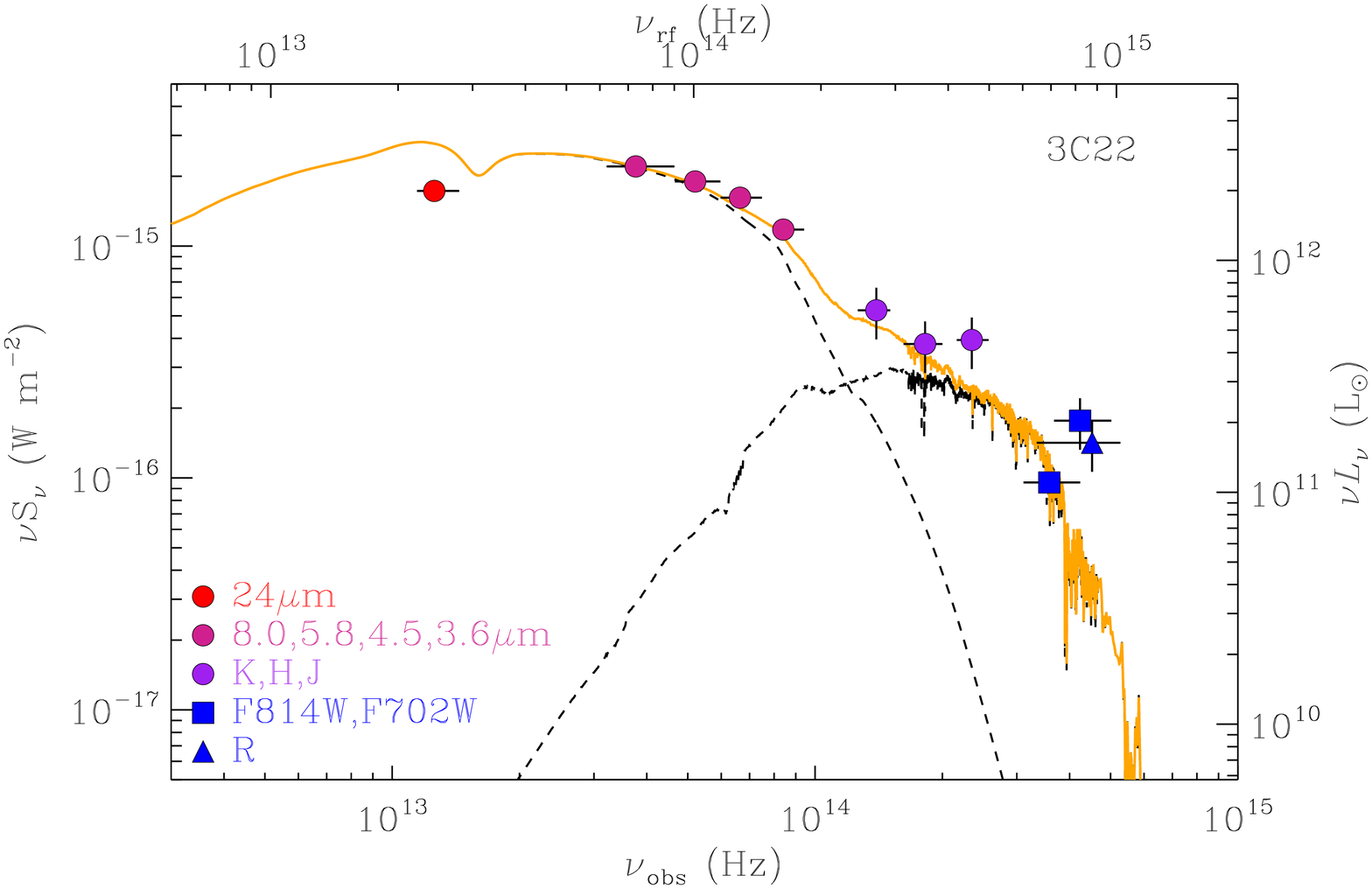} \\
\includegraphics[width=0.99\columnwidth]{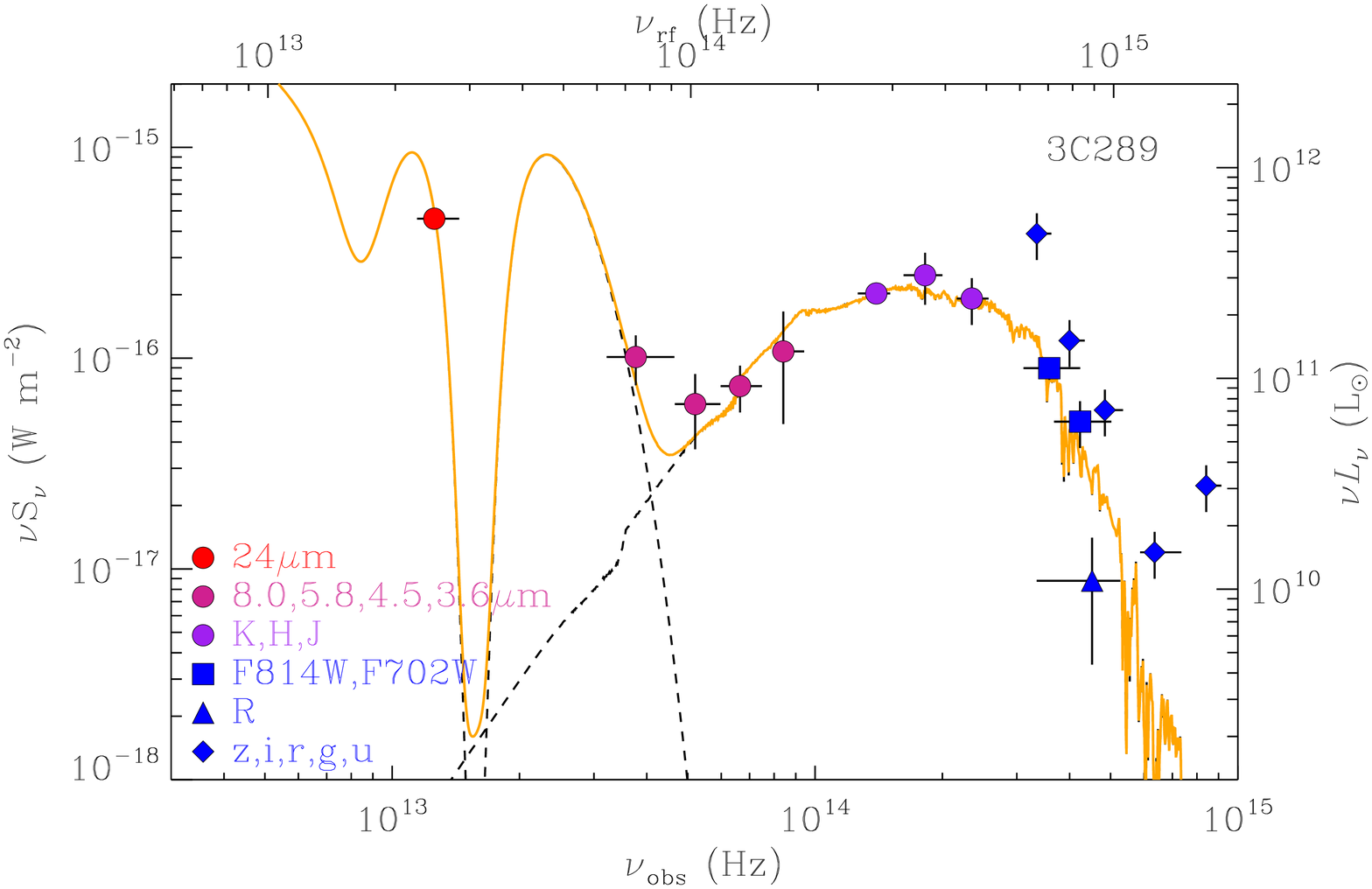}  
\includegraphics[width=0.99\columnwidth]{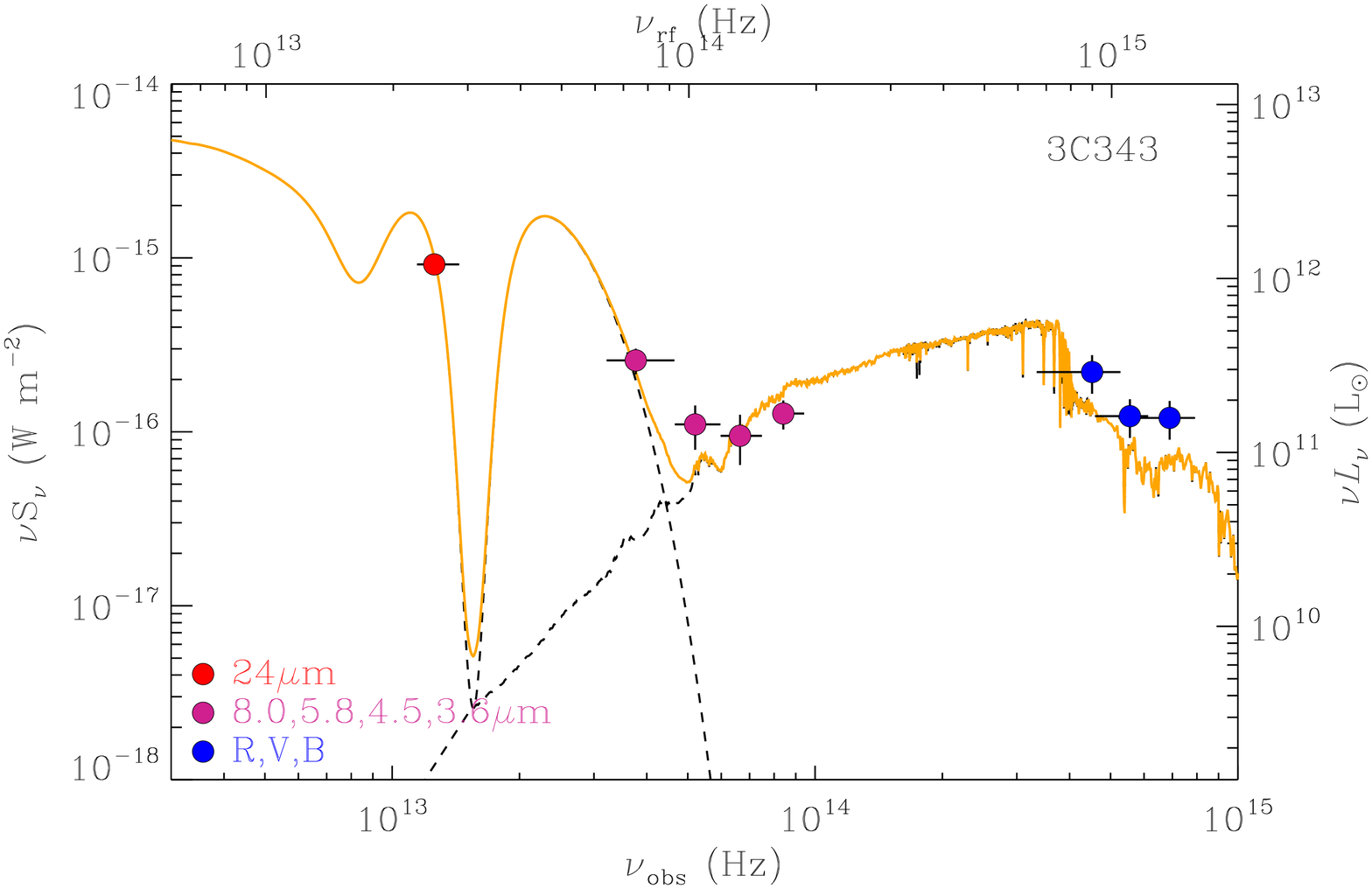} \\
\end{figure*}

\begin{figure*}
\includegraphics[width=0.99\columnwidth]{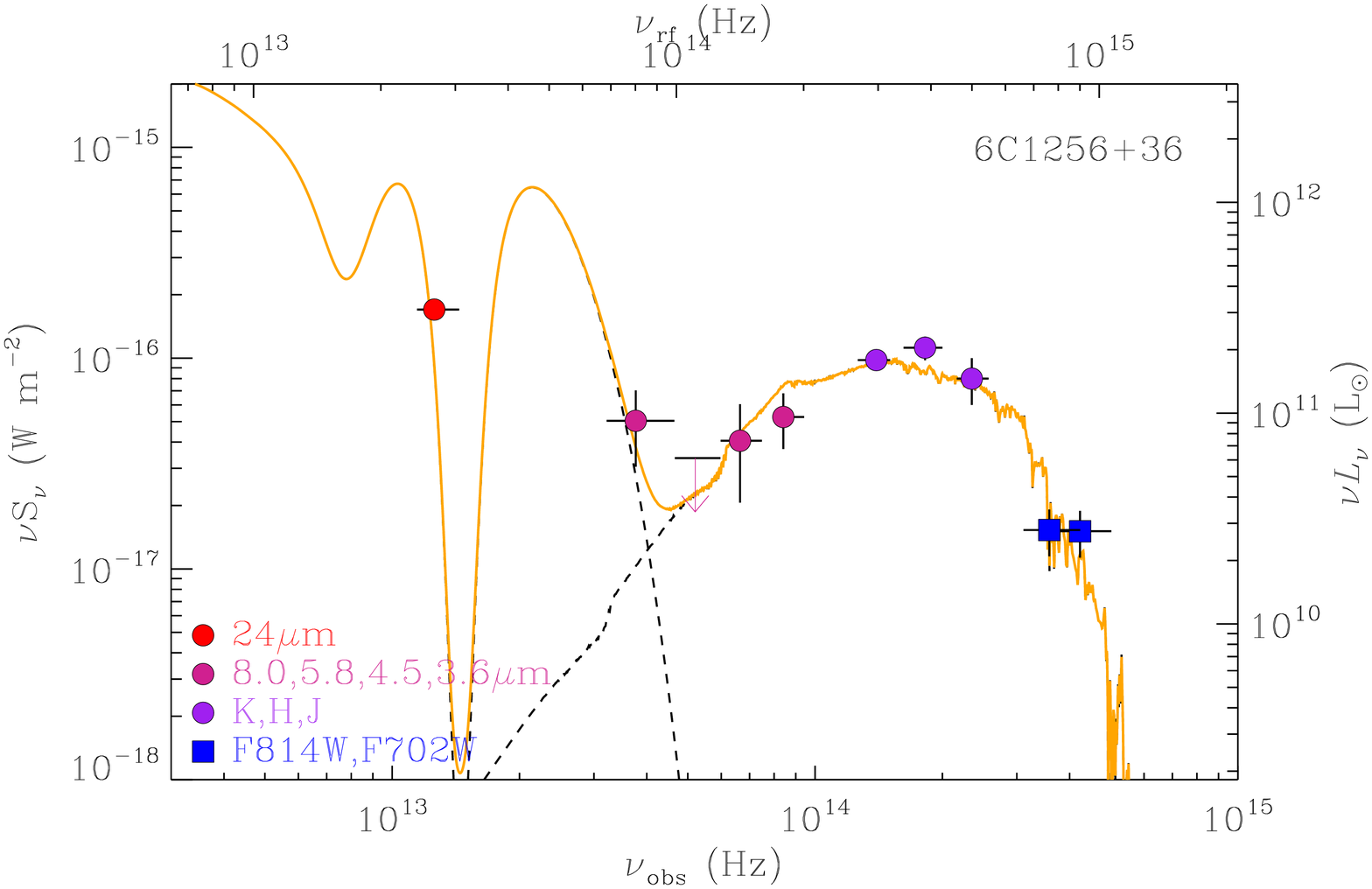}  
\includegraphics[width=0.99\columnwidth]{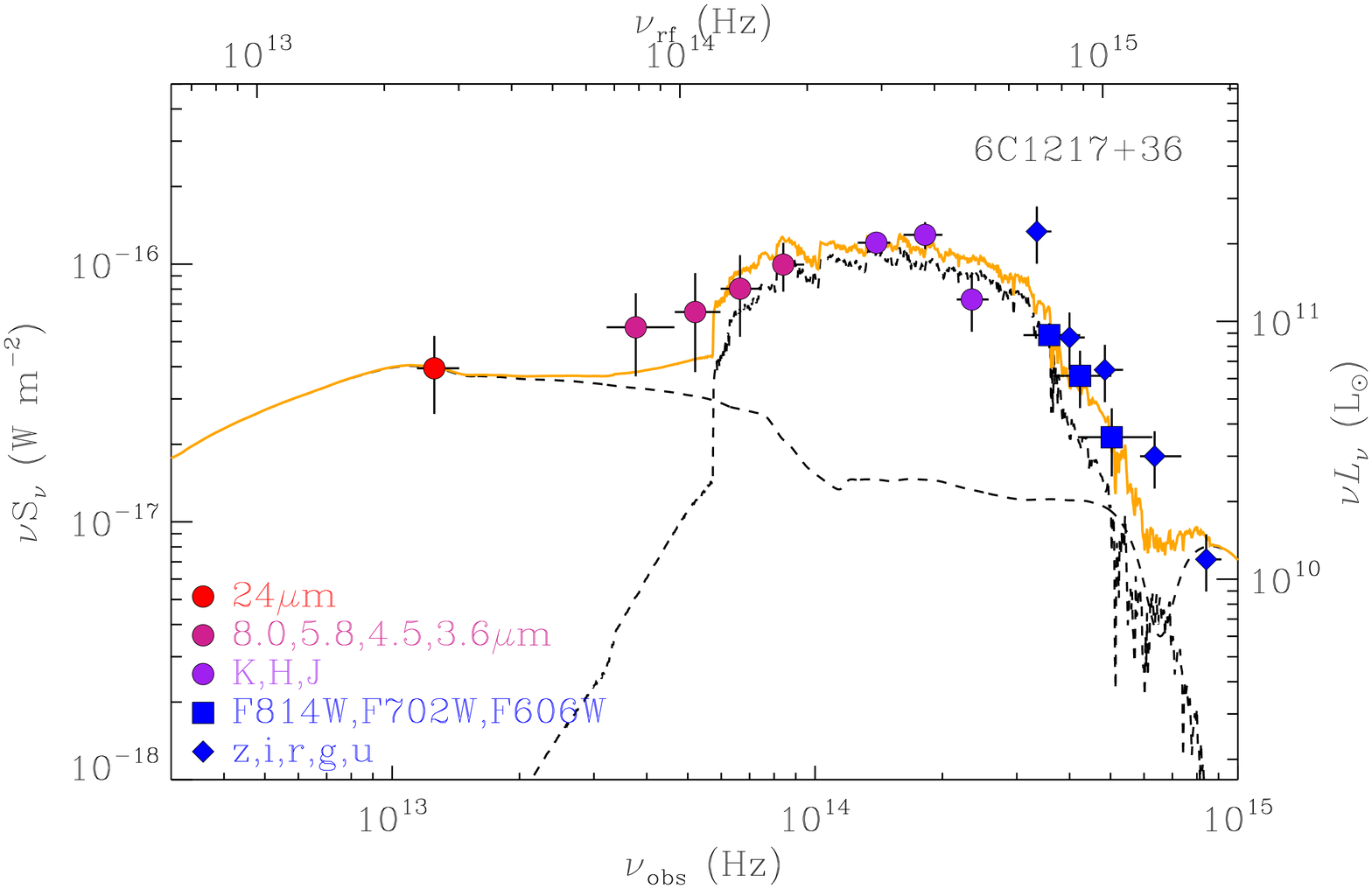} \\
\includegraphics[width=0.99\columnwidth]{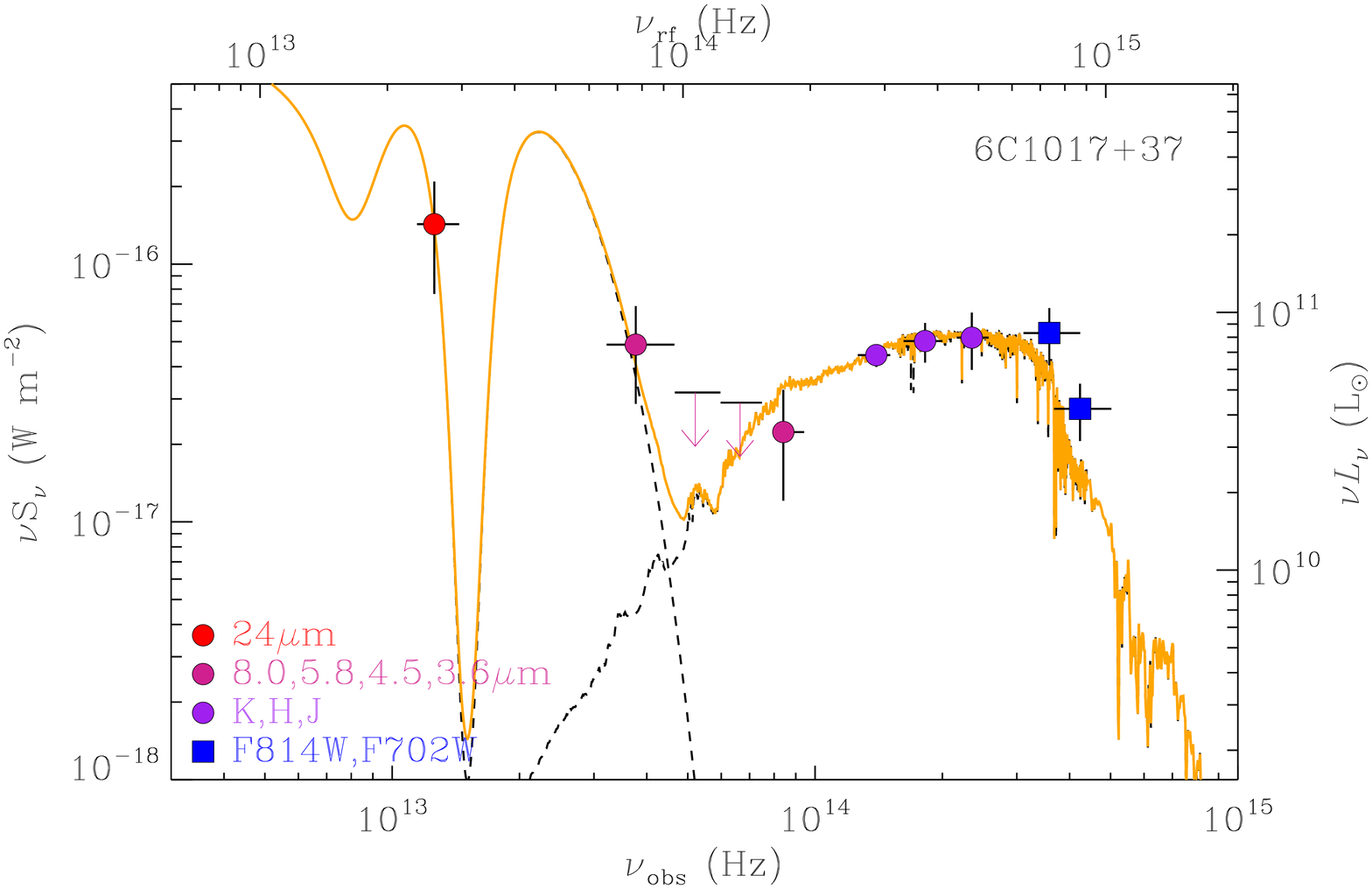}  
\includegraphics[width=0.99\columnwidth]{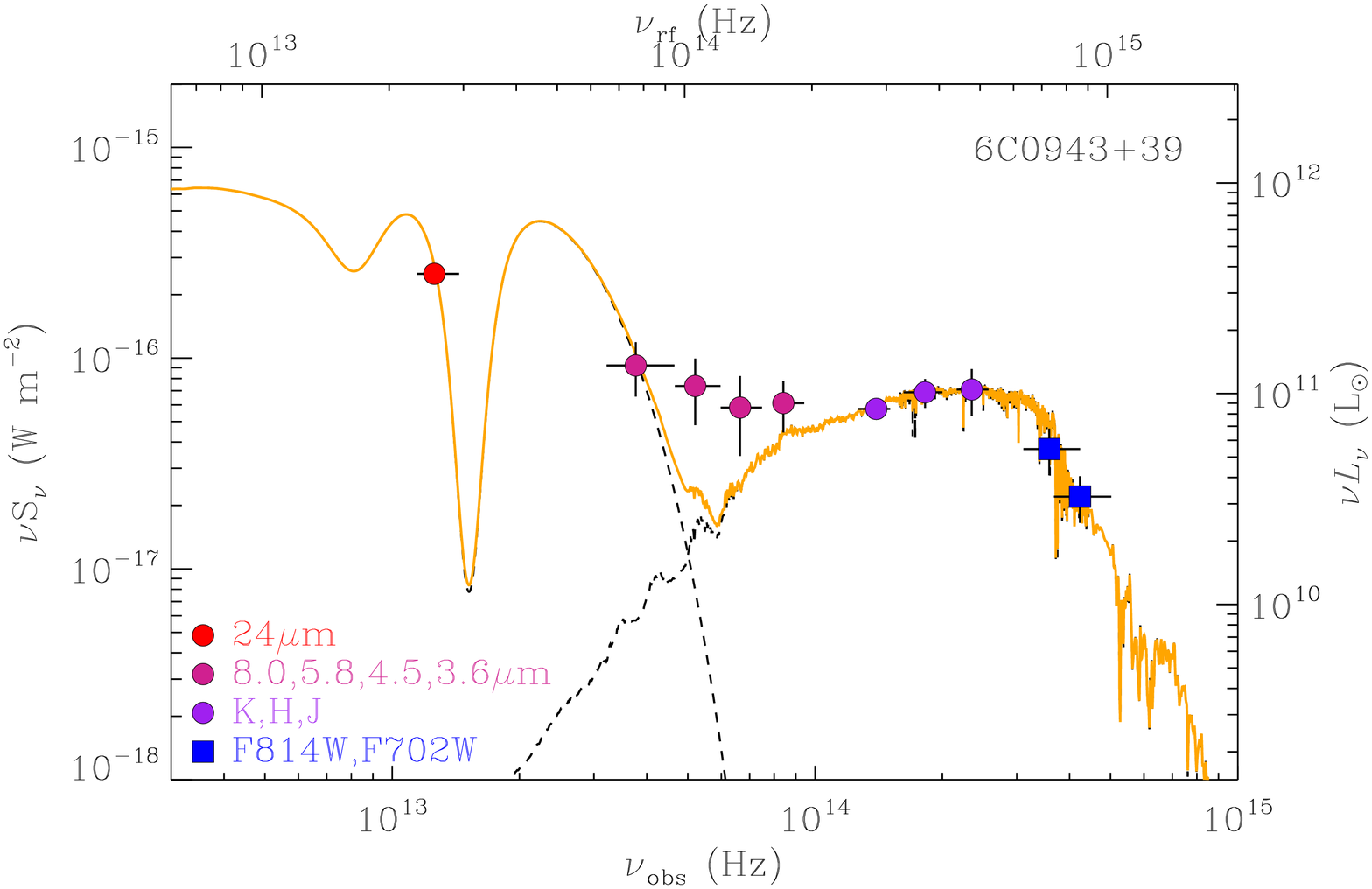} \\
\includegraphics[width=0.99\columnwidth]{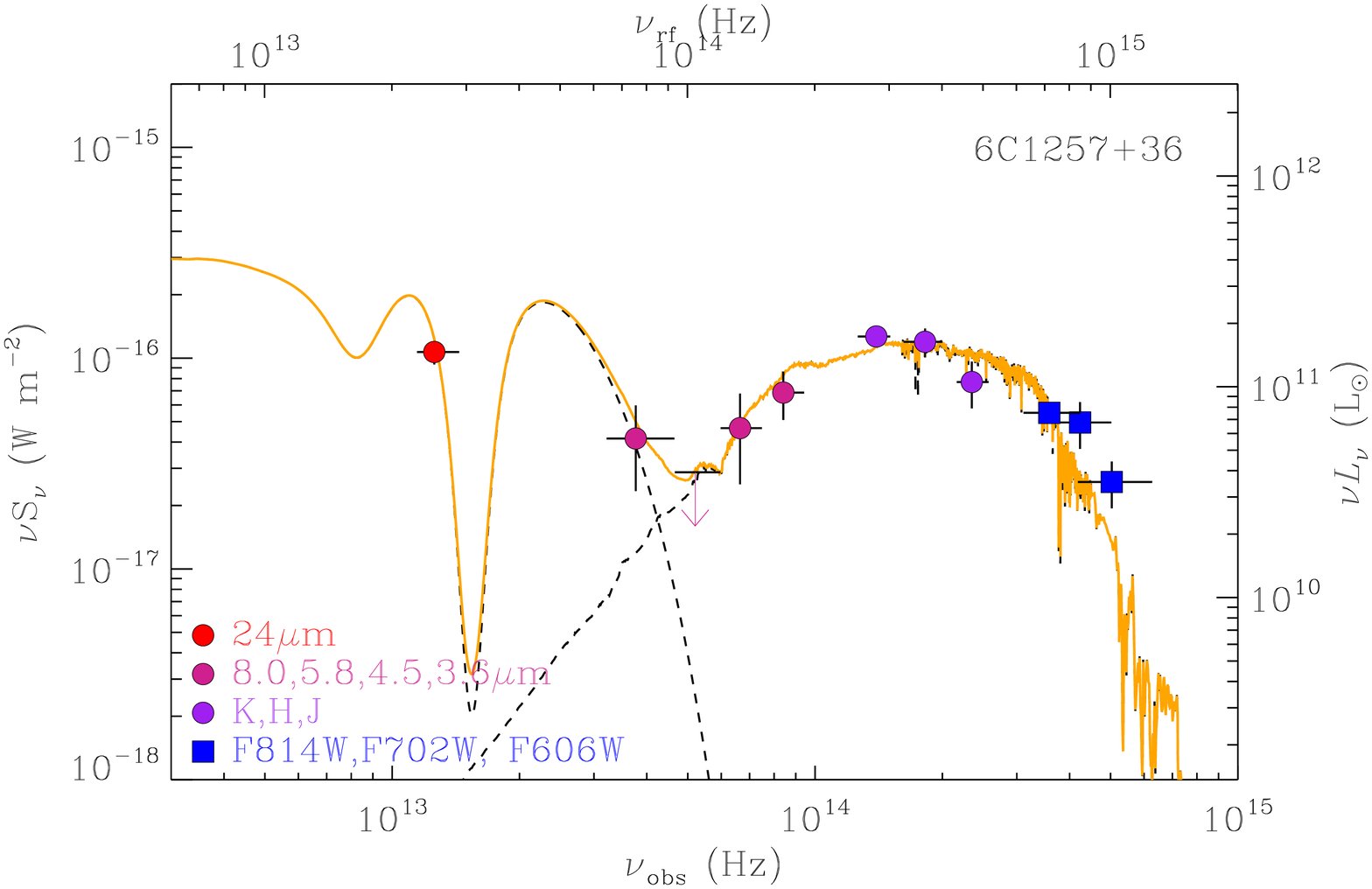}  
\includegraphics[width=0.99\columnwidth]{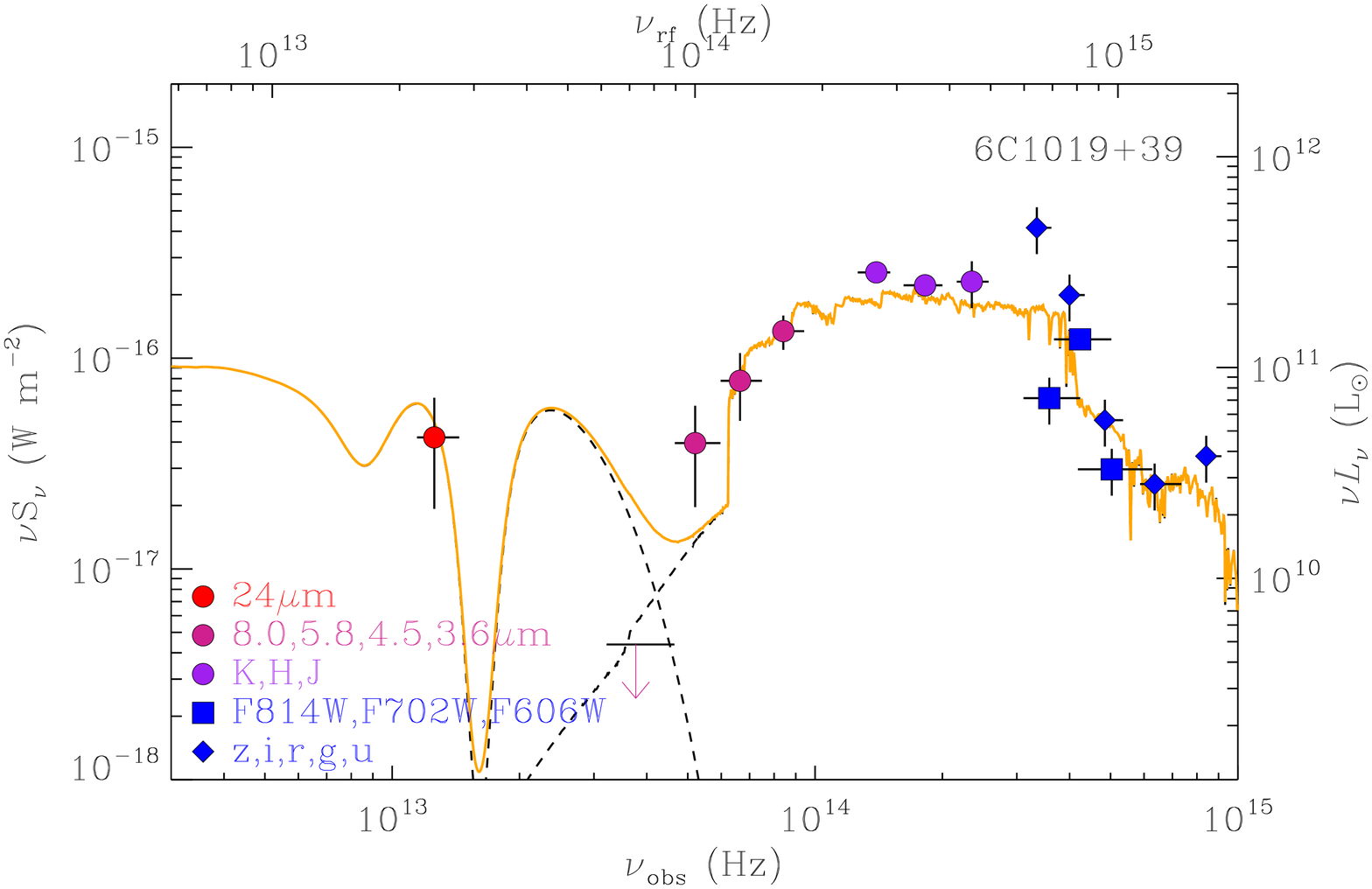} \\
\includegraphics[width=0.99\columnwidth]{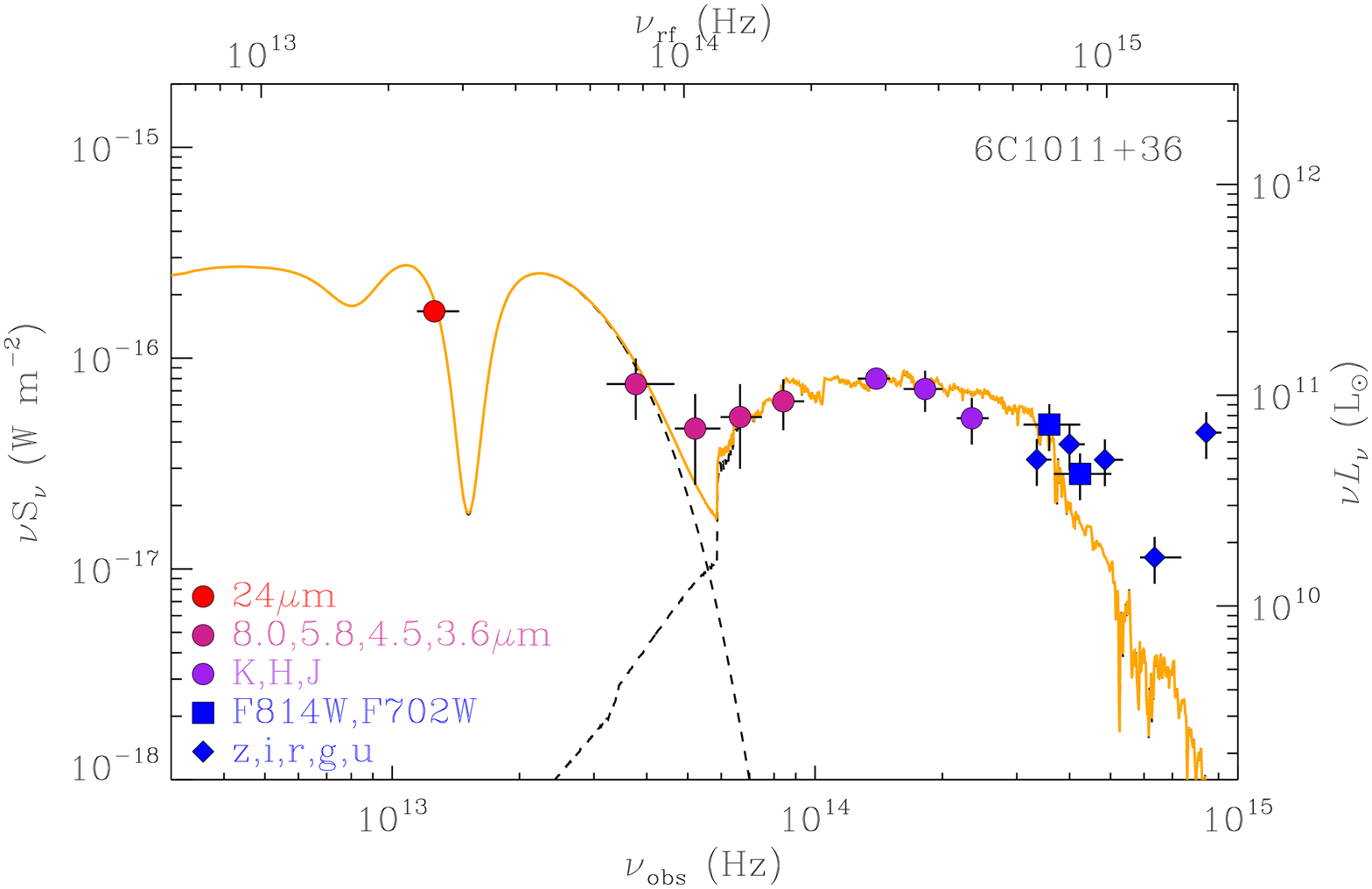}   
\includegraphics[width=0.99\columnwidth]{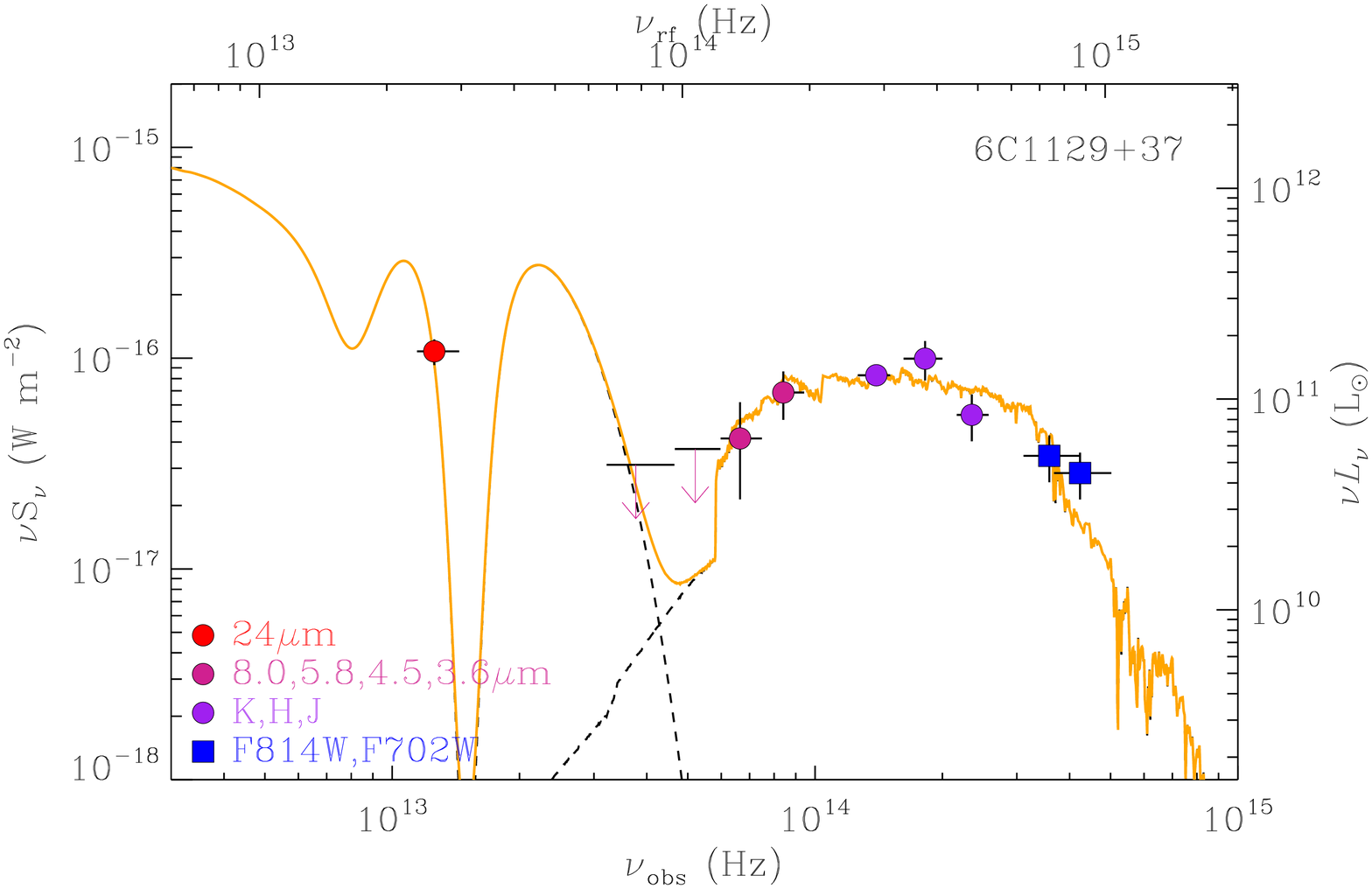} \\
\end{figure*}

\begin{figure*}
\includegraphics[width=0.99\columnwidth]{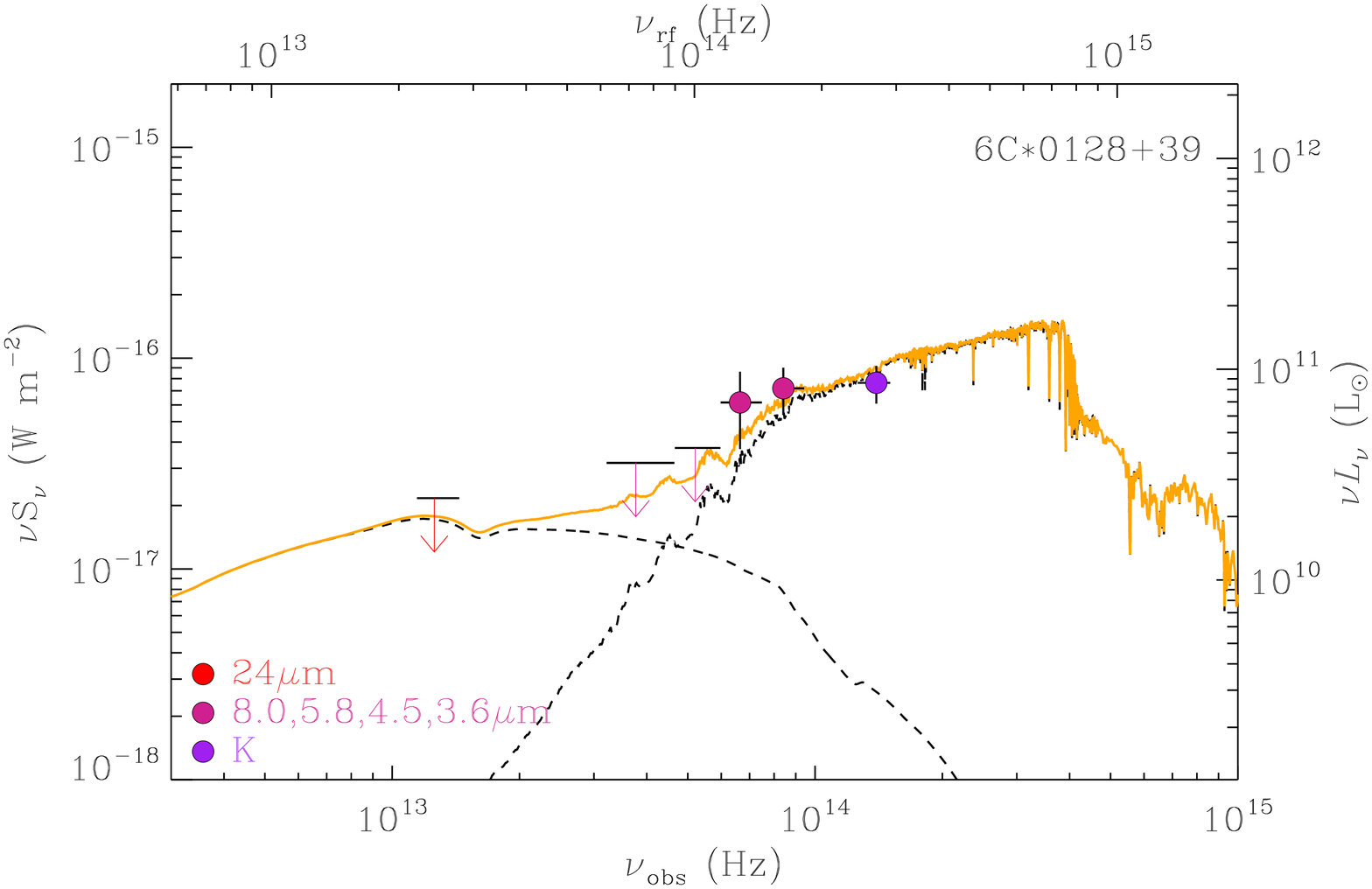}   
\includegraphics[width=0.99\columnwidth]{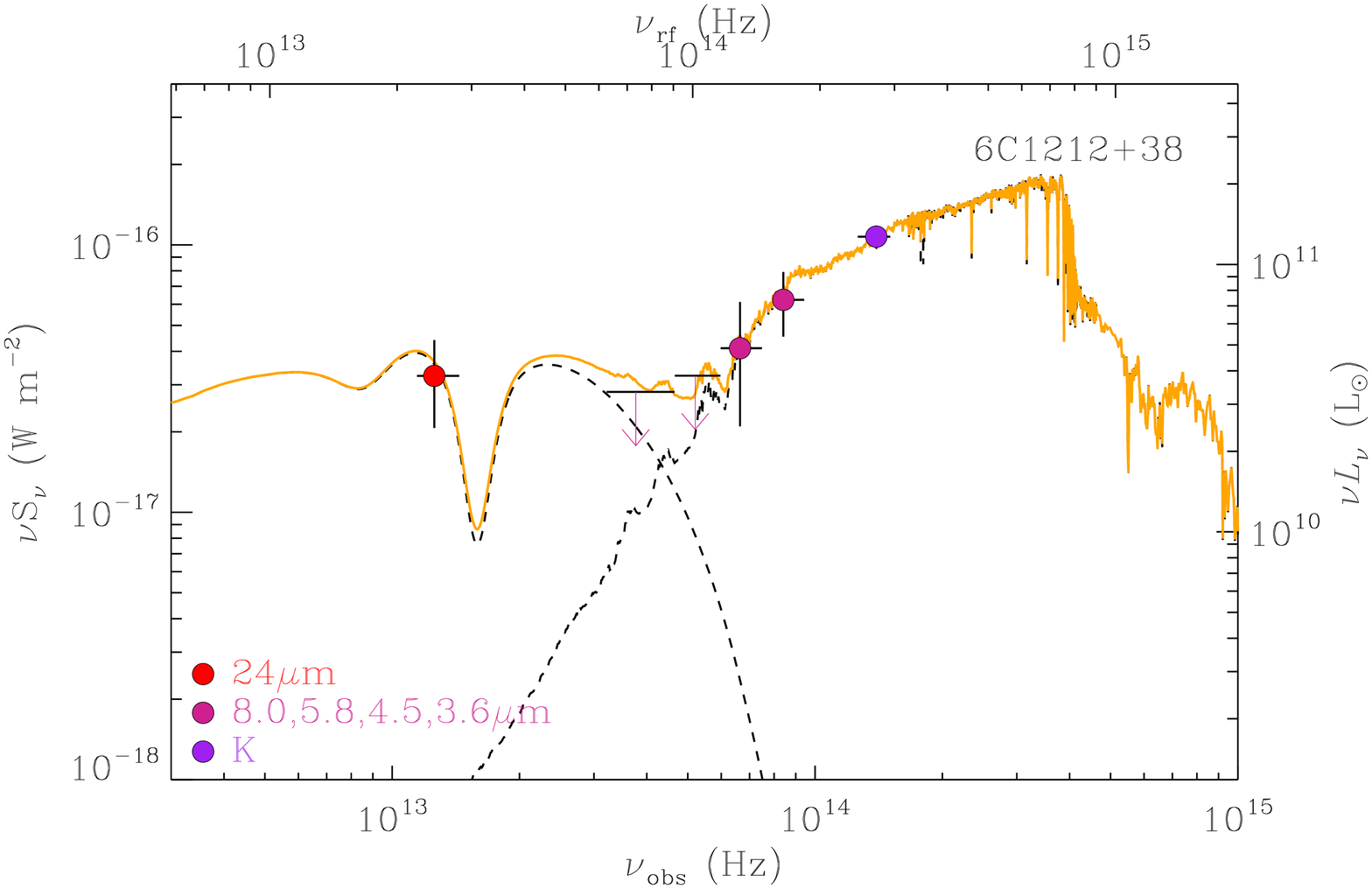} \\
\includegraphics[width=0.99\columnwidth]{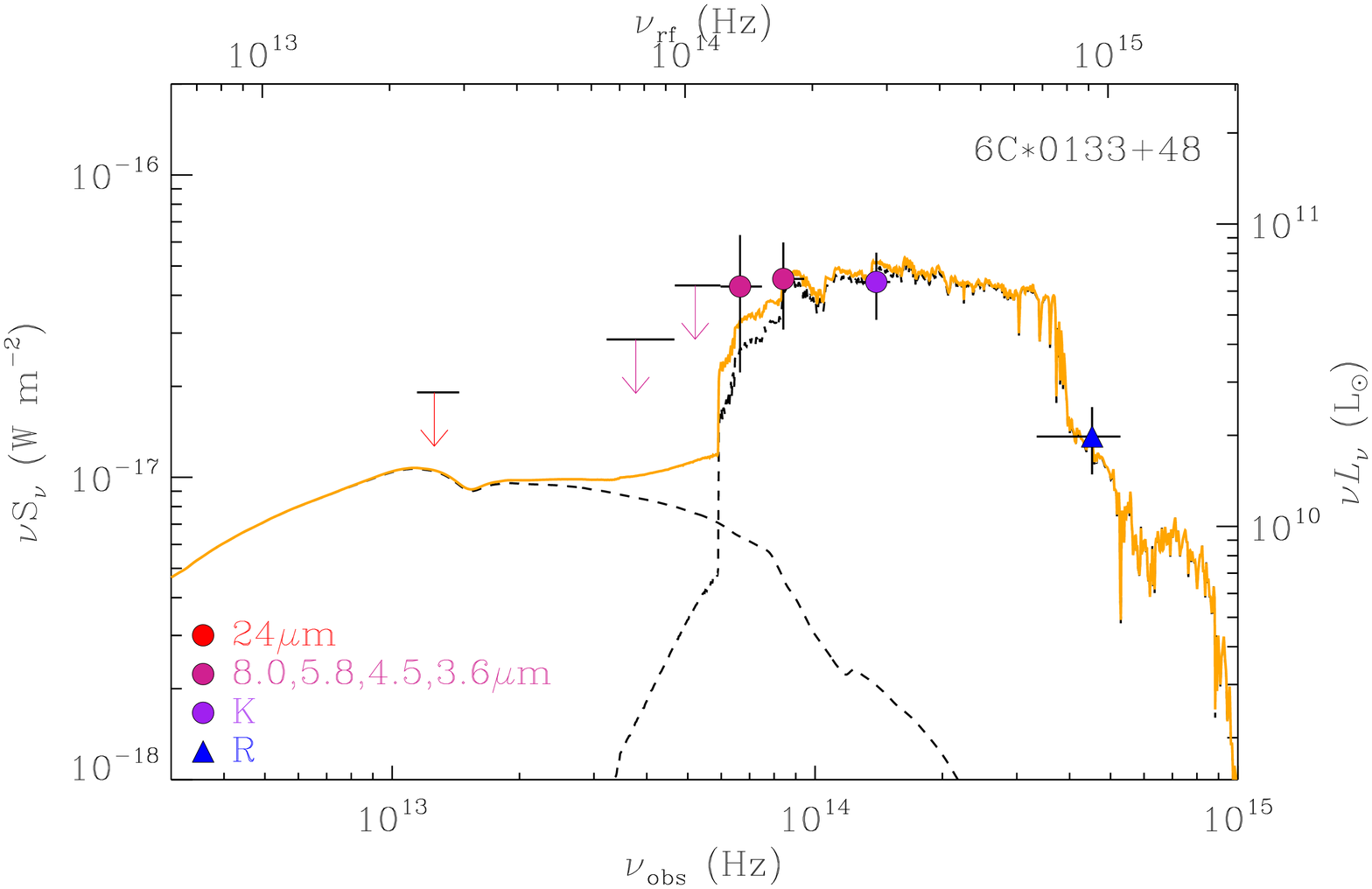}   
\includegraphics[width=0.99\columnwidth]{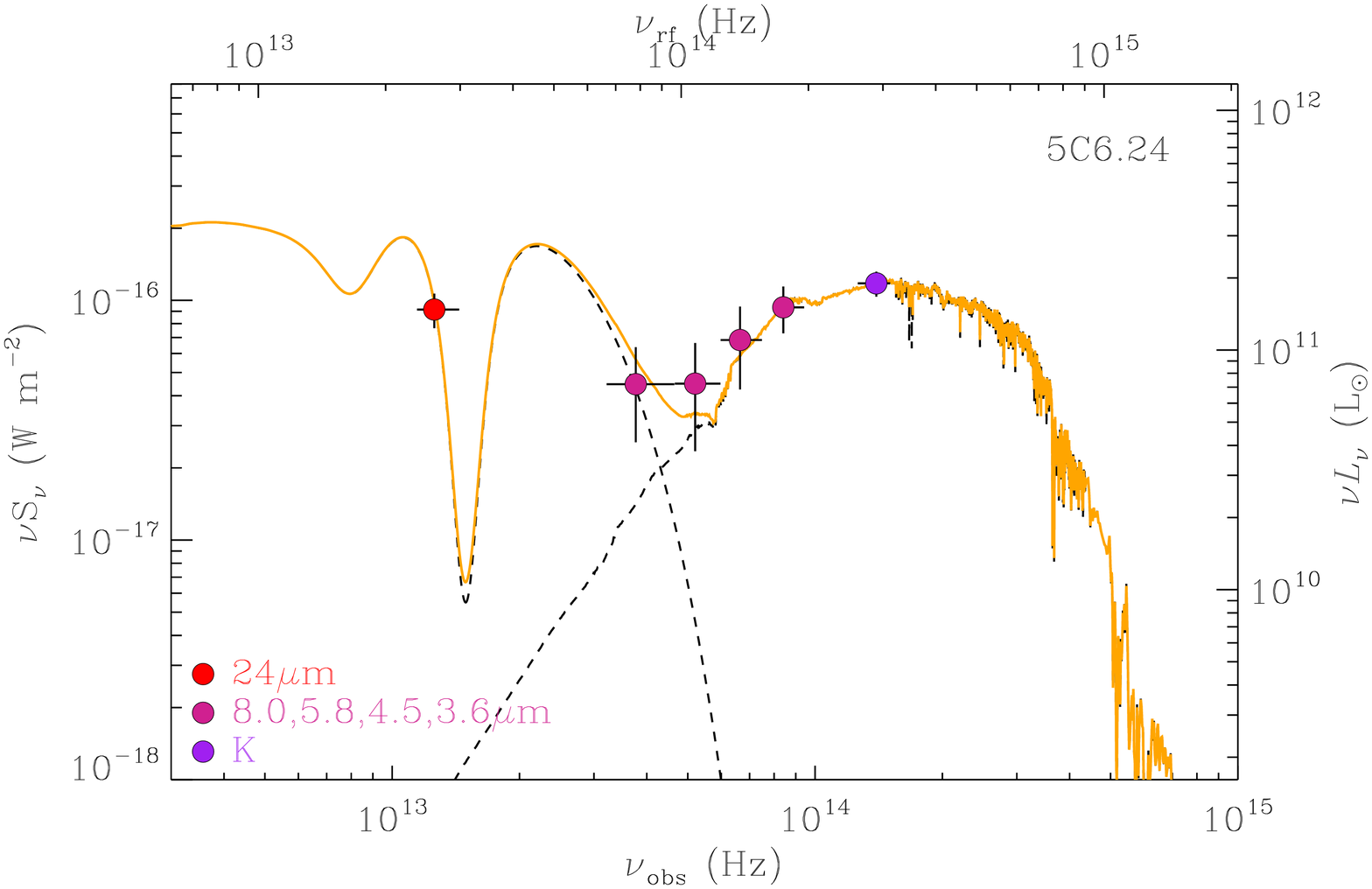} \\
\includegraphics[width=0.99\columnwidth]{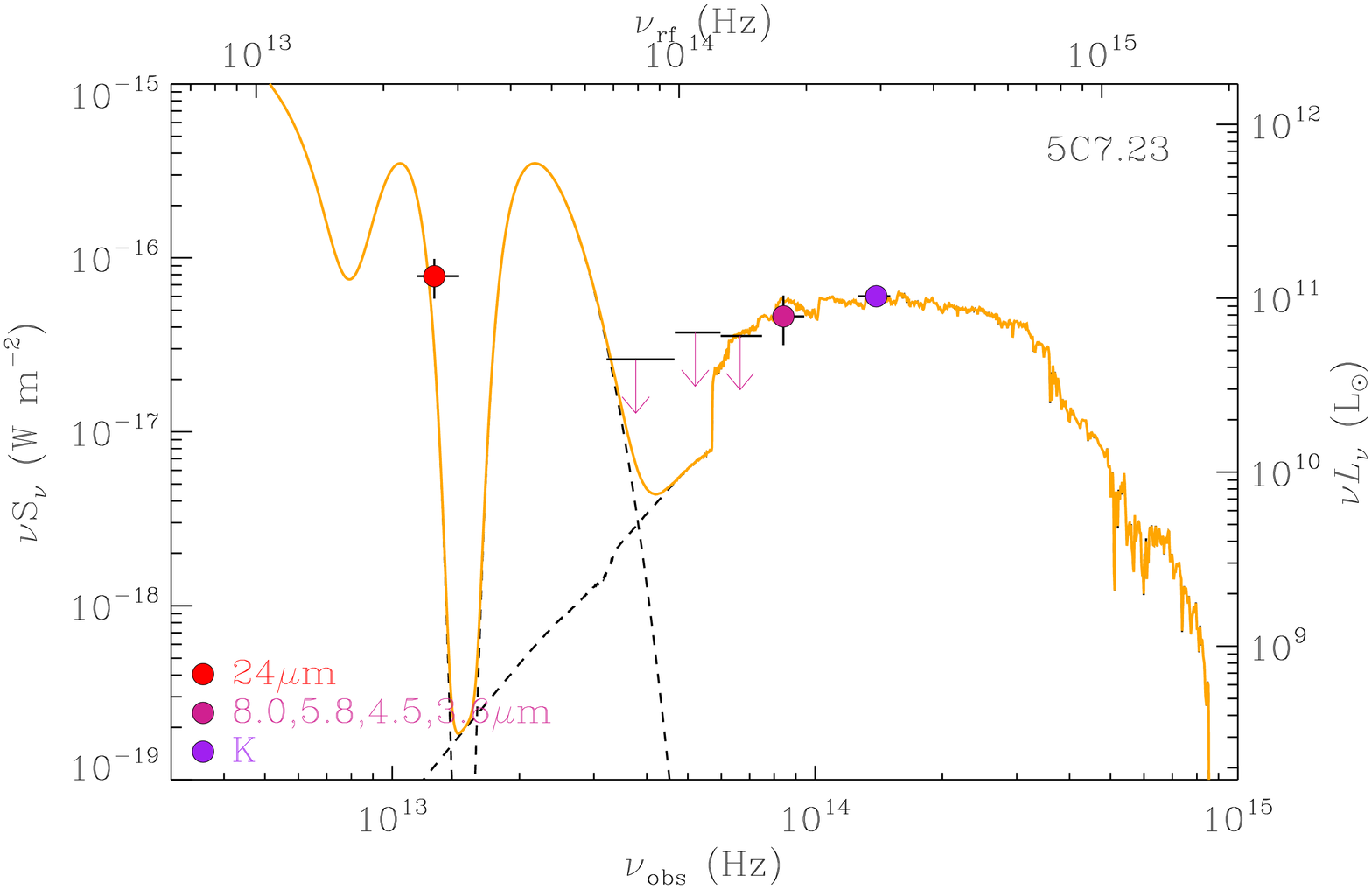}   
\includegraphics[width=0.99\columnwidth]{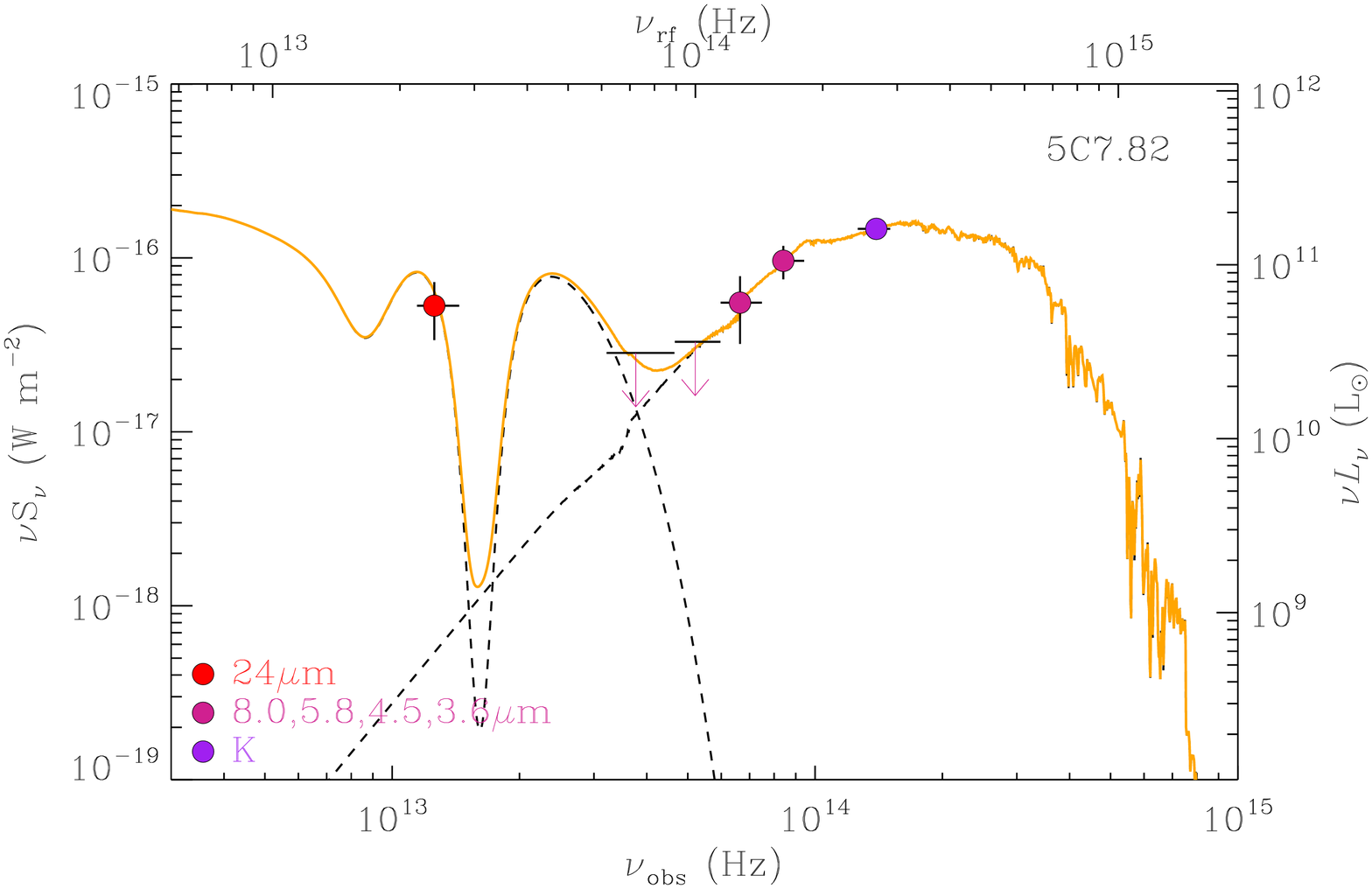} \\
\includegraphics[width=0.99\columnwidth]{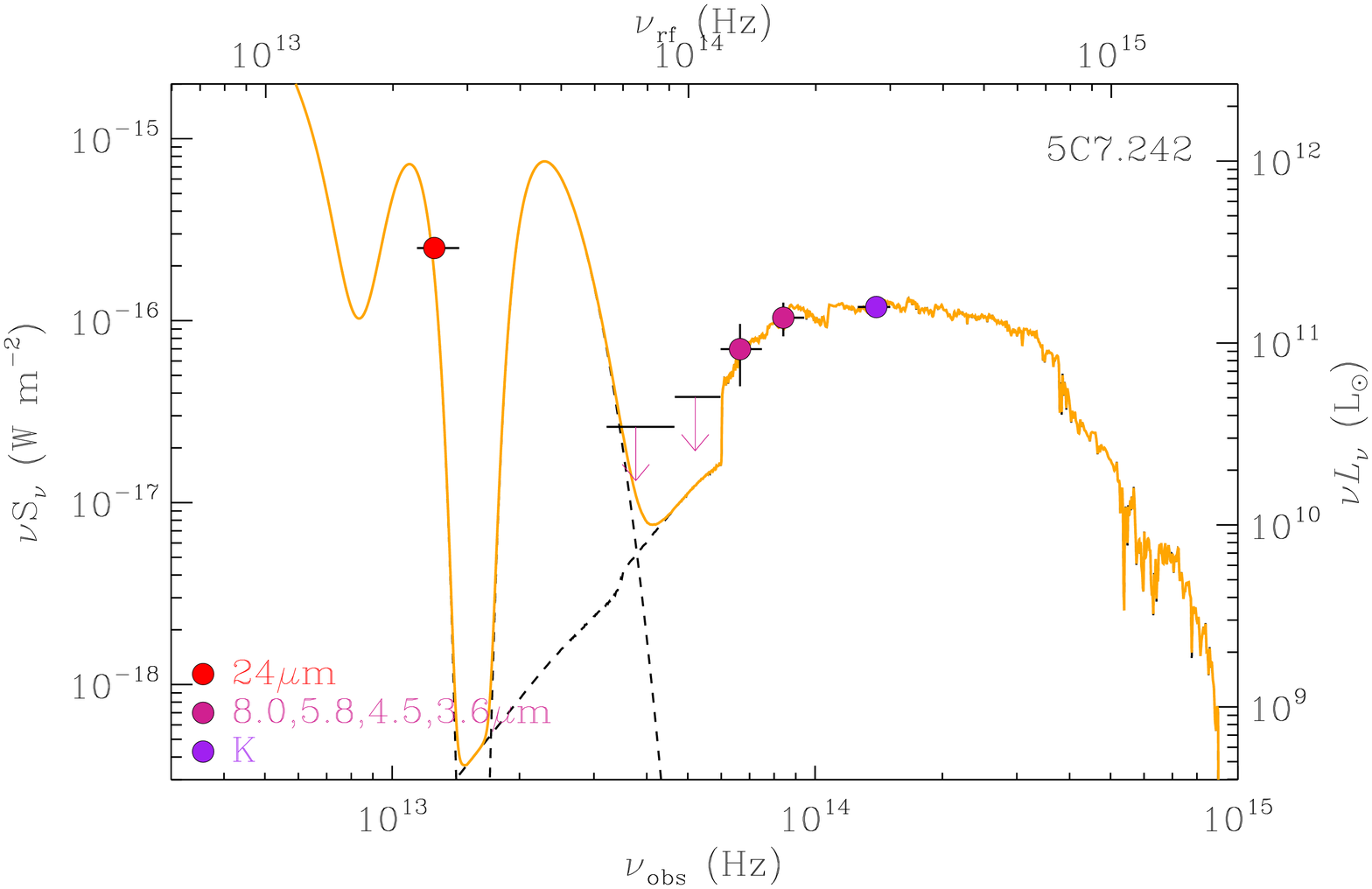}   
\includegraphics[width=0.99\columnwidth]{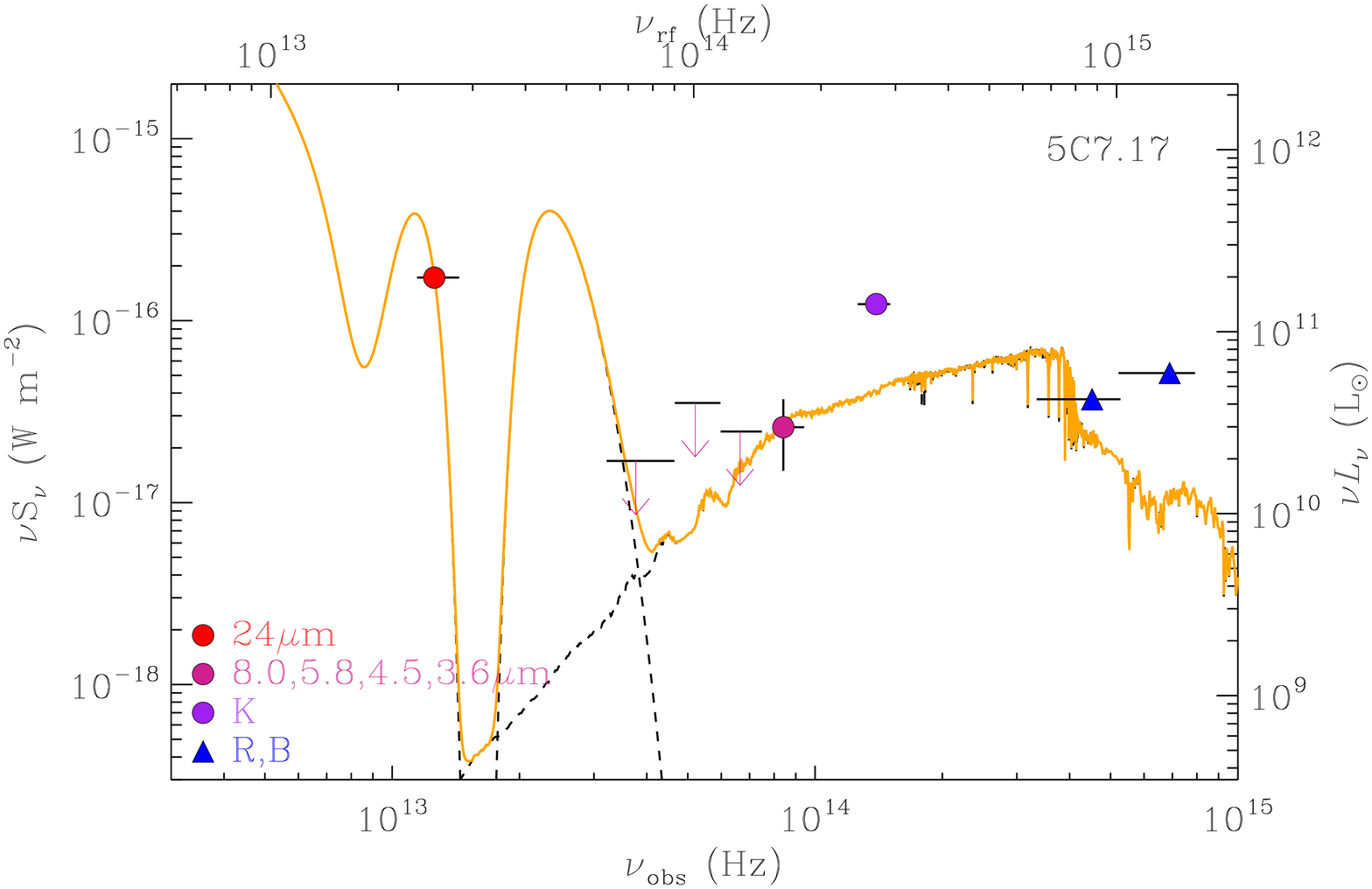} \\
\end{figure*}

\begin{figure*}
\includegraphics[width=0.99\columnwidth]{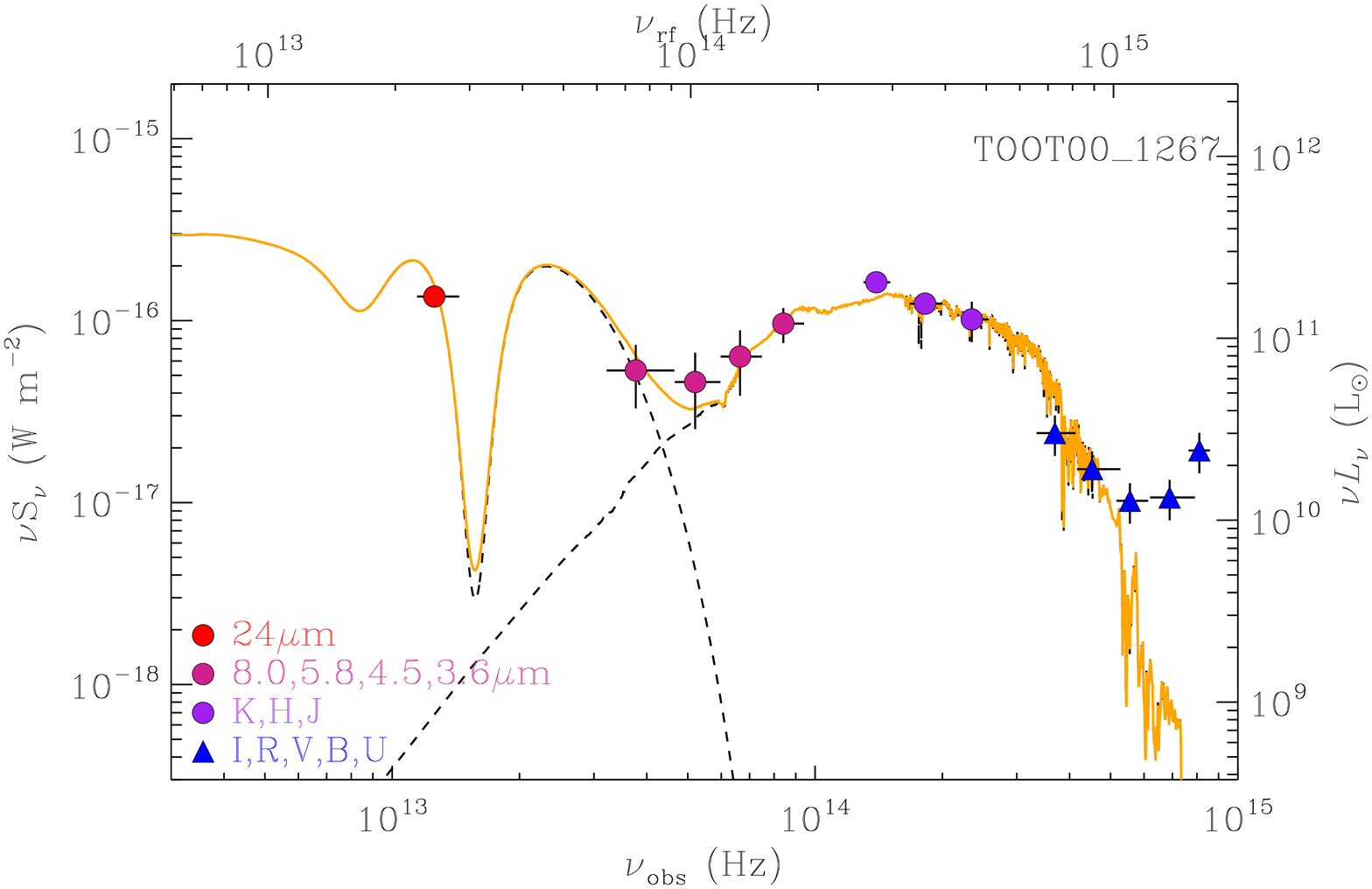}  
\includegraphics[width=0.99\columnwidth]{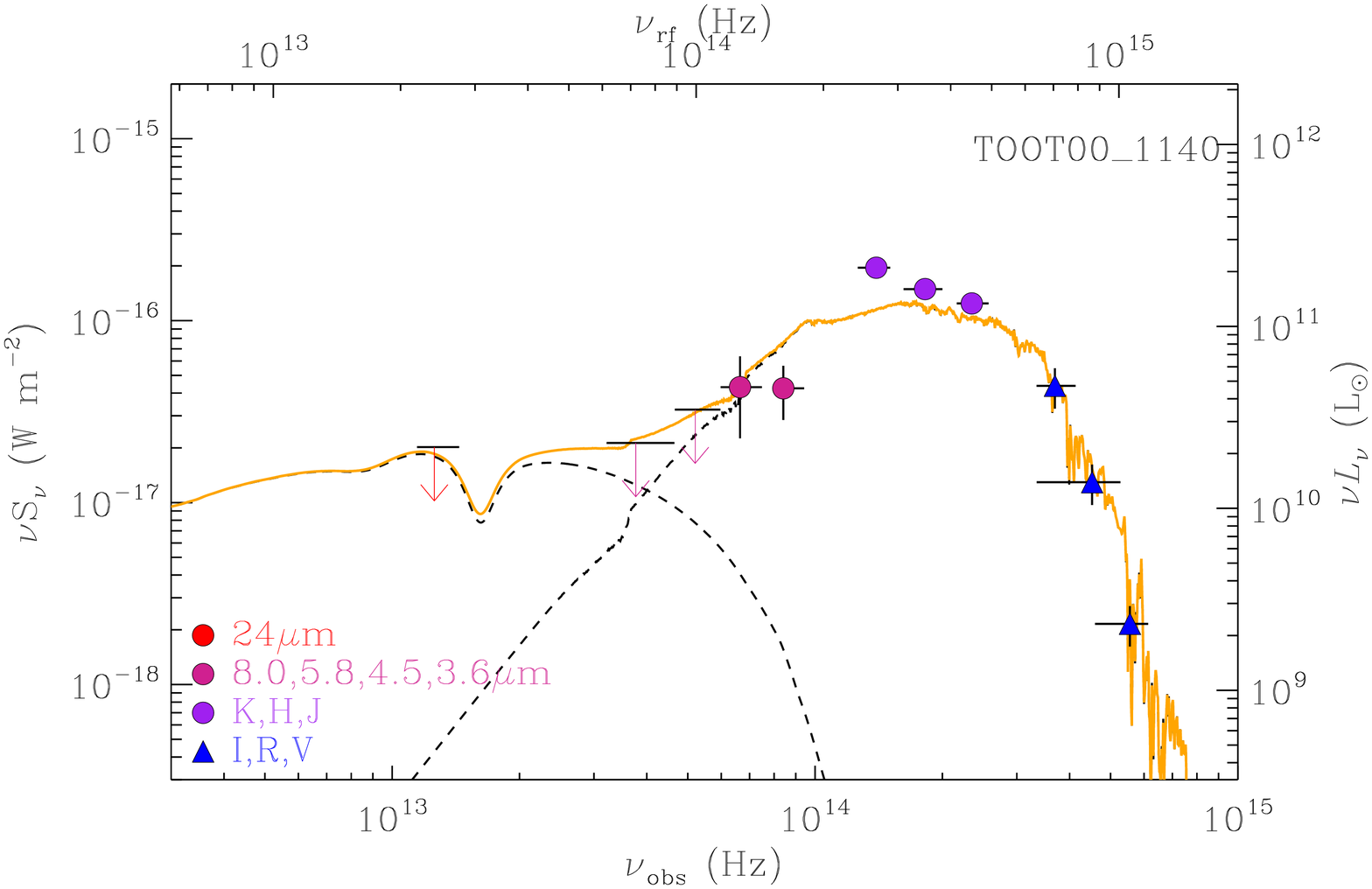} \\
\includegraphics[width=0.99\columnwidth]{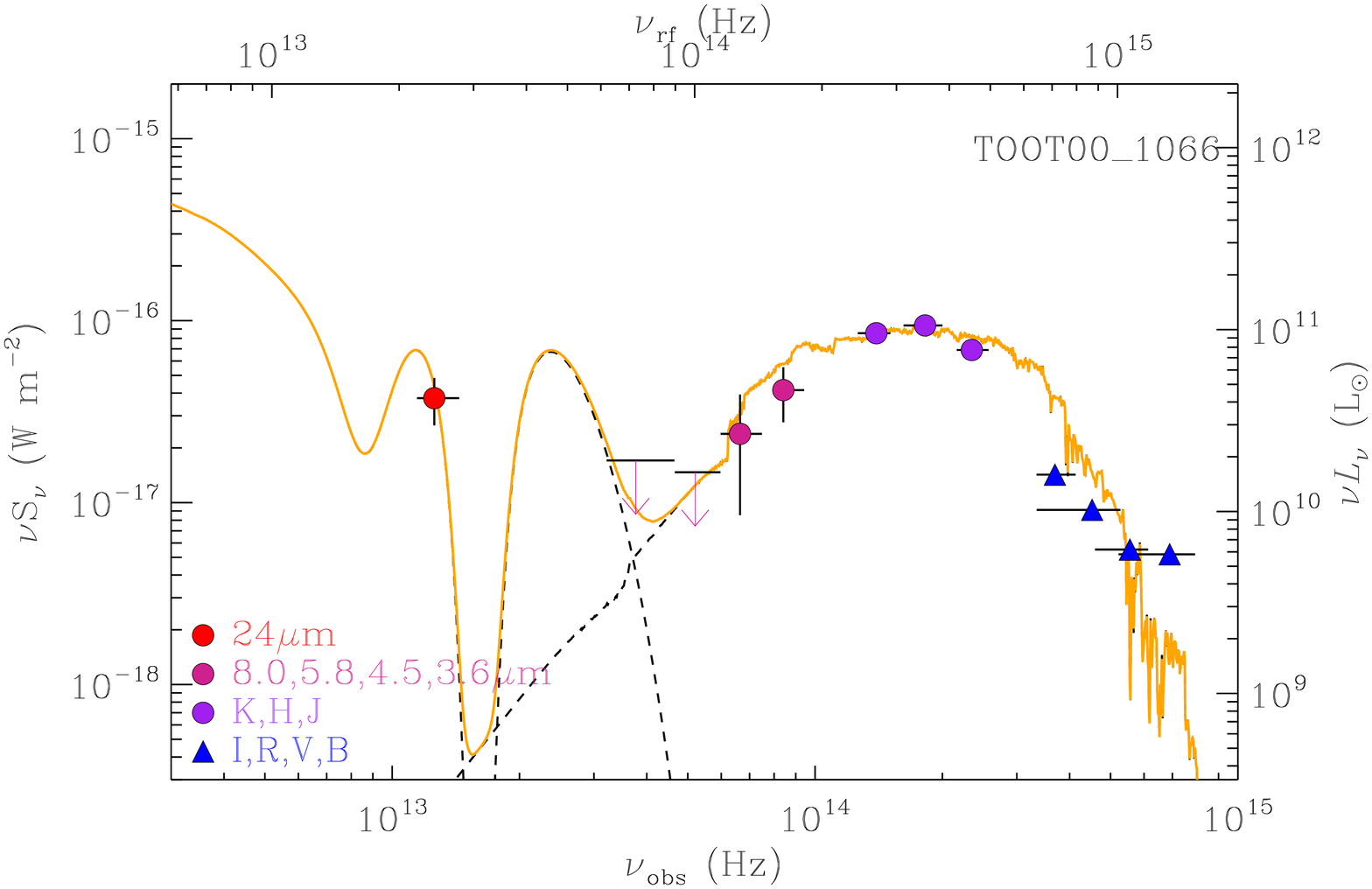}\\
\caption{SED fitting of the radio galaxies in our sample. The dashed
  lines represent both the quasar template affected by dust and the
  stellar template, whose sum is the model that best fits the
  data. The solid orange line represents the model that best fits the
  data. The points represent the data and their error bars. Symbols
  are as follows: red circles are for $12\rm \mu m$ MIPS data; violet
  circles are for $\rm 8\mu m$, $\rm 5.8\mu m$,$\rm 4.5\mu m$, and
  $\rm 3.6\mu m$ IRAC data; purple circles are for K, H, and J band
  data; blue squares are for Hubble F814W, F702W, and F606W band data;
  blue triangles are for I, R, V, B, and U band data; blue diamonds
  are for z, i, r, g, and u band data. The undetections are
  represented as downward arrows. }\label{fig:sed}
\end{figure*}

In Table~\ref{table:Av_Mgal} we present the best-fit values for
$A_{\rm V}$, $M_{\rm gal}$ and the stellar age of the SSP template, and in Figure~\ref{fig:sed} we show these best-fit SEDs
overlaid on the data. 

\begin{table*}
\caption{Best-fit parameters for our sample of radio galaxies. We note
  that the fits to 3C268.1 are unreliable due to a lack of data. \textbf{Column~1} gives the object name; \textbf{Column~2}
  gives the value of $A_{\rm V}$ of the best fit model for the SED of
  the object; \textbf{Column~3} gives the $1\sigma$ error associated
  with the $A_{\rm V}$ value; \textbf{Column~4} gives the value of
  $\log_{10}(M_{\rm gal})$ of the best fit model for the SED of the
  object; \textbf{Column~5} gives the $1\sigma$ error associated with
  the $\log_{10}(M_{\rm gal})$ value; \textbf{Column~6} gives the
  stellar age of the stellar synthesis population model that best
  fitted the data (BC03 - \citealt{2003MNRAS.344.1000B}; M05 -
  \citealt{2005MNRAS.362..799M}) \textbf{Column~7} gives the reduced
  $\chi^{2}$ of the best-fit model.} \centering
\begin{tabular}{l l l c c l l}
\hline\hline
 \multicolumn{1}{c}{Object} & \multicolumn{1}{c}{$A_{\rm V}$} & \multicolumn{1}{c}{$\sigma_{A_{\rm V}}$} & $\log_{10}(M_{\rm gal}/\rm M_{\sun})$ & $\sigma_{\log_{10}(M_{\rm gal})}$ & SSP model & $\chi_{\rm red}^{2}$ \\[0.5ex]
 \multicolumn{1}{c}{(1)}  &   \multicolumn{1}{c}{(2)}  & \multicolumn{1}{c}{(3)} & (4) & (5) & (6) & (7) \\[0.5ex]
\hline
3C280 & 19 & 2 & 11.20 & 0.04 & 0.5\,Gyr (M05) & 2.006\\
3C268.1 & 17 & 6 & 10.33 &0.42 & 0.5\,Gyr (M05) & 0.190\\
3C356 & 127 & 16 & 11.60 &0.04 & 1\,Gyr (BC03) & 3.114\\
3C184 & 1.5 & 0.4 & 11.82 &0.11 & 5\,Gyr (M05) & 2.655\\
3C175.1 & 123 & 117 & 11.58 &0.04 & 2\,Gyr (BC03) & 3.725\\
3C22 & 7 & 1.3 & 12 &0.04 & 6\,Gyr (BC03) & 2.669\\
3C289 & 209 & 48 & 11.95 &0.03 & 4\,Gyr (M05) & 4.943\\
3C343 & 158 & 26 & 11.63 &0.05 & 0.5\,Gyr (BC03) & 2.683\\
6C1256+36 & 180 & 66 & 11.89 &0.03 & 6\,Gyr (M05) & 1.837\\
6C1217+36 & 1 & 0.1 & 11.30 &0.03 & 1\,Gyr (M05) & 1.387\\
6C1017+37 & 141 & 63 & 11.06 &0.03 & 1\,Gyr (BC03) & 1.227\\
6C0943+39 & 99 & 46 & 11.16 &0.03 & 1\,Gyr (BC03) & 1.192\\
6C1257+36 & 112 & 16 & 11.57 &0.03 & 2\,Gyr (BC03) & 1.560\\
6C1019+39 & 110 & 125 & 11.20 &0.02 & 0.5\,Gyr (M05) & 5.025\\
6C1011+36 & 65 & 37 & 11.13 &0.03 & 1\,Gyr (M05) & 3.664\\
6C1129+37 & 164 & 17 & 11.16 &0.03 & 1\,Gyr (M05) & 0.858\\
6C*0128+39 & 4 & 85 & 11.10 &0.08 & 0.5\,Gyr (BC03) & Inf.\\
6C1212+38 & 39 & 34 & 11.20 &0.04 & 0.5\,Gyr (BC03) & 0.161\\
6C*0133+48 & 3 & 42 & 10.70 &0.08 & 0.5\,Gyr (M05) & 0.542\\
5C6.24 & 84.0 & 22 & 11.76 &0.04 & 3\,Gyr (BC03) & 0.351\\
5C7.23 & $>5$ &  & 11 & 0.05 & 1\,Gyr (M05) & Inf.\\
5C7.82 & 146 & 134 & 11.83 &0.04 & 5\,Gyr (M05) & 0.012\\
5C7.242 & 355 & 14 & 11.26 &0.04 & 1\,Gyr (M05) & 0.060\\
5C7.17 & 353 & 23 & 10.78 &0.17 & 0.5\,Gyr (BC03) & 0.179\\
TOOT00\_1267 & 103 & 15 & 11.93 &0.03 & 6\,Gyr (BC03) & 4.744\\
TOOT00\_1140 & 20 & 5 & 11.77 &0.03 & 6\,Gyr (M05) & 8.256\\
TOOT00\_1066 & 232 & 46 & 11.29 &0.03 & 2\,Gyr (M05) & 0.570\\
\hline
\label{table:Av_Mgal}
\end{tabular}
\end{table*}

We note that for several objects (3C289, 6C1019+39, 5C7.17,
6C1011+36, TOOT00\_1267, and TOOT00\_1066) there is an excess of emission at
the highest frequencies of the optical region of the SED that rises
above the galaxy template. This is thought to be either due to
scattered light from the central AGN or blue light from young
stars \citep[see e.g. ][ and references therein]{2010MNRAS.406.1841H}. This does not adversely impact our fit for the stellar masses
however, as the bulk of the stellar mass is traced by the
high-mass-to-light-ratio red stars.

To determine the error associated with each normalisation parameter of
the best fitting model we marginalised over the free parameters,
$A_{\rm V}$ and $M_{\rm gal}$, 
choosing a uniform (i.e. constant) prior of $0\leq A_{\rm
  V}\leq 400$ and $10.0\leq \log_{10}(M_{\rm gal})\leq 13.0$.
We marginalised the free parameters by integrating the
likelihood over the full range of one parameter, for instance $M_{\rm
  gal}$, in order to get the dependence of the total likelihood on the
other parameter, $A_{\rm V}$.
This then allows us to determine the uncertainty from the likelihood function in
relation to $M_{\rm gal}$, by determining the minimal width of the
likelihood function that contains $68$ per cent of all the likelihood. 
We then repeat the same
procedure to estimate $\sigma_{A_{\rm V}}$.


\section{Notes on individual objects}\label{sec:ind}

In this section we provide notes on the individual sources and their
best fit model.

\noindent
\textbf{3C280} is well fitted with a model with an $A_{\rm V}$ of $\sim19$ and a
stellar mass of $M_{\rm gal}\simeq 1.6\times10^{11}\,\rm M_{\sun}$.\\[-2ex]

\noindent
\textbf{3C268.1} lacks optical or near-IR data in the literature. The
fit of the stellar mass of the galaxy is thus not well-constrained,
and the best-fit value of $M_{\rm gal}\simeq2.1\times 10^{10}\,\rm
M_{\sun}$ is not reliable due to the lack of data points constraining
the fit. The SED fit of this galaxy yields a value for visual dust
extinction of $A_{\rm V}\simeq17$.\\[-2ex]

\noindent
\textbf{3C356} has two bright infrared galaxies at $z=1.079$
coincident with two radio cores, one northern and one southern,
approximately 5\,arcsec apart from each other. The identification of
the nucleus is a matter of debate in the literature
\citep{1990MNRAS.243P...1E,1992ApJ...385...61R,1997MNRAS.292..758B,2000MNRAS.311....1B}. We
assume the northern component to be the identification of the host
galaxy and radio jet. We found a model with a high $A_{\rm V}$ of
$\sim127$ and a stellar mass of $M_{\rm gal}\simeq 4\times
10^{11}\,\rm M_{\sun}$ to be a good fit for the SED of this
galaxy.\\[-2ex]

\noindent
\textbf{3C184} is well fit by a model with a stellar mass of $M_{\rm
  gal}\simeq 6.6\times 10^{11}\,\rm M_{\sun}$.\\[-2ex]


\noindent
\textbf{3C175.1} has a best fit model with a high $A_{\rm V}$ of $\sim123$,
with a rather high uncertainty of $\sigma_{A_{\rm V}}=116.8$, and a
stellar mass of $M_{\rm gal}\simeq 3.8\times10^{11}\,\rm M_{\sun}$.\\[-2ex]

\noindent
\textbf{3C22} is a reddened quasar
\citep{1995MNRAS.274..428R,1999MNRAS.306..828S}. As expected for this
type of object, it is well fit by a model with a low dust extinction,
$A_{\rm V}\sim 7$, and the SED is mainly dominated by the quasar light
component. We find that SMC-type dust fits the SED of 3C22 as well
as the MW type. We chose the MW type template for consistency.\\[-2ex]

\noindent
\textbf{3C289} has a best fit model with a very high $A_{\rm V}$ of
$\sim209$, and a stellar mass of $M_{\rm gal}\simeq 8.9\times10^{11}\,\rm
M_{\sun}$.\\[-2ex]

\noindent
\textbf{3C343:} The best fit model for the SED of 3C343 was found with
a visual extinction of $A_{\rm V}\sim 158$ and a stellar mass of $M_{\rm
  gal}\simeq 4.3\times10^{11}\,\rm M_{\sun}$. \\[-2ex]


\noindent
\textbf{6C1256+36} has a best fit model with a high $A_{\rm V}$ of
$\sim180$, and a stellar mass of $M_{\rm gal}=7.8\times10^{11}\,\rm
M_{\sun}$.\\[-2ex]

\noindent
\textbf{6C1217+36} is well fit by a model with a very low dust
extinction, $A_{\rm V}\approx1$, and the SED is mainly dominated by the
galaxy light component, with a mass of $M_{\rm
  gal}\simeq 2\times10^{11}\,\rm M_{\sun}$.\\[-2ex]

\noindent
\textbf{6C1017+37} has a best fit model with a high $A_{\rm V}$ of
$\sim141$, and a stellar mass of $M_{\rm gal}\simeq 4\times10^{11}\,\rm
M_{\sun}$. The $5.8$ and $4.5\,\rm\mu m$ fluxes are not detected at the
$2\sigma$ level and represent only limits.\\[-2ex]

\noindent
\textbf{6C0943+39} has a best fit model with a visual extinction of
$A_{\rm V}\sim 99$, and a stellar mass of $M_{\rm
  gal}\simeq 1.4\times10^{11}\,\rm M_{\sun}$. \\[-2ex]

\noindent
\textbf{6C1257+36} has a best fit model with a visual extinction of
$A_{\rm V}\sim112$, and a stellar mass of $M_{\rm
  gal}=3.7\times10^{11}\,\rm M_{\sun}$. \\[-2ex]

\noindent
\textbf{6C1019+39} is not detected at $8.0\,\rm\mu m$ or 24~$\mu$m at the $2\sigma$
level, which makes the fit of the quasar template more problematic. A
visual extinction of $\sim110$ is our preferred model, with a very high associated uncertainty of
$\sim125$. Despite this, the galaxy is well fit by a template with a stellar mass
of $1.6\times10^{11}\,\rm M_{\sun}$. \\[-2ex]


\noindent
\textbf{6C1011+36} has a best fit model with $A_{\rm V}\sim65$, and
a stellar mass of $M_{\rm gal}\simeq1.3\times10^{10}\,\rm M_{\sun}$. The
high flux of the optical data points compared to the galaxy 
model suggests that these shorter wavelengths might be contaminated by scattered light as
discussed in Section~\ref{sec:sed_fits}.\\[-2ex]

\noindent
\textbf{6C1129+37} is not detected at $8.0$ and $5.8\,\rm\mu m$ at a
$2\sigma$ level. The best fit model is found for $A_{\rm V}\sim 164$
and a stellar mass of $M_{\rm gal}=1.4\times10^{11}\,\rm
M_{\sun}$.\\[-2ex]

\noindent
\textbf{6C*0128+39} is not detected at $24$, $8.0$ or $5.8\,\rm\mu m$
at the $2\sigma$ level, which makes the fitting of the quasar light
difficult. We find the best fit quasar model to converge to
an $A_{\rm V}=4$, with a likelihood that decreases up to values of
$A_{\rm V}\sim 50$ but then maintains a constant likelihood
for higher values of $A_{\rm V}$. The estimated dispersion for this
parameter is thus extremely high $\sigma_{A_{\rm V}}\sim85$ and the
$\chi^{2}$ for the combined model diverges to infinity. Even though
there are not many data points constraining the stellar emission in
the  SED, the best fit model appears to provide a good fit to
the data, with a stellar mass of $M_{\rm
  gal}\simeq1.3\times10^{11}\,\rm M_{\sun}$, and thus we do not
exclude this object from our analysis.\\[-2ex]

\noindent
\textbf{6C1212+38} is not detected at $8.0$ and $5.8\,\rm\mu m$ at a
$2\sigma$ level. The best fit model is found for a high $A_{\rm
  V}\sim39$, with a standard deviation of the same order of magnitude,
$\sigma_{A_{\rm V}}\sim 34$, and for a stellar mass of $M_{\rm
  gal}\simeq1.6\times10^{11}\,\rm M_{\sun}$.\\[-2ex]

\noindent
\textbf{6C*0133+48} is not detected at $24$, $8.0$ or $5.8\,\rm\mu m$
at the $2\sigma$ level, and thus the quasar light component fit is not
reliable, with a likelihood very similar to that for 6C*0128+39. The best fit model was found for $A_{\rm
  V}=3.0$ with a much higher standard deviation of $\sigma_{A_{\rm
    V}}\sim42$. However, the stellar dominated part of the SED appears
well fitted by a stellar mass of $M_{\rm gal}\simeq 5 \times10^{10}\,\rm
M_{\sun}$.\\[-2ex]

\noindent
\textbf{5C6.24} has a best fit model with $A_{\rm V}\sim 84$, and a
stellar mass of $M_{\rm gal}\simeq5.7\times10^{11}\,\rm M_{\sun}$.\\[-2ex]

\noindent
\textbf{5C7.23} is not detected at $8.0$, $5.8$ and $4.5\,\rm\mu m$ at
a $2\sigma$ level, and the quasar light component fit does not
converge to a specific value of $A_{\rm V}$. We determine a lower
limit for the visual dust extinction of $A_{\rm V}>5$. The stellar
mass is constrained mainly with the $3.6\,\rm\mu m$ and K-band data
points and a best value of $M_{\rm gal}\simeq1.1\times10^{11}\,\rm
M_{\sun}$ is found.\\[-2ex]

\noindent
\textbf{5C7.82} is not detected at $8.0$ and $5.8\,\rm\mu m$ at a
$2\sigma$ level, and the quasar light component fit has a rather high
error associated. The best fit parameter values is $A_{\rm
  V}=146$. The stellar dominated part of the SED looks well
constrained by a model with stellar mass $M_{\rm
  gal}=6.8\times10^{11}\,\rm M_{\sun}$.\\[-2ex]

\noindent
\textbf{5C7.242} is not detected at $8.0$ and $5.8\,\rm\mu m$ at the
$2\sigma$ level, and thus the quasar light component fit is not
entirely reliable. The best fit model was found for an extremely high value
of visual extinction, $A_{\rm V}=355$, and a stellar mass of $M_{\rm
  gal}\simeq1.8\times10^{11}\,\rm M_{\sun}$.\\[-2ex]

\noindent
\textbf{5C7.17} is not detected at $8.0$, $5.8$ and $4.5\,\rm\mu m$ at
the $2\sigma$ level, and the quasar light component fit is quite
problematic. The best model was found with an extremely high value of
visual extinction, $A_{\rm V}=353$. The stellar mass value that
produces the best model is $M_{\rm gal}\simeq6\times10^{10}\,\rm
M_{\sun}$, with a set of data that is also rather hard to fit, and
thus with a large associated error. The high fluxes of the
optical data points compared to the stellar population model suggests
that these might be contaminated by scattered light as discussed in
Section~\ref{sec:sed_fits} or have ongoing star formation.\\[-2ex]

\noindent
\textbf{TOOT00\_1267} has a best fit model with a visual extinction
value of $A_{\rm V}=103$, and a stellar mass of $M_{\rm
  gal}\simeq8.5\times10^{11}\,\rm M_{\sun}$. The high fluxes of the optical
data points compared to the galaxy light model suggest that
these might be contaminated by scattered light or ongoing star formation.\\[-2ex]

\noindent
\textbf{TOOT00\_1140} is not detected at $8.0$, $5.8$ and $4.5\,\rm\mu
m$ at the $2\sigma$ level. The best fit model has $A_{\rm V}\sim
20$. The stellar component is well constrained by a model with a
stellar mass of $M_{\rm gal}\simeq 5.9\times10^{11}\,\rm M_{\sun}$
.\\[-2ex]

\noindent
\textbf{TOOT00\_1066:} is not detected at $8.0$ and $5.8\,\rm\mu m$ at
the $2\sigma$ level. The best fit model is found with a high $A_{\rm
  V}\sim 232$, and for a stellar mass of $M_{\rm
  gal}\simeq1.9\times10^{11}\,\rm M_{\sun}$.\\

\section{Discussion}\label{sec:discuss}


The SED fitting provides a good estimation for important physical
properties of the radio galaxies such as bolometric luminosity,
extinction properties and stellar mass. The stellar mass, in particular,
grants, for elliptical galaxies, an accurate estimation of its
supermassive black hole mass. Together with the bolometric luminosity,
the Eddington weighted accretion rates of the sample can be
inferred. In this section, first the extracted physical properties of
our sample of radio galaxies are discussed and then the classification
of each galaxy into HEG or LEG is used to investigate the
HEG/LEG dichotomy at $z\sim1$.



\subsection{Physical properties of the sample}


\subsubsection{Visual extinction}\label{sec:discussion_av}

Our visual extinction estimations are complicated by the fact that at
$z\sim1$, the $24\,\mu$m data point lies on the edge of the $10\,\mu$m
silicate feature and thus more data points, or spectroscopic data,
would be desirable to better constrain the values of $A_{\rm
  V}$. Nonetheless, previous work in the literature, such as
\citet{2007ApJ...660..117C}, agree reasonably well with the values
found. In their study of galaxies and quasars from the 3CRR survey at
$0.4\leq z\leq 1.2$, \citet{2007ApJ...660..117C} fit IRS and MIPS
Spitzer data, and other measurements from the literature, using models
with a synchrotron and a dust component. They consider two variations
of the dust component, one with a screen of cooler dust, and another
with a mixture of warm dust emitting in the MIR and cooler dust. The
visual extinction of their sample ranges from $0<A_{\rm V}<40$ for the
screen dust component model, and $0<A_{\rm V}<150$ for the mixed dust
component model. In particular, for the common objects in our sample
and their sample, we find that the values of $A_{\rm V}$ for 3C268.1,
3C280 and 3C22 agree well with those of \citet{2007ApJ...660..117C}
for a screen dust component. For 3C343 our model is best fit with a
dust component with an $A_{\rm V}=158$, whereas
\citet{2007ApJ...660..117C} find an $A_{\rm V}\sim31$ for a mixed dust
component, and $A_{\rm V}\sim22$ for a screen dust component, both of
them much lower than our value. However, we note that this object does
exhibit a steeply rising slope towards mid-infrared wavelengths in the
{\em Spitzer}-IRS data, similar to what we find with our
photometry. Our estimated high AGN extinction is inconsistent with the
classification of 3C343 as a quasar by
\citet{2007ApJ...660..117C}. However, it is possible that this
difference results from an $A_{\rm V}$ overestimation of our model due
to the proximity of the $24\,\mu m$ data point to the $10\,\mu m$
feature.

%


\noindent
\subsubsection{Stellar mass and Black hole mass}

Pioneering studies like those of \citet{1995ARA&A..33..581K},
\citet{1997AJ....114.1771F} and \citet{1998AJ....115.2285M}
established that the hot stellar component of galaxies - i.e. the
bulge - is proportional to their black hole mass. This relation became
known as the $M_{\rm BH}$-$M_{\rm bulge}$ relation, and
\citet{1998AJ....115.2285M} found the black hole-to-bulge mass ratio
to be approximately 0.006. More recent studies of this relation
\citep[e.g.][]{2004ApJ...604L..89H} find that the median black hole
mass is $0.14$ per cent of the bulge mass for nearby galaxies, $M_{\rm
  BH}=0.0014 M_{\rm bulge}$. We note that including the intrinsic
  scatter of the $M_{\rm BH}-M_{\rm bulge}$ relation does not make any
  difference to our results, as the uncertainties in our stellar mass
  estimates dominate the error budget.



We can use the stellar mass of the galaxy, $M_{\rm gal}$, provided by
the SED fitting, along with the $M_{\rm BH}$-$M_{\rm bulge}$ relation
to estimate the black hole mass, $M_{\rm BH}$. We assume that there is
no significant evolution of the $M_{\rm BH}$-$M_{\rm bulge}$ relation
at $z\sim1$ from the local $M_{\rm BH}/M_{\rm bulge}$. In fact,
\cite{2006MNRAS.368.1395M} shows that for low redshift ($z \lesssim
1$) radio-loud AGN the black hole to spheroid mass ratio lies within
the uncertainties of that found in the local
Universe. \cite{2010A&A...522L...3S} also find that obscured AGNs at
$z \sim 1\--2$ are fully consistent with the local $M_{\rm BH}$-$M_{\rm
  bulge}$ relation. We thus use the local Universe relation $M_{\rm
  BH}\sim 0.0014 M_{\rm bulge}$. The values found for $M_{\rm gal}$
and $M_{\rm BH}$ are presented in Table~\ref{table:results}.

\begin{table*}
  \caption{\textbf{Column~1} gives the object name; \textbf{Column~2}
    gives the black hole masses; \textbf{Column~3} gives the
    bolometric luminosity; \textbf{Column~4} gives the
  accretion rate; \textbf{Column~5} gives the Eddington ratio;
  \textbf{Column 6} gives the optical classification of the object
  according to the literature. 'HEG' stands for high-excitation radio
  galaxy, 'LEG' stands for low-excitation radio galaxy, RQ stands for
  reddened quasar; \textbf{Column 7} gives the reference for the
  optical classification in Column~6 (for the 3CRR, 6CE and 6C*
  objects) or the reference for the optical spectra (for the 7CRS and
  TOOT00 objects). References are as follows: Grimes -
  www-astro.physics.ox.ac.uk/$\sim$sr/grimes; J01 -
  \citealt{2001MNRAS.326.1585J}; JR97 - \citealt{1997MNRAS.286..241J};
  REL01 - \citealt{2001MNRAS.322..523R}; W03 -
  \citealt{2003MNRAS.339..173W}; V10 - \citealt{2010MNRAS.401.1709V}}
    \centering
\begin{tabular}{l c c c c c c}
\hline\hline
 \multicolumn{1}{c}{Object} & $\log_{10}(M_{\rm BH}/\rm M_{\sun})$  & $\log_{10}(L_{\rm bol} /\rm W)$ & $\log_{10}(\dot{M}/\rm~M_{\sun}/yr)$ & $\lambda$ & Opt. Class &  Ref \\[0.5ex]
 \multicolumn{1}{c}{(1)}  &   \multicolumn{1}{c}{(2)}  & (3) & (4) & (5) & (6) & (7)\\[1ex]
\hline
3C280 & 8.346$\pm$0.425 & 39.707$\pm$0.012 & 0.953$\pm$0.012 & 1.765$^{+0.882}_{-0.710}$ & HEG & JR97 \\
3C268.1 & 7.476$\pm$4.204 & 38.689$\pm$0.057 & -0.065$\pm$0.057 & 1.257$^{+104.822}_{-1.245}$ & HEG & JR97 \\
3C356 & 8.746$\pm$0.425 & 39.435$\pm$0.021 & 0.681$\pm$0.021 & 0.375$^{+0.178}_{-0.155}$ & HEG & JR97 \\
3C184 & 8.966$\pm$1.062 & 38.608$\pm$0.110 & -0.146$\pm$0.110 & 0.034$^{+0.065}_{-0.024}$ & HEG & JR97 \\
3C175.1 & 8.726$\pm$0.425 & 38.578$\pm$0.088 & -0.176$\pm$0.088 & 0.055$^{+0.027}_{-0.022}$ & HEG & JR97 \\
3C22 & 9.366$\pm$0.425 & 39.813$\pm$0.010 & 1.059$\pm$0.010 & 0.215$^{+0.103}_{-0.088}$ & RQ & JR97 \\
3C289 & 9.096$\pm$0.255 & 39.271$\pm$0.020 & 0.517$\pm$0.020 & 0.115$^{+0.026}_{-0.034}$ & HEG & JR97 \\
3C343 & 8.776$\pm$0.467 & 39.594$\pm$0.014 & 0.840$\pm$0.014 & 0.506$^{+0.286}_{-0.219}$ & HEG & Grimes \\
6C1256+36 & 9.036$\pm$0.297 & 39.005$\pm$0.040 & 0.251$\pm$0.040 & 0.072$^{+0.020}_{-0.024}$ & HEG? & REL01 \\
6C1217+36 & 8.446$\pm$0.297 & 38.331$\pm$0.151 & -0.423$\pm$0.151 & 0.059$^{+0.022}_{-0.020}$ & HEG? & REL01 \\
6C1017+37 & 8.206$\pm$0.340 & 38.855$\pm$0.051 & 0.101$\pm$0.051 & 0.343$^{+0.127}_{-0.118}$ & HEG & REL01 \\
6C0943+39 & 8.306$\pm$0.297 & 39.082$\pm$0.033 & 0.328$\pm$0.033 & 0.459$^{+0.123}_{-0.154}$ & LEG? & REL01 \\
6C1257+36 & 8.718$\pm$0.263 & 38.678$\pm$0.056 & -0.076$\pm$0.056 & 0.070$^{+0.016}_{-0.021}$ & HEG & REL01 \\
6C1019+39 & 8.346$\pm$0.212 & 38.182$\pm$0.263 & -0.572$\pm$0.263 & 0.053$^{+0.017}_{-0.019}$ & LEG? & REL01 \\
6C1011+36 & 8.276$\pm$0.297 & 38.911$\pm$0.049 & 0.157$\pm$0.049 & 0.332$^{+0.096}_{-0.108}$ & HEG & REL01 \\
6C1129+37 & 8.306$\pm$0.340 & 38.740$\pm$0.061 & -0.014$\pm$0.061 & 0.209$^{+0.072}_{-0.074}$ & HEG? & REL01 \\
6C*0128+39 & 8.246$\pm$0.849 & 37.795$\pm$0.328 & -0.959$\pm$0.328 & 0.027$^{+0.040}_{-0.018}$ & HEG? & J01 \\
6C1212+38 & 8.346$\pm$0.425 & 38.097$\pm$0.164 & -0.657$\pm$0.164 & 0.043$^{+0.023}_{-0.019}$ & LEG & REL01 \\
6C*0133+48 & 7.846$\pm$0.849 & 37.690$\pm$0.702 & -1.064$\pm$0.702 & 0.054$^{+0.108}_{-0.038}$ & LEG? & J01 \\
5C6.24 & 8.906$\pm$0.425 & 38.682$\pm$0.072 & -0.072$\pm$0.072 & 0.046$^{+0.023}_{-0.019}$ & HEG & W03 \\
5C7.23 & 8.196$\pm$0.500 & 38.640$\pm$0.115 & -0.114$\pm$0.115 & 0.214$^{+0.134}_{-0.098}$ & HEG & W03 \\
5C7.82 & 8.976$\pm$0.382 & 38.278$\pm$0.167 & -0.476$\pm$0.167 & 0.015$^{+0.007}_{-0.006}$ & LEG? & W03 \\
5C7.242 & 8.406$\pm$0.382 & 39.036$\pm$0.035 & 0.282$\pm$0.035 & 0.328$^{+0.144}_{-0.122}$ & HEG? & W03 \\
5C7.17 & 7.926$\pm$1.699 & 38.811$\pm$0.056 & 0.057$\pm$0.056 & 0.590$^{+2.798}_{-0.503}$ & HEG & W03 \\
TOOT00\_1267 & 9.076$\pm$0.255 & 38.743$\pm$0.049 & -0.011$\pm$0.049 & 0.036$^{+0.007}_{-0.011}$ & HEG & V10 \\
TOOT00\_1140 & 8.916$\pm$0.255 & 37.779$\pm$0.294 & -0.975$\pm$0.294 & 0.006$^{+0.002}_{-0.002}$ & LEG & V10 \\
TOOT00\_1066 & 8.436$\pm$0.340 & 38.136$\pm$0.130 & -0.618$\pm$0.130 & 0.039$^{+0.016}_{-0.014}$ & LEG? & V10 \\
\hline
\label{table:results}
\end{tabular}
\end{table*}

The SED models of the observed photometry bands provide a very good
fit for the stellar mass of our objects. In fact, in similar studies
to this, \citet{2007ApJS..171..353S} found that, for a sample of 70
radio galaxies at $1<z<5.2$, their broad-band photometry implied
stellar masses in the range $10^{11}-10^{12}\,\rm M_{\odot}$, with a
mean mass of $\sim 10^{11.55}\,\rm M_{\odot}$ up to redshifts of
$z=3$. The radio galaxies in our sample have an average stellar mass
of $10^{11.59}\,\rm M_{\odot}$ (excluding 3C268.1, for which the
stellar mass estimation is poorly constrained), consistent with the
study of \citet{2007ApJS..171..353S}.

\citet{2006MNRAS.368.1395M} estimated the black hole mass of powerful
radio quasars from the 3CRR survey with redshifts between
$0<z<2$ using the virial mass estimator for
Mg{\sc ii} emission line \citep{2002MNRAS.337..109M,2004MNRAS.352.1390M},
and estimated black hole masses in the range $10^{8.3}\,\rm
M_{\odot}-10^{10.1}\,\rm M_{\odot}$. The radio galaxies in our sample at
$z\sim1$ have black hole masses in the range  $10^{7.8}\,\rm
M_{\odot}-10^{9.4}\,\rm M_{\odot}$ if we exclude 3C268.1. These are 
consistent with the range of values found in the literature for
similar objects. This reinforces the view that powerful radio sources
reside in the most massive galaxies with the most massive black holes.\\


\subsubsection{Accretion rate}\label{sec:accretion}

The bolometric radiative power of an AGN, $L_{\rm bol}$, is proportional to
the accretion rate of the black hole, $\dot{M}$, and to the fraction
of accreted mass that is radiated, i.e. the radiative efficiency,
$\epsilon$, through the expression:
\begin{equation}\label{eq:lbol}
L_{\rm bol}={\epsilon\dot{M}c^{2}}.
\end{equation}
Assuming that $\epsilon$ takes the fiducial value of $0.1$
\citep[e.g.][]{2004MNRAS.351..169M,2004MNRAS.354.1020S,2009ApJ...692..964M},
we can determine the accretion rate of our sources using their
estimated bolometric luminosity, $L_{\rm bol}$. We calculate the values of $L_{\rm bol}$ from the rest-frame 12$\mu$m
luminosity, as in \citet{2011MNRAS.411.1909F}, using a bolometric
correction of 8.5 \citep[e.g.][fig.12]{2006ApJS..166..470R}, 
i.e. $L_{\rm bol}=8.5\lambda L_{12\mu\rm m}$. Our estimated values for
the accretion rate  are shown
in Table~\ref{table:results}.\\


As reviewed by \citet[sec.3.1]{1997iagn.book.....P}, according to the
most widely accepted model, radiatively efficient AGNs are powered by
gravitational infall of material onto a supermassive black hole, with
this material achieving high temperatures in the dissipative accretion
disk (e.g. \citealt{1964ApJ...140..796S}). For the material in the
galaxy to be in hydrostatic equilibrium, the inward gravitational
force needs to be balanced by the outwards radiation pressure. The
Eddington luminosity, $L_{\rm Edd}$, is the maximum luminosity that a
body needs to be radiating to remain in hydrostatic equilibrium in the
case of spherical accretion, assuming a pure ionized hydrogen
plasma. This energy is a function of the mass of the system and is
given by $L_{\rm Edd}=1.3\times 10^{31} \left( \frac{M_{\rm
    BH}}{M_{\odot}} \right) \rm \,W$. The Eddington ratio, $\lambda$
is therefore simply,
\begin{equation}\label{eq:edd}
\lambda\equiv \frac{L_{\rm bol}}{L_{\rm Edd}}.
\end{equation}
This gives an estimate of the actual accretion rate of the
AGN, compared to the maximal Eddington accretion rate.

Having the black hole mass and the accretion rate, we use
equations~\ref{eq:lbol} and \ref{eq:edd} to estimate the Eddington ratio of the sources in our sample. The values obtained are shown in
Table~\ref{table:results}. The accretion rate properties of our sample
are detailed in the following section.

\subsection{The HEG/LEG dichotomy}\label{sec:heg_leg}

Table~\ref{table:results} shows how the radio galaxies in our sample
are distributed in terms of their optical/near-IR spectral
classification. This classification was either taken from the
literature, where available (see references in
Table~\ref{table:results}), or determined by inspecting the optical
spectra of the objects. We label the reddened quasar 3C22
\citep{1995MNRAS.274..428R,1999MNRAS.306..828S}, and distinguish it as
a different class (RQ).


\subsubsection{Relation between $L_{\rm 151MHz}$ and $\lambda$}\label{sec:l151_eddr}

\begin{figure}
  \begin{center}
    \includegraphics[width=0.99\columnwidth]{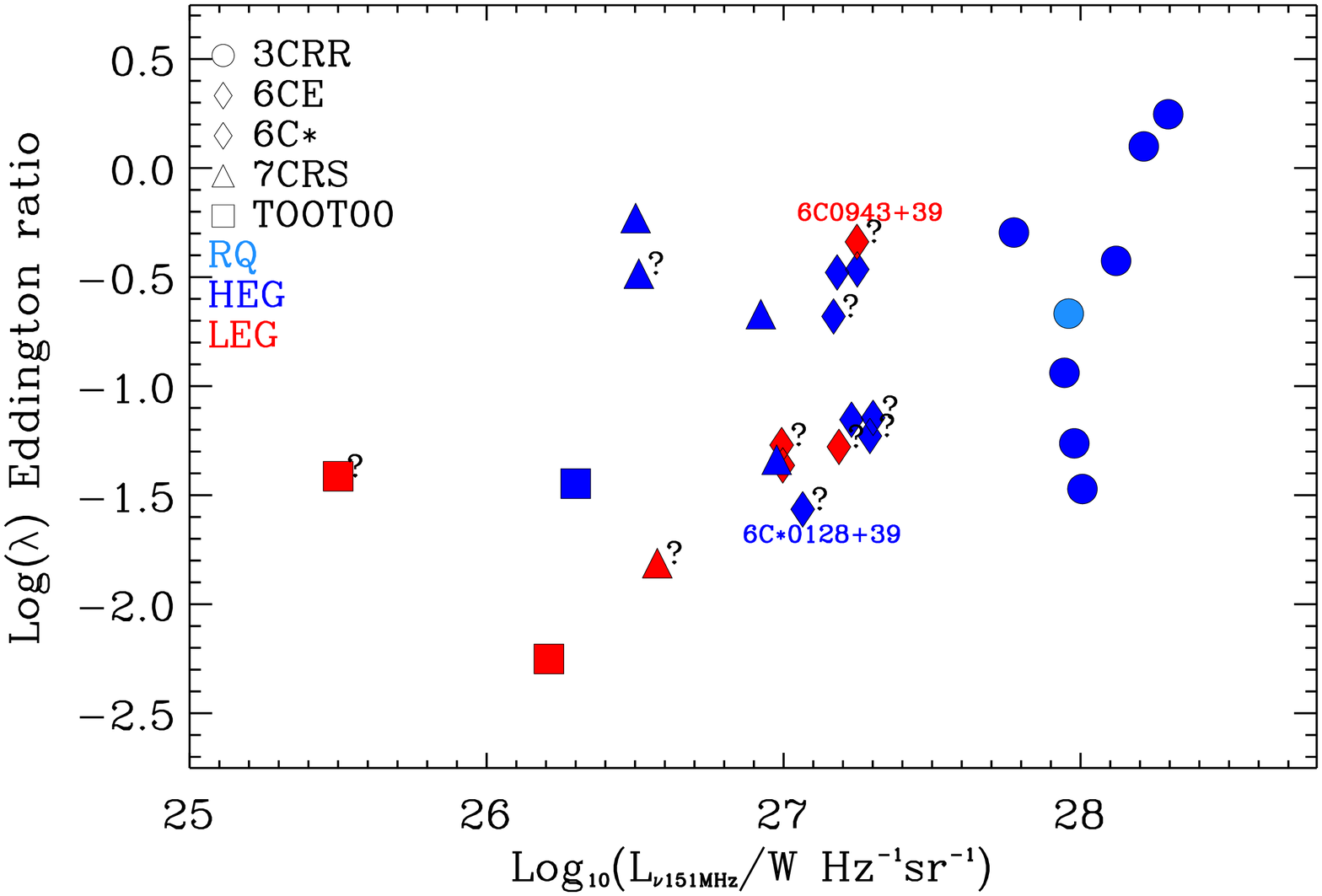}
  \end{center}
  \caption[Relation between $L_{\rm 151MHz}$ and $\lambda$.]{Relation
    between low frequency radio luminosity and Eddington
    ratio. Symbols are: circles for the 3CRR survey objects, diamonds
    for the 6CE and the 6C* surveys, triangles for the 7CRS, and
    squares for the TOOT00 survey objects. Objects coloured in red are
    classified as LEGs, objects in dark blue are HEGs, and those in light
    blue are reddened quasars. `?' denotes objects whose
    classification is uncertain. The two outliers to the general trend
    are identified with their names. Both of them have an uncertain
    classification.}
  \label{fig:l151_eddr}
\end{figure}

First we discuss the relationship between the jet activity and the
accretion rate. In
Figure~\ref{fig:l151_eddr} we show the distribution of the Eddington
ratio versus $\rm 151\,MHz$ radio luminosity, where objects of
distinct optical classifications are displayed in different colours.


The full sample shows a modest positive correlation between $\rm 151\,MHz$
radio luminosity and Eddington ratio. The Spearman coefficient for
this correlation is $\rho=0.42$ and the standard
deviation in the Spearman Rank correlation is $\sigma_{\rho}=0.03$.

Apart from objects 6C0943+39 and 6C*0128+39, which have the most
uncertain classification into HEG/LEGs, there is a clear trend for the
most luminous radio galaxies, with the highest Eddington ratios to be
HEGs, or reddened quasars, and, conversely, for LEGs to have lower
$\lambda$ and $L_{\rm 151MHz}$ values. The complete HEG population in
our sample have a mean Eddington ratio of $\lambda\approx 0.33$
whereas LEGs have a lower mean of $\lambda\approx 0.09$. Excluding the
two outliers that have uncertain classification, 6C0943+39 and
6C*0128+39, though, the mean ratio for HEGs is $\lambda\approx 0.35$
and LEGs is $\lambda\approx 0.03$. We used a Kolmogorov-Smirnov (K-S)
test to evaluate whether the HEGs have a statistically different
distribution in their accretion rate compared to LEGs, and found, for
the full sample, $D=0.66$ with an associated probability of $p=0.01$,
rejecting the null hypothesis that the two populations are drawn from
the same underlying distribution at a 99 per cent level. A
Mann-Whitney test also yields a 97 per cent probability that the null
hypothesis that the two populations HEGs and LEGs come from the same
distribution is not supported by the data. The transition from LEG to
HEG appears to occur around $\lambda\sim 0.04$, which is in excellent
agreement with the theoretical expectations of where accretion rate
becomes radiatively inefficient
\citep[e.g.][]{1982Natur.295...17R,1995ApJ...452..710N,1997ApJ...489..865E}.

The distribution of HEGs and LEGs with respect to radio luminosity
seems indistinguishable up to 151\,MHz luminosities of 
$\sim 3\times 10^{27} \rm \,W\,Hz^{-1}\,sr^{-1}$. For higher luminosities
HEGs fully dominate the sample. Even though the literature generally
states a bias for FRI radio galaxies
to be predominantly LEGs, and FRIIs a mix of both HEGs and LEGs
\citep[e.g.][]{1979MNRAS.188..111H,1994ASPC...54..201L}, there have
been recent studies showing that HEGs and LEGs span a similar range of
radio luminosities \citep[e.g.][]{2012MNRAS.421.1569B} in the local
Universe. If the radio properties of HEGs and LEGs are indeed
indistinguishable, and the HEG and LEG distinction is due to the
accretion process, this means that the
mechanism that generates the jets in both HEGs and LEGs is not
solely related to the accretion rate, and other physical processes
must also play an important role, such as black hole spin.

\subsubsection{Relation between $M_{\rm BH}$ and $\dot{M}$}

From equations~\eqref{eq:lbol} and \eqref{eq:edd}, the black hole mass
and accretion rate can be related through:
\begin{equation}
\dot{M}=\frac{\lambda}{\epsilon c^{2}} 1.3\times 10^{31} \left(
\frac{M_{\rm BH}}{M_{\odot}} \right) \rm \,W.
\end{equation}

\begin{figure}
  \begin{center}
    \includegraphics[width=0.99\columnwidth]{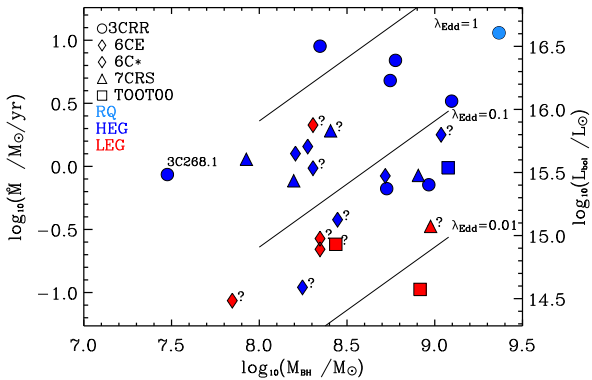}

  \caption[HEGs and LEGs $M_{\rm BH}$ and $\dot{M}$
    distribution.]{Black hole mass and accretion rate distribution for
    HEGs and LEGs. Symbols are as in Figure~\ref{fig:l151_eddr}. The
    right axis shows $L_{\rm bol}$ estimated from $\dot{M}$ using
    equation \eqref{eq:lbol}. Object 3C268.1 is identified as it has
    an unreliable mass estimation (see Section~\ref{sec:ind}).}
  \label{fig:Mbh_Mdot}
  \end{center}
\end{figure}

If we assume the radiative efficiency to be constant, then the slope
of the relationship between black hole mass and accretion rate should
provide information on the Eddington ratio. The Eddington ratio is
thus proportional to the radiative efficiency times $\dot{M}/M_{\rm
  BH}$. The distribution of the inferred accretion rate and bolometric
luminosity versus black hole mass is shown in
Figure~\ref{fig:Mbh_Mdot}.

We observe a general trend for more massive black holes to have higher
accretion rates. The sources in our sample of galaxies have Eddington
ratios of $0.005 < \lambda < 1.7$ with only 3C280 measured to be
accreting above Eddington, albeit with a large uncertainty.

The distribution seen in Eddington ratios could either be due to the
accretion rate, or to the black hole mass. Figure~\ref{fig:Mbh_Mdot}
shows that accretion onto the black hole is the dominant factor for
the separation between the two classes, given the large overlap in
black hole mass. Therefore, acknowledging HEGs and LEGs as being
powered by different modes of accretion - i.e. radiatively efficient
and radiatively inefficient, respectively - the rate at which matter
is being accreted onto the supermassive black hole is determinant for
the accretion mode. Furthermore, the HEGs in our sample have similar
accretion rates to quasars at $z\sim 1$ selected from the SDSS survey
which have $\log_{10}(\lambda)\sim -0.5$ \citep{2004MNRAS.352.1390M},
compared to a mean of $\log_{10}(\lambda) = -0.47^{+0.009}_{-0.03}$ for the HEG radio
galaxies in our sample. This is consistent with unified models in
which radio-loud quasars and HEGs are similar objects viewed
at different orientations \citep[e.g.][]{1989ApJ...336..606B}.


\subsubsection{Host galaxies of HEGs and LEGs}

\begin{figure}
  \begin{center}
    \includegraphics[width=0.99\columnwidth]{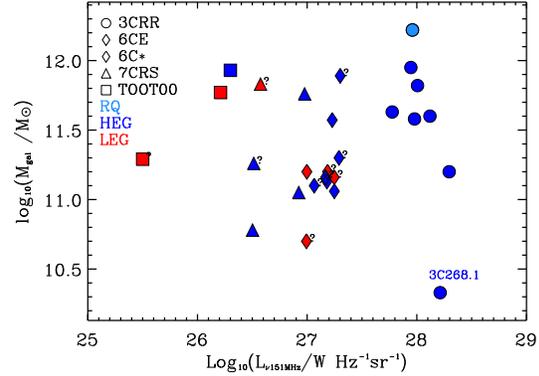}
  \end{center}
  \caption[Host galaxy properties]{Distribution of stellar mass for
    the two populations HEGs and LEGs. Symbols are as in
    Figure~\ref{fig:l151_eddr}. Object 3C268.1 is identified as it has
    an unreliable mass estimation (see Section~\ref{sec:ind}).}
  \label{fig:l151_mgal}
\end{figure}


For a large sample of radio galaxies in the local Universe,
\citet{2012MNRAS.421.1569B} find evidence that HEGs are hosted by less
massive galaxies relative to LEGs. At $z\sim 1$, we do not observe any
clear difference between the host galaxy masses of HEGs and LEGs in
our much smaller sample (see Figure~\ref{fig:l151_mgal}). Given that
our sample was selected to include the most powerful radio sources, we
can expect it to be biased towards the largest masses. 

\citet{2004MNRAS.351..347M} and \citet{2011MNRAS.410.1360H} have found
that for a similar sample of radio galaxies at $z\sim0.5$, HEGs
demonstrate a significant correlation between $M_{\rm BH}$ and $L_{\rm
  151MHz}$. \citet{2004MNRAS.351..347M} also note that the TOOT
objects deviate from the $M_{\rm BH}$ and $L_{\rm 151MHz}$ correlation, and have higher mean galaxy masses than
other subsamples, such as 7CRS and 6CE, which probe higher radio
luminosities. With a larger redshift span, however,
\citet{2007ApJS..171..353S} find only a weak correlation between
stellar mass and radio luminosity.

In our sample we do not find a strong correlation between $M_{\rm
  gal}$ and $L_{\rm 151MHz}$ for the full sample of HEGs. Even
dismissing 3C268.1, which has an unreliable mass estimation (see
Section~\ref{sec:ind}), the correlation between $M_{\rm gal}$ and
$L_{\rm 151MHz}$ has a Spearman rank correlation coefficient of
$\rho=0.39$ and a standard deviation of $\sigma_{\rho}=0.09$, not
highly significant.


\subsubsection{Accounting for the jet power in the accretion rate}

Most studies in the literature define the Eddington ratio as in
section \ref{sec:accretion}, however some studies consider a more
physical approach is to include the contribution of the jet mechanical
energy in the output of the accretion energy, i.e. the total energy
produced due to the accretion process should equal the sum of the
radiative luminosity and the jet mechanical luminosity
\citep[e.g.][]{2007MNRAS.376.1849H,2012MNRAS.421.1569B,2014MNRAS.440..269M}. Considering
the Eddington luminosity as a normalisation factor to compare systems
over a wide range of luminosities, the total Eddington ratio is in
fact an Eddington-scaled accretion rate. The radiative output alone
does not account for the total Eddington accretion, hence neglecting
the contribution of the jet kinetic energy would underestimate the
energy output from the accretion process. This may be particularly
important in LEGs, for which the radiative energy is much lower and
the jet power is an important component of the total energy budget.

\begin{figure}
  \begin{center}
     \includegraphics[width=0.99\columnwidth]{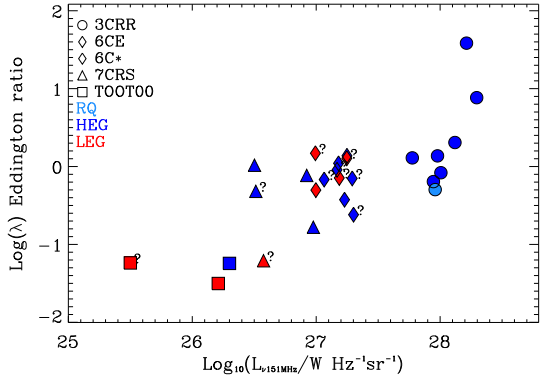} 
  \end{center}
  \caption[L151 vs Eddington ratio]{Relation between low frequency
    radio luminosity and Eddington ratio, including the jet power
    contribution in the Eddington ratio calculation as detailed in the
    text. Symbols are as in Figure~\ref{fig:l151_eddr}.}
  \label{fig:l151_eddr_wQjet}
\end{figure}

Figure~\ref{fig:l151_eddr_wQjet} shows the Eddington ratio
distribution with radio 151\,MHz luminosity, as in
Figure~\ref{fig:l151_eddr}, but including the contribution of $Q_{\rm
  jet}$, i.e.:
\begin{equation}\label{eq:edd2}
\lambda_{\rm rad+mec}\equiv \frac{L_{\rm bol}+Q_{\rm jet}}{L_{\rm Edd}},
\end{equation}
where $\lambda_{\rm rad+mec}$ is the Eddington ratio accounting for
both the radiative energy and the jet mechanical energy. We estimate
the jet power using the relation $Q_{\rm jet}\simeq 3\times
10^{38}f^{3/2}( L_{\nu \rm 151MHz} / 10^{28} )^{6/7} \rm W $
\citep{1999MNRAS.309.1017W}, where $1\leq f\leq 20$ represents several
uncertainties associated with estimating $Q_{\rm jet}$ from $L_{\nu
  151MHz}$. Following Fernandes et al (2011), we chose $f=10$ as this
is the expectation value of a flat prior in natural space.

The Spearman rank test, with a Spearman coefficient of $\rho \sim
0.59$ and $\sigma_{\rho}=0.001$, reflects a tighter correlation
between 151\,MHz radio luminosity and Eddington ratio with the
inclusion of $Q_{\rm jet}$ than when accounting solely for the $L_{\rm
  bol}$.

The separation between HEGs and LEGs is not as clear as when we
consider the Eddington scaled accretion rate to be solely dictated by
the radiated luminosity $L_{\rm bol}$. Indeed, the K-S test now shows
a much lower $64\%$ probability that the two populations come from
distinct distributions ($p=0.36$ and $D=0.38$). Given that our
sample contains the strongest radio galaxies at $z\sim 1$, the jet power
is not negligible when compared to the radiative energy. 

\begin{figure}
  \begin{center}
    \includegraphics[width=0.99\columnwidth]{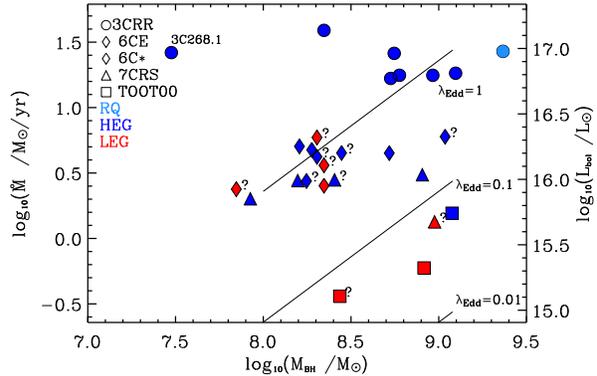}
  \caption[HEGs and LEGs $M_{\rm BH}$ and $\dot{M}$
    distribution.]{Black hole mass and accretion rate distribution for
    HEGs and LEGs, including the jet power contribution in the
    accretion rate calculation. Symbols are as in
    Figure~\ref{fig:l151_eddr}. The right axis shows $L_{\rm bol}$
    estimated from $\dot{M}$ using equation \eqref{eq:lbol}.}
  \label{fig:Mbh_Mdot_wQjet}
  \end{center}
\end{figure}

Figure~\ref{fig:Mbh_Mdot_wQjet} shows the distribution of accretion
rate with black hole mass, where the contribution of the jet
mechanical energy is taken into account in the calculation of the accretion
rate. The separation between HEGs and LEGs is not as clear as when
considering only the radiated energy for the calculation of the
accretion rate, however, it is still clear that HEGs have on average
higher accretion rates than LEGs. The inclusion of the contribution of
$Q_{\rm jet}$ has given rise to a transition range between
$0.1\lesssim \log_{10}(\dot{M}/\,M_{\odot}yr^{-1})\lesssim 0.8$, with
both HEGs and LEGs found at these accretion rates. Only LEGs are found
at lower accretion rates $\log_{10}(\dot{M}/\,M_{\odot}yr^{-1})<0.1$,
and only HEGs above $\log_{10}(\dot{M}/\,M_{\odot}yr^{-1})>0.8$. 

As expected, an Eddington ratio that only takes into account the
radiated luminosity to balance the gravitational pull shows a more
clear separation between HEGs and LEGs. Due to the fact that the
sources in our sample span a wide range in radio luminosities, and
include some of the strongest radio sources at $z\sim 1$, it is
expected for the jet mechanical energy of these sources to
significantly contribute to the energy balance. This is particularly
true for LEGs, where the radiated emission is weaker. 

\begin{figure}
  \begin{center}
    \includegraphics[width=0.99\columnwidth]{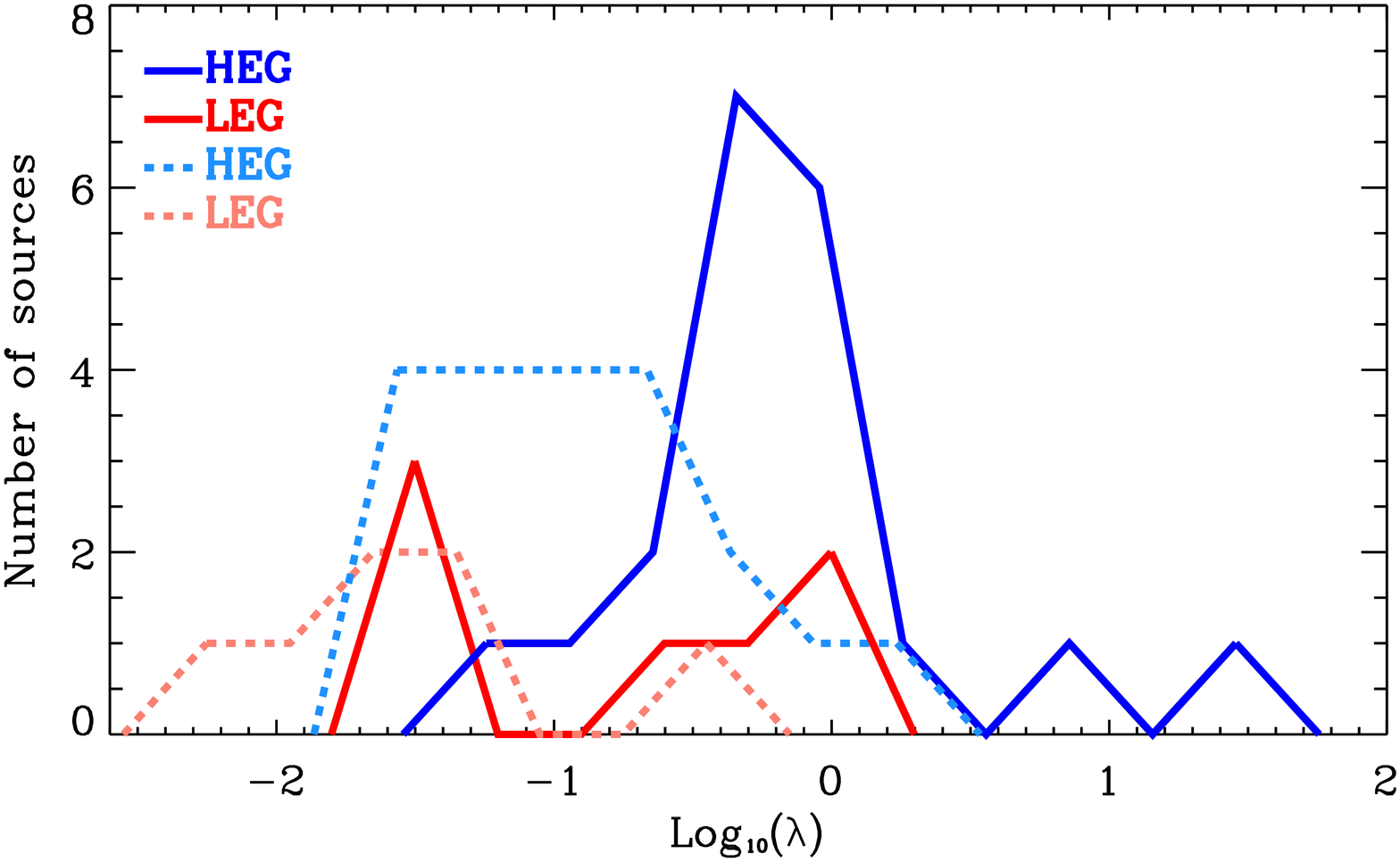}
  \caption[Eddington ratio histogram]{Eddington ratio distribution of
    HEGs and LEGs. Solid lines are for the Eddington ratio estimated
    using $\lambda=(L_{\rm bol}+Q_{\rm jet})/L_{\rm Edd}$; dashed
    lines are for the Eddington ratio estimated using
    $\lambda=L_{\rm bol}/L_{\rm Edd}$.}
  \label{fig:eddr_histo_wQjet}
  \end{center}
\end{figure}

Figure~\ref{fig:eddr_histo_wQjet} shows the Eddington ratio
distribution of HEGs and LEGs obtained with both methods of calculating
the Eddington ratio. Solid lines are for $\lambda=(L_{\rm bol}+Q_{\rm
  jet})/L_{\rm Edd}$ and dashed lines for $\lambda=L_{\rm bol}/L_{\rm
  Edd}$. The total accretion energy for both HEGs and LEGs is significantly
increased, however, the trend for HEGs to show higher Eddington ratios than
LEGs remains.

\section{Conclusions}\label{sec:conc}

We have used {\em Spitze}r $24\mu\rm m$ MIPS and $3.6$, $4.5$, $5.8$,
and $8\mu\rm m$ IRAC data, as well as near-IR and optical data from
the literature, to constrain the SEDs of the sample of radio
galaxies previously studied by \citet{2011MNRAS.411.1909F}. We applied
a dust extinction law to a quasar template to approximate the mid-IR
region of the SED, and a template of an elliptical galaxy.

From these fits we are able to determine the black-hole masses through
$M_{\rm bulge}-M_{\rm BH}$ relation, the quasar bolometric luminosity,
the accretion rate and the Eddington ratio. Our analysis revealed
the following:

\begin{enumerate}

\item We find a significant correlation between the Eddington ratio and
  the radio luminosity, with more luminous radio sources yielding
  higher Eddington ratios, and thus higher accretion rates.


\item We find that HEGs tend to have higher Eddington ratios and radio
  luminosities, whereas the LEGs in our sample predominantly gather
  towards the lowest values of $\lambda$ and $L_{\rm 151MHz}$. We find
  the HEG/LEG division to lie approximately at $\lambda\sim 0.04$,
  which is in excellent agreement with theoretical predictions of
  where the accretion rate becomes radiatively inefficient and also
  with other studies
  \citep{2012MNRAS.421.1569B,2014MNRAS.440..269M}. This result further
  confirms the suspicion that the differences between HEGs and LEGs
  arise due to the different modes of accretion, which result in
  different accretion rates. Higher accretion rates are observed with
  the radiatively efficient `cold mode' accretion, whereas lower
  accretion rates can be explained with the radiatively inefficient
  `hot mode' of accretion, such as ADAFs.


\item By including the contribution of the jet power to the Eddington
  ratio instead of solely accounting for the radiated power, the HEG/LEG dichotomy becomes less clear. Indeed, the
  contribution of the jet power is expected to have a significant
  weight for LEGs, which have a weaker radiated power. Moreover, given
  that our sample consists of some of the most powerful radio sources,
  the jet power contribution appreciably increases the Eddington
  ratios of HEGs as well. The trend for HEGs to have higher accretion
  rates than LEGs remains.

\item We find that the more massive black holes have higher
  accretion rates for the HEGs population. HEGs and LEGs are similarly
  distributed in terms of black hole mass. This relation further
  displays the dichotomy between HEGs and LEGs in accretion rate,
  meaning that the dichotomy seen in Eddington ratio is due to the
  accretion rate and not a dichotomy in black hole mass. 


\item We do not find a strong correlation between the host galaxy
  mass, or equivalently the black hole mass, and the radio luminosity
  of the sample. Similarly to \citet{2007ApJS..171..353S}, there is
  only a slight trend for the highest mass galaxies to host the most
  powerful radio sources.


\item Recent studies in the local Universe find evidence that LEGs are
  hosted by more massive galaxies than HEGs. However, at $z\sim 1$ we
  do not observe any difference in terms of host galaxy mass
  distribution for HEGs and LEGs, although we note that our sample is
  much smaller than those used in the local Universe.

\item It is important to ascertain whether the radio properties of
  HEGs and LEGs are indistinguishable or not. Our study hints that, at
  least up to radio luminosities of $\sim 3\times 10^{27} \rm
  \,W\,Hz^{-1}\,sr^{-1}$, HEGs and LEGs span the same range of radio
  luminosities. If HEGs and LEGs' radio properties are indeed
  indistinguishable, and the HEG and LEG distinction is due to the
  accretion process, this means that the power being channeled into
  the jets is most likely not solely dependent on the accretion rate and an
  additional process must be influencing the jet
  formation. Understanding the HEG/LEG dichotomy may thus bring
  important new constraints on the relation between accretion process
  and jet formation.

\end{enumerate}


\section*{Acknowledgments}
We thank the anonymous referee for providing constructive comments and
raising questions that have helped improving the contents of this
paper and their presentation. CACF is currently supported by CNPq
(through PCI-DA grant 302388/2013-3 associated with the PCI/MCT/ON
program) and gratefully acknowledges past financial support from the
Foundation for Science and Technology (FCT Portugal) through project
grant PTDC/FIS/100170/2008 and doctoral grant SFRH/BD/30486/2006. MJJ
acknowledges the continued support of the South African SKA project
and the NRF. A.M.-S. gratefully acknowledges a Post-Doctoral
Fellowship from the United Kingdom Science and Technology Facilities
Council, reference ST/G004420/1 as well as support from SEPnet. This
work is based (in part) on observations made with the Spitzer Space
Telescope, which is operated by the Jet Propulsion Laboratory,
California Institute of Technology under a contract with NASA. This
research has made use of the NASA/IPAC Extragalactic Database (NED)
which is operated by the Jet Propulsion Laboratory, California
Institute of Technology, under contract with the National Aeronautics
and Space Administration.

{}

\appendix

\begin{table*}
  \caption{J, H, and K band magnitudes found in the
  literature. ``aper. corr'', ``line cont.'', and ``Gal. ext.'' stand
  for the magnitude values after being corrected for aperture,
  emission-line contamination, and Galactic extinction,
  respectively. An error of $25\%$ was assumed for each magnitude value
  subjected to any of these corrections. The references to the
  literature magnitudes are: B97 - \citealt{1997MNRAS.292..758B}; DP93 -
  \citealt{1993MNRAS.263..936D}; E97 - \citealt{1997MNRAS.291..593E};
  I03 - \citealt{2003MNRAS.345.1365I}; I06 -
  \citealt{2006MNRAS.367..693I}; J01 - \citealt{2001MNRAS.326.1585J};
  L84 - \citealt{1984MNRAS.211..833L}; Leb81 -
  \citealt{1981ApJ...245L..59L}; LRL83 -
  \citealt{1983MNRAS.204..151L}; W98 - \citealt{1998MNRAS.300..625W};
  W03 - \citealt{2003MNRAS.339..173W}; V10 -
  \citealt{2010MNRAS.401.1709V} }\label{table:JHK}\centering
\begin{tabular}{l l c c c c c c}
\noalign{\smallskip}
\hline\hline
\noalign{\smallskip}
Object &  correction & J  & Ref  & H  & Ref  & K & Ref\\[1.2ex]
\noalign{\smallskip}
\hline
\noalign{\smallskip}
3C280 & literature  & 18.07 $\pm$ 0.05 & B97 & 18.10 $\pm$ 0.15 (4)& I06 & 16.70 $\pm$ 0.04 & B97\\
      & aper. corr. &                  &     & 17.84          &    &        &   \\
      & line conta. & 18.36            &    &      &          &      & \\ [1.2ex]
3C356 & literature & - & - & 17.35 $\pm$ 0.13 & L84 & 16.75 $\pm$ 0.13 & L84\\[1.2ex]
3C184 & literature & - & - & 17.87 $\pm$ 0.12 & Leb81 & 16.80 $\pm$ 0.10 & Leb81\\
      & Gal. ext. &  &   & 17.85         &    & 16.79                 &   \\[1.2ex]
3C175.1 & literature & - & - &  - & - & 17.56 $\pm$ 0.30 (5)& DP93 \\
        & aper. corr.&   &   &   &   & 17.35                &    \\
        & Gal. ext.&   &   &   &   & 17.32                  &    \\[1.2ex]
3C22 & literature & 17.53 $\pm$ 0.08 (4) & I06 & 16.95 $\pm$ 0.09 (4) & I06 & 15.66 $\pm$ 0.05 (4) & I06\\
     & aper. corr.& 17.32                &       & 16.79              &   & 15.62                  &   \\
     & line corr. & 17.45                &       &                      &                            &   \\[1.2ex]
3C289 & literature & 18.19 $\pm$ 0.06 & B97 & 17.25 $\pm$ 0.33 & L84 & 16.66 $\pm$ 0.07 & B97\\[1.2ex]
      & line cont. & 18.23            &    &              &         &        & \\[1.2ex]
6CE1256+3648 & literature & 19.11 $\pm$ 0.16 & I03  & 18.11 $\pm$ 0.14 & I03 & 17.45 $\pm$ 0.07 & I03\\*
             & line cont. & 19.18           &    & \\[1.2ex]
6CE1217+3645 & literature & 19.26 $\pm$ 0.16 & I03 & 17.95 $\pm$ 0.13 & I03 & 17.22 $\pm$ 0.08 & I03\\*
             & line cont. & 19.28 &    & \\[1.2ex]
6CE1017+3712 & literature & 19.35 $\pm$ 0.2 & I03 & 18.98 $\pm$ 0.19 & I03 & 18.31 $\pm$ 0.08 & I03\\*
             & line cont. & 19.65 &    & \\[1.2ex]
6CE0943+3958 & literature & 19.14 $\pm$ 0.12 & I03 & 18.64 $\pm$ 0.17 & I03 & 18.03 $\pm$ 0.11 & I03\\*
             & line cont. & 19.31   &    & \\[1.2ex]
6CE1257+3633 & literature & 19.15 $\pm$ 0.17 & I03 & 18.04 $\pm$ 0.17 & I03 & 17.17 $\pm$ 0.05 & I03\\*
             & line cont. & 19.22 $\pm$ 1.92 &    & \\[1.2ex]
6CE1019+3924 & literature  & 18.00 $\pm$ 0.06 & I03 & 17.37 $\pm$ 0.09 & I03 & 16.41 $\pm$ 0.04 & I03\\*
             & line cont.  & 18.03  &     & \\[1.2ex]
6CE1011+3632 & literature  & 19.59 $\pm$ 0.28 & I03 & 18.6 $\pm$ 0.24 & I03 & 17.67 $\pm$ 0.09 & I03\\*
             & line cont.  & 19.65  &    \\[1.2ex]
6CE1129+3710 & literature  & 19.38 $\pm$ 0.2 & I03 & 18.24 $\pm$ 0.23 & I03 & 17.63 $\pm$ 0.11 & I03\\*
             & line cont.  & 19.61  &    \\[1.2ex]
6C*0128+394 & literature  & - & - & - & - & 17.72 $\pm$ 0.22 & J01\\[1.2ex]
6CE1212+3805 & literature  & - & - & - & - & 17.35 $\pm$ 0.08  & E97 \\[1.2ex]
6C*0133+486 & literature  & - & - & - & - & 18.69 $\pm$ 0.27 (5) & J01\\*
            & aper. corr. &   &   &   &   & 18.51 & \\[1.2ex]
5C6.24 & literature  & - & - & - & - & 17.25 $\pm$ 0.13 & W03\\[1.2ex]
5C7.23 & literature  & - & - & - & - & 17.98 $\pm$ 0.09 & W03\\[1.2ex]
5C7.82 & literature  & - & - & - & - & 17.01 $\pm$ 0.09 & W03\\[1.2ex]
5C7.242 & literature  & - & - & - & - & 17.24 $\pm$ 0.09 & W03\\[1.2ex]
5C7.17 & literature  &  - & - & - & - & 17.37 $\pm$ 0.07 (5) & W98\\*
       & aper. corr. &    &   &   &   & 17.20  & \\[1.2ex]
TOOT1267 & literature  & 18.90 $\pm$ 0.13 & V10 & 18.00 $\pm$ 0.10 & V10 & 16.90 $\pm$ 0.05 & V10\\*
         & line cont.  & 18.92  & \\[1.2ex]
TOOT1140 & literature  & 18.70 $\pm$ 0.12 & V10 & 17.8 $\pm$ 0.09 & V10 & 16.7 $\pm$ 0.04 & V10\\*
         & line cont.  & 18.70 & \\[1.2ex]
TOOT1066 & literature  & 19.30 $\pm$ 0.15 & V10 & 18.30 $\pm$ 0.09 & V10 & 17.60 $\pm$ 0.07 & V10\\*
         & line cont.  & 19.34 &     & \\[1.2ex]
\noalign{\smallskip}
\hline
\end{tabular}
\end{table*}

\begin{table*}
  \caption{HST F606W, F702W and F814W-band magnitudes found in the
  literature. ``aper. corr'' and ``line cont.'' stand for the
  magnitude values after being corrected for aperture, and
  emission-line contamination, respectively. An error of $25\%$ was
  assumed for each magnitude value subjected to any of these
  corrections. The references to the literature magnitudes
  are: B97~-~\citealt{1997MNRAS.292..758B}; I03~-~
  \citealt{2003MNRAS.345.1365I}; I06~-~
  \citealt{2006MNRAS.367..693I}}\label{table:HST}\centering
\begin{tabular}{l l c c c c c c}
\noalign{\smallskip}
\hline\hline
\noalign{\smallskip}
Object & correction & F606W & Ref & F702W & Ref & F814W  & Ref\\[1.2ex]
\noalign{\smallskip}
\hline
\noalign{\smallskip}
  3C280 & literature & - & - & 20.92 $\pm$ 0.06 (4)& I06 & 19.78 $\pm$0.04 & B97 \\
        & aper. corr.&   &   & 20.41   &    &              & \\
        & line conta. &  &   & 20.53    &    &           &     \\[1.2ex]
  3C356 & literature & - &  - & 21.22$\pm$0.09 & I06 & 20.60$\pm$0.03 & I06\\
        & aper. corr. &  &    & 20.68 &    & 20.22 &   \\
        & line cont. &   &   &  20.73 &           \\[1.2ex]
  3C22 & literature & - & - & 20.00 $\pm$ 0.15 (4)& I06  & 19.95$\pm$0.03 & B97\\
       & aper. corr. &   &  & 19.57     &  &                &    \\
       & line cont.  &   &  & 19.62   &    &                &   \\[1.2ex] 
  3C289 & literature & - & - & 21.49 $\pm$ 0.07 (4) & I06 & 20.02 $\pm$ 0.13 & B97\\
        & aper. corr. & &   & 20.92  &       &      &   \\
        & line cont. &   &  & 20.99  &       &      &   \\[1.2ex]
  6CE1256+3648 & literature & - & -  & 22.23 $\pm$ 0.16 & I03 & 22.60 $\pm$ 0.16 (4) & I06\\
               & aper. corr. &   &   &                  &      & 21.94  &    \\
               & line cont.  &   &   & 22.29  &    & \\[1.2ex]
  6CE1217+3645 & literature  & 22.26 $\pm$ 0.32 & I03 & 21.91$\pm$0.25 & I06    & 20.59 $\pm$ 0.12 & I03\\
               & aper. corr. &                  &     & 21.31  &    & \\
               & line cont.  & 22.26  &   & 21.32  &    & \\[1.2ex]
  6CE1017+3712 & literature  & - & - & 21.53 $\pm$ 0.16 & I03 & 21.10$\pm$0.14 (4) & I06\\*
               & aper. corr. &   &   &                  &     & 20.57  &   \\*
               & line cont.  &   &   & 21.64  &    & \\[1.2ex]
  6CE0943+3958 & literature  & - & - & 21.78 $\pm$ 0.38 & I03 & 21.55 $\pm$ 0.20 (4)& I06\\*
               & aper. corr. &   &  &                     &    & 20.98  &   \\*
               & line cont. &    &   & 21.88  &    & \\[1.2ex]
  6CE1257+3633 & literature  & 22.04 $\pm$ 0.26 & I03 & 21.55 $\pm$ 0.30 (4) & I06 & 20.55 $\pm$ 0.12 & I03\\*
               & aper. corr. &                 &       & 20.98  &       & \\*
               & line cont.  & 22.05  &    & 21.00  &    & \\[1.2ex]
  6CE1019+3924 & literature  & 21.89 $\pm$ 0.25 & I03 & 21.04 $\pm$ 0.29 (4)& I06 & 19.86 $\pm$ 0.1 & I03\\*
               & aper. corr. &                  &     & 20.52  &   \\*
               & line cont.  & 21.90  &     & 20.53  & \\[1.2ex]
  6CE1011+3632 & literature  & -  & - & 21.59 $\pm$ 0.19 & I03 & 21.23 $\pm$ 0.15 (4) & I06\\*
               & aper. corr. &    &   &               &     & 20.69  & \\*
               & line cont.  &     &   & 21.61  & \\[1.2ex]
  6CE1129+3710 & literature  & - & - & 21.52 $\pm$ 0.14 & I03 & 21.64 $\pm$ 0.16 (4)  & I06\\*
               & aper. corr. &   &   &                  &     & 21.06  & \\*
               & line cont.  &   &   & 21.60  & \\[1.2ex]
\noalign{\smallskip}
\hline
\end{tabular}
\end{table*}

\begin{table*}
  \caption{U, B, V, R and I band magnitudes found in the
    literature. ``aper. corr'', ``line cont.'', and ``Gal. ext.''
    stand for the magnitude values after being corrected for aperture,
    emission-line contamination, and Galactic extinction,
    respectively. An error of $25\%$ was assumed for each magnitude
    value subjected to any of these corrections. The references to the
    literature magnitudes are: J01 - \citealt{2001MNRAS.326.1585J};
    LRL83 - \citealt{1983MNRAS.204..151L}; W98 -
    \citealt{1998MNRAS.300..625W}; V10 - \citealt{2010MNRAS.401.1709V}}\label{table:UBVRI}\centering
\setlength{\tabcolsep}{3pt}
\begin{tabular}{l l c c c c c c c c c c}
\noalign{\smallskip}
\hline\hline
\noalign{\smallskip}
Object &  correction & U  & Ref & B  & Ref & V & Ref & R  & Ref & I & Ref \\[1.2ex]
\noalign{\smallskip}
\hline
\noalign{\smallskip}
  3C280 & literature & - & - & - & - & 22.00$\pm$ 1.00 & LRL83 & 21.50$\pm$0.50 & LRL83 & - & - \\
        & line cont. &   &   &   &   & 22.12 &      & 21.76 &     \\
        & Gal. ext.  &   &   &   &   & 22.08 &    &   21.72 &     \\[1.2ex]
  3C356 & literature & - & - & - & - & - & - & 21.50$\pm$1.00 & LRL83 & - & - \\
        & line cont. &   &   &   &   &   &   & 21.59 &     &   & \\
        & Gal.ext.   &   &   &   &   &   &   & 21.50 &    &   & \\[1.2ex]
  3C184 & literature & - & - & - & - & - & - & 22.00$\pm$0.50 & LRL83 & - & - \\
        & line cont. &   &   &   &   &   &   & 22.30 &    &  & \\
        & Gal. ext.  &   &   &   &   &   &   & 22.22 &    &  & \\[1.2ex]
  3C175.1 & literature & - & - & - & - & - & - & 22.00$\pm$0.50 & LRL83  & - & -\\
          & line cont. &   &   &   &  &   &    & 22.20 &    &  & \\
          & Gal. ext.  &   &    &   &  &  &   &  21.96 &    & & \\[1.2ex]
  3C22 & literature & - & -  & - & - & - & - & 20.50$\pm$ 1.00 & LRL83 & - & - \\
        & line cont. &   &   &   &    &   &   & 20.59 &    & \\
        & Gal. Ext. &    &   &   &   &    &   & 19.98 &    & \\[1.2ex]
  3C289 & literature & - & - & - & - & - & - & 23.00$\pm$0.50 & LRL83 & - & - \\*
        & line cont. &   &   &   &   &   &   & 23.42 &     &   & \\*
        & Gal. ext.  &   &   &   &   &   &   & 23.39 &     &   & \\[1.2ex]
  3C343 & literature & - & - & 21.06$\pm$0.50 & LRL83 & 20.61$\pm$0.50 & LRL83 & 19.55$\pm$0.50 & LRL83 & - & -\\*
        & line cont. &   &   & 21.06 &     & 20.61 &     & 19.55 &    & \\*
        & Gal. ext.  &   &   & 20.97 &     & 20.54 &     & 19.50 &    & \\[1.2ex]
  6C*0133+486 & literature  & - & - & - & - & - & - & 23.39 (5) & J01 & - & -\\*
              & aper. corr.  &   &   &   &   &   &   & 22.66  & \\*
              & line cont.  &   &   &   &   &   &   & 22.72  & \\[1.2ex]
  5C7.17 & literature  & - & - & 22.08 $\pm$ 0.17 (5) & W98 & - & - & 21.59 $\pm$ 0.14 & W98 & - & -\\*
         & aper. corr. &   &   & 21.88     &     &   &   & 21.40  & \\*
         & line cont.  &   &   & 21.89    &     &   &   & 21.44  & \\[1.2ex]
TOOT1267 & literature  & 23.50 $\pm$ 0.23 & V10 & 24.60 $\pm$ 0.23 & V10 & 24.00 $\pm$ 0.23 & V10 & 23.00 $\pm$ 0.23 & V10 & 21.70 $\pm$ 0.23 & V10\\*
         & line cont.  &                  &     & 24.64  &     & 24.03 &     & 23.04 & \\*
         & Gal. ext.   & 22.20  &     & 23.60  &     & 23.24  &     & 22.40  &     & 21.48  &   \\[1.2ex]
TOOT1140 & literature  & - & - & - & - & 25.20 $\pm$ 0.38 & V10 & 22.80 $\pm$ 0.38 & V10 & 21.00 $\pm$ 0.38 & V10\\*
         & line cont.  &   &   &   &   & 25.22  &     & 22.81  &     & \\*
         & Gal. ext.   &   &   &   &   & 24.93  &     & 22.58  &     & 20.83  & \\[1.2ex]
TOOT1066 & literature  & - & - & 25.30 $\pm$ 0.29 & V10 & 24.60 $\pm$ 0.05 & V10 & 23.50 $\pm$ 0.05 & V10 & 22.50 $\pm$ 0.05 & -\\*
         & line cont.  &   &   & 25.39  &     & 24.68  &     & 23.58  &     &     \\*
         & Gal. ext.   &   &   & 24.38  &     & 23.91  &     & 22.96  &     & 22.05  & \\[1.2ex]
\noalign{\smallskip}
\hline
\end{tabular}
\end{table*}

\begin{table*}
\caption{\small{u, g, r, i, z band magnitudes found in the literature. ``line
  cont.'' and ``Gal. ext.'' stand for the magnitude values after being
  corrected for emission-line contamination and Galactic extinction,
  respectively. An error of $25\%$ was assumed for each magnitude
  value subjected to any of these corrections. The references to the
  literature magnitudes are: L84 - \citealt{1984MNRAS.211..833L}; SDSS -
  http://www.sdss.org/DR7/access/index.html}}\label{table:ugriz}\centering
\setlength{\tabcolsep}{4pt}
\begin{tabular}{l l c c c c c c c c c c}
\noalign{\smallskip}
\hline\hline
\noalign{\smallskip}
Object & correction &  u  & Ref & g & Ref & r  & Ref & i & Ref & z & Ref \\[1.2ex]
\noalign{\smallskip}
\hline
\noalign{\smallskip}
3C280 & literature & - & - & - & - & 21.57 $\pm$ 0.07 & L84 & - & - & - & -\\[1.2ex]
3C356 & literature & - & - & - & - & 21.84 $\pm$ 0.22 & L84 & - &  - & - & -\\[1.2ex]
3C184 & literature & - & - & - & - & 21.94 $\pm$ 0.14 & L84 & 21.14 $\pm$ 0.3 & L84 & -  & -\\[1.2ex]
3C175.1 & literature  & - & - & - & - & 21.59 $\pm$ 0.15 & L84 & - & - & - & -\\[1.2ex]
3C289 & literature& 23.0 $\pm$ 0.92 & SDSS & 23.07 $\pm$ 0.32 & SDSS & 21.65 $\pm$ 0.14 & SDSS & 20.86 $\pm$ 0.1 & SDSS & 20.14 $\pm$ 0.2 & SDSS\\
      & line cont.&                &      & 23.19  &    &   \\
      & Gal. ext. & 22.93 &    & 23.15  &    & 21.62  &    & 20.83 &    & 20.12  &   \\[1.2ex]
6CE1217+3645 & literature  & 23.32 $\pm$ 0.64 & SDSS & 22.77 $\pm$ 0.19 & SDSS & 22.08 $\pm$ 0.12 & SDSS & 21.84 $\pm$ 0.15 & SDSS & 20.77 $\pm$ 0.19 & SDSS\\
             & line cont.  &                  &      & 22.78  &    & \\
             & Gal. ext.   & 23.23  &    & 22.71  &    & 22.03  &    & 21.80  &    & 20.74  &    \\[1.2ex]
6CE1019+3924 & literature  & 22.66 $\pm$ 0.51 & SDSS & 22.37 $\pm$ 0.40 & SDSS & 21.78 $\pm$ 0.15 & SDSS & 20.37 $\pm$ 0.06 & SDSS & 19.54 $\pm$ 0.09 & SDSS\\
             & line cont.  &                  &      & 22.39  &      &\\
             & Gal. ext.   & 22.58  &      & 22.34  &      & 21.74  &      & 20.34  &      & 19.51  &   \\[1.2ex]
6CE1011+3632 & literature  & 22.37 $\pm$ 0.31 & SDSS & 23.24 $\pm$ 0.23 & SDSS & 22.25 $\pm$ 0.16 & SDSS & 22.14 $\pm$ 0.26 & SDSS & 21.54 $\pm$ 0.46 & SDSS\\
             & line cont.  &                  &      & 23.26 & \\
             & Gal. Ext.   & 22.30 &      & 23.21  &      & 22.21 &      & 22.11  &      & 21.51  & \\[1.2ex]    
\noalign{\smallskip}
\hline
\end{tabular}
\end{table*}

\clearpage

\end{document}